\documentclass[aps,prd,reprint,preprintnumbers,showpacs,floatfix,nofootinbib,superscript address,longbibliography]{revtex4-1}
\pdfoutput=1

\usepackage[utf8x]{inputenc}
\usepackage{scalerel,stackengine}
\usepackage{titlesec}
\usepackage{parskip}
\usepackage{stix}
\usepackage{amssymb}
\usepackage{amsmath}
\usepackage{amsfonts}
\usepackage{hhline}
\usepackage{mathtools}
\usepackage[dvipsnames]{xcolor}

\usepackage{pifont}
\newcommand{\cmark}{\hspace{5pt}\color{black}\ding{51}\hspace{5pt}}%
\newcommand{\xmark}{\hspace{5pt}\color{black}\ding{55}\hspace{5pt}}%
\newcommand{\UnitaryTensor}{\color{SeaGreen}$\mathbf{2}$\color{black}}%
\newcommand{\UnitaryVector}{\color{SeaGreen}$\mathbf{3}$\color{black}}%
\newcommand{\GhostVector}{\color{red}$\mathbf{3_{\mathrm{\mathbf{g}}}}$\color{black}}%
\newcommand{\UnitaryStrongVector}{\color{red}$\mathbf{3}_{\mathrm{\mathbf{s}}}$\color{black}}%
\newcommand{\GhostStrongVector}{\color{red}$\mathbf{3}_{\mathrm{\mathbf{g,s}}}$\color{black}}%

\usepackage{xspace}
\usepackage{multirow,tabularx}
\usepackage{siunitx}
\usepackage{multirow}
\usepackage{graphicx}
\usepackage{xstring}
\usepackage{etoolbox}
\usepackage{notoccite}
\usepackage{natbib}
\usepackage{mathrsfs}
\usepackage{lineno}
\usepackage{tensor}
\usepackage{accents}
\usepackage{tikz}

\allowdisplaybreaks

\newcommand*{\ie}{i.e., }
\newcommand*{\eg}{e.g., }

\newcommand*{\eq}{eq.\@\xspace}

\newcommand*{\wrt}{w.r.t.\@\xspace}
\newcommand*{\dof}{d.o.f.\@\xspace}

\newrobustcmd{\pea}[1]{%
	\emph{#1}\textbf{\ \ \ ---}
}
\titleformat{\paragraph}[runin]{\normalfont\normalsize\bfseries}{\emph\theparagraph}{1em}{\pea}

\newrobustcmd{\alp}[1]{%
       {\alpha}	
}

\newrobustcmd{\bet}[1]{%
  \tensor[^{(#1)}]{\mu}{}
}

\stackMath 
\newcommand\OmitIndices[1]{%
\savestack{\tmpbox}{\stretchto{%
\scaleto{%
\scalerel*[\widthof{\ensuremath{#1}}]{\kern-.6pt\curlywedge\kern-.6pt}%
{\rule[-\textheight/2]{1ex}{\textheight}}
}{\textheight}%
}{0.5ex}}%
\stackon[1pt]{#1}{\tmpbox}%
}

\newrobustcmd{\LPV}[1]{%
	\IfEqCase{#1}{%
	{}{L_{\text{PV}}}%
	}%
	[{L_{\text{PV}}\left(#1\right)}]%
}

\newrobustcmd{\F}[2][placeholder]{%
	\IfEqCase{#1}{%
	{placeholder}{\tensor{T}{#2}}%
	{1}{\tensor[^{(1)}]{F}{#2}}%
	{2}{\tensor[^{(2)}]{F}{#2}}%
	{3}{\tensor[^{(3)}]{F}{#2}}%
	}%
	[\packageError{cosmicclass}{Symbol #1 is not an irreducible part!}{}]%
}

\newrobustcmd{\GenericVector}[1]{%
	\smash{{#1}_{\tensor[^{{(2)}}]{\hspace{-1pt}\lambda}{}}^{J^P}}%
}
\newrobustcmd{\GenericTensor}[1]{%
	\smash{{#1}_{\tensor[^{{(1)}}]{\hspace{-1pt}\lambda}{}}^{J^P}}%
}

\newrobustcmd{\qe}[1]{%
	\tensor{q}{_e}%
}
\newrobustcmd{\dqe}[1]{%
	\tensor{\dot{q}}{_e}%
}
\newrobustcmd{\qw}[1]{%
	\tensor{q}{_\omega}%
}
\newrobustcmd{\dqw}[1]{%
	\tensor{\dot{q}}{_\omega}%
}

\newrobustcmd{\chie}[1]{%
	\tensor{\chi}{_e}%
}
\newrobustcmd{\chiw}[1]{%
	\tensor{\chi}{_\omega}%
}
\newrobustcmd{\phil}[1]{%
	\tensor{\phi}{_\lambda}%
}
\newrobustcmd{\phie}[1]{%
	\tensor{\phi}{_e}%
}
\newrobustcmd{\phiw}[1]{%
	\tensor{\phi}{_\omega}%
}
\newrobustcmd{\pil}[1]{%
	\tensor{p}{_\lambda}%
}
\newrobustcmd{\pie}[1]{%
	\tensor{p}{_e}%
}
\newrobustcmd{\piw}[1]{%
	\tensor{p}{_\omega}%
}

\newrobustcmd{\g}[1]{%
	\tensor{g}{#1}%
}
\newrobustcmd{\rcCon}[1]{%
	\tensor{\Gamma}{#1}%
}
\newrobustcmd{\rCon}[1]{%
	\tensor*{\mathring{\Gamma}}{#1}%
}
\newrobustcmd{\B}[1]{%
	\tensor{B}{#1}%
}
\newrobustcmd{\PD}[1]{%
	\tensor{\partial}{#1}%
}
\newrobustcmd{\rD}[1]{%
	\tensor{\mathring{\nabla}}{#1}%
}
\newrobustcmd{\rcD}[1]{%
	\tensor{\nabla}{#1}%
}
\newrobustcmd{\rR}[1]{%
	\tensor{\mathring{R}}{#1}%
}
\newrobustcmd{\rcR}[1]{%
	\tensor{R}{#1}%
}

\renewrobustcmd{\F}[2][placeholder]{%
	\IfEqCase{#1}{%
	{placeholder}{\tensor{T}{#2}}%
	{1}{\tensor[^{(1)}]{F}{#2}}%
	{2}{\tensor[^{(2)}]{F}{#2}}%
	{3}{\tensor[^{(3)}]{F}{#2}}%
	}%
	[\packageError{cosmicclass}{Symbol #1 is not an irreducible part!}{}]%
}

\newrobustcmd{\Si}[2][placeholder]{%
	\IfEqCase{#1}{%
	{placeholder}{\tensor{T}{#2}}%
	{1}{\tensor[^{(1)}]{S}{#2}}%
	{2}{\tensor[^{(2)}]{S}{#2}}%
	{3}{\tensor[^{(3)}]{S}{#2}}%
	}%
	[\packageError{cosmicclass}{Symbol #1 is not an irreducible part!}{}]%
}

\newrobustcmd{\T}[2][placeholder]{%
	\IfEqCase{#1}{%
	{placeholder}{\tensor{T}{#2}}%
	{1}{\tensor[^{(1)}]{T}{#2}}%
	{2}{\tensor[^{(2)}]{T}{#2}}%
	{3}{\tensor[^{(3)}]{T}{#2}}%
	}%
	[\packageError{cosmicclass}{Symbol #1 is not an irreducible part!}{}]%
}

\newrobustcmd{\mass}[2][placeholder]{%
	\IfEqCase{#1}{%
	{placeholder}{\tensor{m}{#2}}%
	{1}{\tensor[^{(1)}]{m}{#2}}%
	{2}{\tensor[^{(2)}]{m}{#2}}%
	{3}{\tensor[^{(3)}]{m}{#2}}%
	}%
	[\packageError{cosmicclass}{Symbol #1 is not an irreducible part!}{}]%
}

\newrobustcmd{\TLambda}[2][placeholder]{%
	\IfEqCase{#1}{%
	{placeholder}{\tensor{\lambda}{#2}}%
	{1}{\tensor[^{(1)}]{\lambda}{#2}}%
	{2}{\tensor[^{(2)}]{\lambda}{#2}}%
	{3}{\tensor[^{(3)}]{\lambda}{#2}}%
	{Q}{\tensor[^{(Q)}]{\lambda}{#2}}%
	}%
	[\packageError{cosmicclass}{Symbol #1 is not an irreducible part!}{}]%
}

\newrobustcmd{\RLambda}[2][placeholder]{%
	\IfEqCase{#1}{%
	{placeholder}{\tensor{\lambda}{#2}}%
	{1}{\tensor[^1]{\lambda}{#2}}%
	{2}{\tensor[^2]{\lambda}{#2}}%
	{3}{\tensor[^3]{\lambda}{#2}}%
	{R}{\tensor[^{(R)}]{\lambda}{#2}}%
	}%
	[\packageError{cosmicclass}{Symbol #1 is not an irreducible part!}{}]%
}

\newrobustcmd{\QLambda}[2][placeholder]{%
	\IfEqCase{#1}{%
		{placeholder}{\tensor{\hat{\lambda}}{#2}}%
	{1}{\tensor[^1]{\hat{\lambda}}{#2}}%
	{2}{\tensor[^2]{\hat{\lambda}}{#2}}%
	{3}{\tensor[^3]{\hat{\lambda}}{#2}}%
	}%
	[\packageError{cosmicclass}{Symbol #1 is not an irreducible part!}{}]%
}

\newrobustcmd{\Mgra}[1]{%
  {\tensor{M}{_{\text{#1}}}}%
}
\newrobustcmd{\Mpro}[1]{%
  {\tensor{\mathproper{M}}{_{\text{#1}}}}%
}
\newrobustcmd{\Malt}[1]{%
  {\tensor{\mathscr{M}}{_{\text{#1}}}}%
}
\newrobustcmd{\Mkom}[1]{%
  {\tensor{\mathfrak{M}}{_{\text{#1}}}}%
}

\newrobustcmd{\Mtotal}{%
  {\tensor{M}{_{\text{T}}}}%
}

\newrobustcmd{\Qtotal}{%
  {\tensor{Q}{_{\text{T}}}}%
}

\newrobustcmd{\Qtotalcal}{%
  {\tensor{\mathcal{  Q}}{_{\text{T}}}}%
}

\newrobustcmd{\action}[1]{%
  {\tensor{S}{_{\text{#1}}}}%
}

\newrobustcmd{\lagrangian}[1]{%
  {\tensor{L}{_{\text{#1}}}}%
}

\newrobustcmd{\lagrangianprop}[1]{%
  {\tensor{\mathproper{L}}{_{\text{#1}}}}%
}

\newrobustcmd{\epl}{%
  {\tensor{\mathsf{e}}{_+}}%
}

\newrobustcmd{\epe}{%
  {\tensor{\mathsf{e}}{_\perp}}%
}

\newrobustcmd{\qz}{%
  {\text{\color{orange}\cmark}}%
}
\newrobustcmd{\jz}{%
  {\text{\color{red}\xmark}}%
}

\newrobustcmd{\projmatrix}[2][placeholder]{%
  {\tensor*{M}{_{#1}^{#2}}}
}
\newrobustcmd{\projorthhum}[2][placeholder]{%
  {\tensor[^#2]{\smash{\check{\mathcal{  P}}}}{#1}}
}
\newrobustcmd{\projorthhumu}[2][placeholder]{%
  {\tensor[^#2]{\smash{\check{\mathcal{  P}}}}{#1}}
}
\newrobustcmd{\projorth}[2][placeholder]{%
  {\tensor[^#2]{\smash{\hat{\mathcal{  P}}}}{#1}}
}
\newrobustcmd{\projlore}[2][placeholder]{%
  {\tensor[^#2]{\hat{\mathcal{  P}}}{#1}}
}

\newrobustcmd{\gensec}[3][placeholder]{%
  {\tensor*[^#1]{\chi}{^{#2}_{\acu{#3}}}}
}

\newrobustcmd{\glfourr}{%
  {\mathrm{GL}(4,\mathbb{R})}%
}

\newrobustcmd{\sltwoc}{%
  {\mathrm{SL}(2,\mathbb{C})}%
}

\newrobustcmd{\poincare}{%
  {\mathbb{R}^{1,3}\rtimes\mathrm{SO}^+(1,3)}%
}

\newrobustcmd{\poincaref}{%
  {\mathrm{P}(1,3)}%
}

\newrobustcmd{\weyl}{%
  {\mathrm{W}(1,3)}%
}

\newrobustcmd{\conformal}{%
  {\mathrm{C}(1,3)}%
}

\newrobustcmd{\diffeomorphism}{%
  {\mathbb{R}^{1,3}}%
}

\newrobustcmd{\soonethree}{%
  {\mathrm{SO}^+(1,3)}%
}

\newrobustcmd{\othree}{%
  {\mathrm{SO}(3)}%
}

\newrobustcmd{\sothree}{%
  {\mathrm{SO}(3)}%
}

\newrobustcmd{\sotwo}{%
  {\mathrm{SO}(2)}%
}

\newrobustcmd{\suthreec}{%
  {\mathrm{SU}(3)_{\text{c}}}%
}

\newrobustcmd{\sutwol}{%
  {\mathrm{SU}(2)_{\text{L}}}%
}

\newrobustcmd{\uoney}{%
  {\mathrm{U}(1)_{\text{Y}}}%
}

\newrobustcmd{\uone}{%
  {\mathrm{U}(1)}%
}

\newrobustcmd{\uoneem}{%
  {\mathrm{U}(1)_{\text{em}}}%
}

\newrobustcmd{\sutwo}{%
  {\mathrm{SU}(2)}%
}

\newrobustcmd{\eplus}{%
  {\tensor{\mathsf{e}}{_{+}}}%
}

\newrobustcmd{\esf}[1]{%
  {\tensor{\mathsf{e}}{_{#1}}}
}%
\newrobustcmd{\esfu}[1]{%
  {\tensor{\mathsf{e}}{^{#1}}}
}%
\newrobustcmd{\gam}[1]{%
  {\tensor{\gamma}{_{#1}}}
}%
\newrobustcmd{\gamu}[1]{%
  {\tensor{\gamma}{^{#1}}}
}%

\newrobustcmd{\planck}{%
  {m_{\text{p}}}%
}

\newrobustcmd{\Planck}{%
	{M_{\text{Pl}}}%
}

\newrobustcmd{\caligR}{%
  {\mathcal{R}}%
}
\newrobustcmd{\caligT}{%
  {\mathcal{T}}%
}

\newrobustcmd{\pgt}{%
  PGT\textsuperscript{q,+}\ %
}

\newrobustcmd{\unl}[1]{%
  {\mathfrak{#1}}%
}
\newrobustcmd{\ovl}[1]{%
\overline{#1}%
}
\newrobustcmd{\acu}[1]{%
\acute{#1}%
}

\newrobustcmd{\indiq}[2][placeholder]{%
\IfEqCase{#1}{%
  {placeholder}{%
    \IfEqCase{#2}{%
      {1}{\ovl{k}}%
      {2}{\ovl{kl}}%
      {3}{\ovl{klm}}%
    }%
  }%
}[#1]%
}%

\newrobustcmd{\indaq}[2][placeholder]{%
\IfEqCase{#1}{%
  {placeholder}{%
    \IfEqCase{#2}{%
      {1}{\overline{k}}%
      {2}{\overline{kl}}%
      {3}{\overline{klm}}%
    }%
  }%
}[#1]%
}%

\newrobustcmd{\indeq}[2][placeholder]{%
\IfEqCase{#1}{%
  {placeholder}{%
    \IfEqCase{#2}{%
      {1}{k}%
      {2}{kl}%
      {3}{klm}%
    }%
  }%
}[#1]%
}%

\newrobustcmd{\indoq}[2][placeholder]{%
\IfEqCase{#1}{%
  {placeholder}{%
    \IfEqCase{#2}{%
      {1}{\alpha}%
      {2}{\alpha\beta}%
      {3}{\alpha\beta\gamma}%
    }%
  }%
}[#1]%
}%

\newrobustcmd{\fcphi}[1]{%
  \tensor[^{\text{FC}}]{\phi}{_{#1}}%
}

\newrobustcmd{\scphi}[1]{%
  \tensor[^{\text{SC}}]{\phi}{_{#1}}%
}

\newrobustcmd{\fcmul}[1]{%
  \tensor[^{\text{FC}}]{\upsilon}{_{#1}}%
}

\newrobustcmd{\arb}{%
  {\tensor{f}{_{\text{lin}}}}%
}

\newrobustcmd{\scmul}[1]{%
  \tensor[^{\text{SC}}]{\upsilon}{_{#1}}%
}

\newrobustcmd{\foli}[1]{%
\tensor{n}{_{#1}}%
}

\newrobustcmd{\foliu}[1]{%
\tensor{n}{^{#1}}%
}

\newrobustcmd{\covderl}[1]{%
\tensor{\mathcal{D}}{^{\flat}_{\indiq[#1]{1}}}%
}

\newrobustcmd{\covder}[1]{%
\tensor{\mathcal{D}}{_{\indiq[#1]{1}}}%
}
\newrobustcmd{\coder}[1]{%
\tensor{D}{_{\indiq[#1]{1}}}%
}

\newrobustcmd{\deltal}[2]{%
  \tensor*{\delta}{_{\phantom{\flat}}^{\flat}_{#1}^{#2}}%
}

\newrobustcmd{\deltaud}[2]{%
  \tensor*{\delta}{^{#1}_{#2}}%
}
\newrobustcmd{\etau}[1]{%
\tensor{\eta}{^{\indiq[#1]{2}}}%
}
\newrobustcmd{\etaul}[1]{%
\tensor{\eta}{^{\flat}^{\indiq[#1]{2}}}%
}
\newrobustcmd{\etad}[1]{%
\tensor{\eta}{_{\indiq[#1]{2}}}%
}
\newrobustcmd{\etadl}[1]{%
\tensor{\eta}{^{\flat}_{\indiq[#1]{2}}}%
}

\newrobustcmd{\epsul}[1]{%
\tensor{\epsilon}{^{\flat}^{\indiq[#1]{3}}^{\perp}}
}
\newrobustcmd{\epsdl}[1]{%
\tensor{\epsilon}{^{\flat}_{\indiq[#1]{3}}_{\perp}}
}
\newrobustcmd{\epsd}[1]{%
\tensor{\epsilon}{_{\indiq[#1]{3}}_{\perp}}
}
\newrobustcmd{\epsu}[1]{%
\tensor{\epsilon}{^{\indiq[#1]{3}}^{\perp}}
}

\newrobustcmd{\hfl}[2]{%
  \tensor{h}{^{\flat}_{#1}^{#2}}
}

\newrobustcmd{\cgalp}{\tensor{\alpha}{_{\text{CG}}}}

\newrobustcmd{\cbet}[1]{%
  \tensor{\bar{\beta}}{_{#1}}
}
\newrobustcmd{\calp}[1]{%
  \tensor{\bar{\alpha}}{_{#1}}
}

\newrobustcmd{\alpg}[1]{%
  \tensor{\check{\alpha}}{_{#1}}
}
\newrobustcmd{\betg}[1]{%
  \tensor{\check{\beta}}{_{#1}}
}
\newrobustcmd{\calpg}[1]{%
  \tensor{\acu{\alpha}}{_{#1}}
}
\newrobustcmd{\cbetg}[1]{%
  \tensor{\acu{\beta}}{_{#1}}
}

\newrobustcmd{\hub}{%
  {\underline{\mathsf{h}}}
}
\newrobustcmd{\hubm}{%
  {\underline{\mathsf{h}}^{-1}}
}
\newrobustcmd{\hob}{%
  {\bar{\mathsf{h}}}
}
\newrobustcmd{\hobm}{%
  {\bar{\mathsf{h}}^{-1}}
}
\newrobustcmd{\hdet}{%
  {\det \mathsf{h}}
}
\newrobustcmd{\hmdet}{%
  {\det \mathsf{h}^{-1}}
}
\newrobustcmd{\Rsf}{%
  {\mathsf{R}}
}

\newrobustcmd{\alpm}[2][placeholder]{%
  {\tensor*{\hat{\alpha}}{_{#1}^{#2}}}
}
\newrobustcmd{\calpm}[2][placeholder]{%
  \tensor*{\bar{\alpha}}{_{#1}^{#2}}
}

\newrobustcmd{\betm}[2][placeholder]{%
  {\tensor*{\hat{\beta}}{_{#1}^{#2}}}
}
\newrobustcmd{\cbetm}[2][placeholder]{%
  \tensor*{\bar{\beta}}{_{#1}^{#2}}
}

\newrobustcmd{\lamr}{%
  {\zeta_{\mathcal{  R}} }
}
\newrobustcmd{\barlamr}{%
  {\bar{\zeta}_{\mathcal{  R}} }
}
\newrobustcmd{\lamt}{%
  {\zeta_{\mathcal{  T}} }
}
\newrobustcmd{\barlamt}{%
  {\bar{\zeta}_{\mathcal{  T}} }
}

\newrobustcmd{\atmp}[1]{%
  \tensor{\hat{a}}{_{#1}}
}
\newrobustcmd{\btmp}[1]{%
  \tensor{b}{_{#1}}
}

\newrobustcmd{\ctmp}[2][placeholder]{%
  {\tensor*{c}{_{#1}^{#2}}}
}
\newrobustcmd{\dtmp}[2][placeholder]{%
  {\tensor*{d}{_{#1}^{#2}}}
}

\newrobustcmd{\etmp}[1]{%
  \tensor{e}{_{#1}}
}

\newrobustcmd{\batmp}[1]{%
  \tensor{\ovl{a}}{_{#1}}
}
\newrobustcmd{\bbtmp}[1]{%
  \tensor{\ovl{b}}{_{#1}}
}
\newrobustcmd{\bctmp}[1]{%
  \tensor{\ovl{c}}{_{#1}}
}
\newrobustcmd{\bdtmp}[1]{%
  \tensor{\ovl{d}}{_{#1}}
}
\newrobustcmd{\betmp}[1]{%
  \tensor{\ovl{e}}{_{#1}}
}

\newrobustcmd{\ptl}[1]{%
  \tensor{\partial}{#1}
}

\newrobustcmd{\etaf}[1]{%
  \tensor{\eta}{#1}
}

\newrobustcmd{\epsf}[1]{%
  \tensor{\epsilon}{#1}
}

\newrobustcmd{\RSO}[2][placeholder]{%
  {\tensor[^{#2}]{\mathcal{  R}}{#1}}
}
\newrobustcmd{\TSO}[2][placeholder]{%
  {\tensor[^{#2}]{\mathcal{  T}}{#1}}
}
\newrobustcmd{\FSO}[2][placeholder]{%
  {\tensor[^{#2}]{\mathcal{  F}}{#1}}
}
\newrobustcmd{\spinSO}[2][placeholder]{%
  {\tensor[^{#2}]{\sigma}{#1}}
}
\newrobustcmd{\RLambdaSO}[2][placeholder]{%
  {\tensor[^{#2}]{\zeta}{#1}}
}
\newrobustcmd{\TLambdaSO}[2][placeholder]{%
  {\tensor[^{#2}]{\zeta}{#1}}
}
\newrobustcmd{\KSO}[2][placeholder]{%
  {\tensor[^{#2}]{\mathcal{  K}}{#1}}
}

\newrobustcmd{\bper}[2][placeholder]{%
\IfEqCase{#2}{%
  {s}{\tensor{\mathfrak{s}}{#1}}%
  {a}{\tensor{\mathfrak{a}}{#1}}%
  {sbar}{\tensor{\bar{\mathfrak{s}}}{#1}}%
}[\packageError{cosmicclass}{Unidentified Critical Case: #1}{}]%
}

\newrobustcmd{\Jl}{%
  {J^{\flat}}%
}%

\newrobustcmd{\Nl}{%
  {N^{\flat}}%
}%

\newrobustcmd{\haml}[2][placeholder]{%
\IfEqCase{#2}{%
{mom0p}{\tensor{\mathcal{H}}{^{\flat}_{\perp}}}%
{mom1m}{\tensor{\mathcal{H}}{^{\flat}_{\indoq[#1]{1}}}}%
{rot1p}{\tensor{\mathcal{H}}{^{\flat}_{\indaq[#1]{2}}}}%
{rot1m}{\tensor{\mathcal{H}}{^{\flat}_{\perp}_{\indaq[#1]{1}}}}%
}[\packageError{cosmicclass}{Unidentified Critical Case: #1}{}]%
}

\newrobustcmd{\arc}[2][placeholder]{%
\IfEqCase{#2}{%
{B1p}{\tensor{\vartheta}{_{\perp\indiq[#1]{2}}}}%
{B2m}{\tensor[^{\text{T}}]{\vartheta}{_{\indiq[#1]{3}}}}%
{A0m}{\tensor[^{\text{P}}]{\vartheta}{}}%
{A1p}{\tensor{\overset{\wedge}{\vartheta}}{_{\perp\indiq[#1]{2}}}}%
{A1m}{\tensor{\overset{\rightharpoonup}{\vartheta}}{_{\indiq[#1]{1}}}}%
{A2p}{\tensor{\overset{\sim}{\vartheta}}{_{\perp\indiq[#1]{2}}}}%
{A2m}{\tensor[^{\text{T}}]{\vartheta}{_{\perp\indiq[#1]{3}}}}%
}[\packageError{cosmicclass}{Unidentified Critical Case: #1}{}]%
}

\newrobustcmd{\pic}[2][placeholder]{%
\IfEqCase{#2}{%
{B0p}{\varphi}%
{B1p}{\tensor{\overset{\wedge}{\varphi}}{_{\indiq[#1]{2}}}}%
{B1m}{\tensor{\varphi}{_{\perp\indiq[#1]{1}}}}%
{B2p}{\tensor{\overset{\sim}{\varphi}}{_{\indiq[#1]{2}}}}%
{A0p}{\tensor{\varphi}{_\perp}}%
{A0m}{\tensor[^{\text{P}}]{\varphi}{}}%
{A1p}{\tensor{\overset{\wedge}{\varphi}}{_{\perp\indiq[#1]{2}}}}%
{A1m}{\tensor{\overset{\rightharpoonup}{\varphi}}{_{\indiq[#1]{1}}}}%
{A2p}{\tensor{\overset{\sim}{\varphi}}{_{\perp\indiq[#1]{2}}}}%
{A2m}{\tensor[^{\text{T}}]{\varphi}{_{\indiq[#1]{3}}}}%
}[\packageError{cosmicclass}{Unidentified Critical Case: #1}{}]%
}

\newrobustcmd{\picu}[2][placeholder]{%
\IfEqCase{#2}{%
{B0p}{\varphi}%
{B1p}{\tensor{\smash{\overset{\wedge}{\varphi}}}{^{\indiq[#1]{2}}}}%
{B1m}{\tensor{\varphi}{^{\perp\indiq[#1]{1}}}}%
{B2p}{\tensor{\smash{\overset{\sim}{\varphi}}}{^{\indiq[#1]{2}}}}%
{A0p}{\tensor{\varphi}{_\perp}}%
{A0m}{\tensor[^{\text{P}}]{\varphi}{}}%
{A1p}{\tensor{\smash{\overset{\wedge}{\varphi}}}{^{\perp\indiq[#1]{2}}}}%
{A1m}{\tensor{\smash{\overset{\rightharpoonup}{\varphi}}}{^{\indiq[#1]{1}}}}%
{A2p}{\tensor{\smash{\overset{\sim}{\varphi}}}{^{\perp\indiq[#1]{2}}}}%
{A2m}{\tensor[^{\text{T}}]{\varphi}{^{\indiq[#1]{3}}}}%
}[\packageError{cosmicclass}{Unidentified Critical Case: #1}{}]%
}

\newrobustcmd{\picl}[2][placeholder]{%
\IfEqCase{#2}{%
{B0p}{\tensor{\varphi}{^{\flat}}}%
{B1p}{\tensor{\smash{\overset{\wedge}{\varphi}}}{^{\flat}_{\indiq[#1]{2}}}}%
{B1m}{\tensor{\varphi}{^{\flat}_{\perp}_{\indiq[#1]{1}}}}%
{B2p}{\tensor{\smash{\overset{\sim}{\varphi}}}{^{\flat}_{\indiq[#1]{2}}}}%
{A0p}{\tensor{\varphi}{_\perp}^{\flat}}%
{A0m}{\tensor[^{\text{P}}]{\varphi}{^{\flat}}}%
{A1p}{\tensor{\smash{\overset{\wedge}{\varphi}}}{^{\flat}_{\perp\indiq[#1]{2}}}}%
{A1m}{\tensor{\smash{\overset{\rightharpoonup}{\varphi}}}{^{\flat}_{\indiq[#1]{1}}}}%
{A2p}{\tensor{\smash{\overset{\sim}{\varphi}}}{^{\flat}_{\perp\indiq[#1]{2}}}}%
{A2m}{\tensor[^{\text{T}}]{\varphi}{^{\flat}_{\indiq[#1]{3}}}}%
}[\packageError{cosmicclass}{Unidentified Critical Case: #1}{}]%
}

\newrobustcmd{\mull}[2][placeholder]{%
\IfEqCase{#2}{%
{B0p}{\tensor{u}{^{\flat}}}%
{B1p}{\tensor{\smash{\overset{\wedge}{u}}}{^{\flat}_{\indiq[#1]{2}}}}%
{B1m}{\tensor{u}{^{\flat}_{\perp}_{\indiq[#1]{1}}}}%
{B2p}{\tensor{\smash{\overset{\sim}{u}}}{^{\flat}_{\indiq[#1]{2}}}}%
{A0p}{\tensor{u}{_\perp}^{\flat}}%
{A0m}{\tensor[^{\text{P}}]{u}{^{\flat}}}%
{A1p}{\tensor{\smash{\overset{\wedge}{u}}}{^{\flat}_{\perp\indiq[#1]{2}}}}%
{A1m}{\tensor{\smash{\overset{\rightharpoonup}{u}}}{^{\flat}_{\indiq[#1]{1}}}}%
{A2p}{\tensor{\smash{\overset{\sim}{u}}}{^{\flat}_{\perp\indiq[#1]{2}}}}%
{A2m}{\tensor[^{\text{T}}]{u}{^{\flat}_{\indiq[#1]{3}}}}%
}[\packageError{cosmicclass}{Unidentified Critical Case: #1}{}]%
}

\newrobustcmd{\mul}[2][placeholder]{%
\IfEqCase{#2}{%
{B0p}{\tensor{u}{}}%
{B1p}{\tensor{\smash{\overset{\wedge}{u}}}{_{\indiq[#1]{2}}}}%
{B1m}{\tensor{u}{_{\perp}_{\indiq[#1]{1}}}}%
{B2p}{\tensor{\smash{\overset{\sim}{u}}}{_{\indiq[#1]{2}}}}%
{A0p}{\tensor{u}{_\perp}}%
{A0m}{\tensor[^{\text{P}}]{u}{}}%
{A1p}{\tensor{\smash{\overset{\wedge}{u}}}{_{\perp\indiq[#1]{2}}}}%
{A1m}{\tensor{\smash{\overset{\rightharpoonup}{u}}}{_{\indiq[#1]{1}}}}%
{A2p}{\tensor{\smash{\overset{\sim}{u}}}{_{\perp\indiq[#1]{2}}}}%
{A2m}{\tensor[^{\text{T}}]{u}{_{\indiq[#1]{3}}}}%
}[\packageError{cosmicclass}{Unidentified Critical Case: #1}{}]%
}

\newrobustcmd{\PiP}[2][placeholder]{%
\IfEqCase{#2}{%
{B0p}{\hat{\pi}}%
{B1p}{\tensor{\overset{\wedge}{\hat{\pi}}}{_{\indiq[#1]{2}}}}%
{B1m}{\tensor{\hat{\pi}}{_{\perp\indiq[#1]{1}}}}%
{B2p}{\tensor{\overset{\sim}{\hat{\pi}}}{_{\indiq[#1]{2}}}}%
{A0p}{\tensor{\hat{\pi}}{_\perp}}%
{A0m}{\tensor[^{\text{P}}]{\hat{\pi}}{}}%
{A1p}{\tensor{\overset{\wedge}{\hat{\pi}}}{_{\perp\indiq[#1]{2}}}}%
{A1m}{\tensor{\overset{\rightharpoonup}{\hat{\pi}}}{_{\indiq[#1]{1}}}}%
{A2p}{\tensor{\overset{\sim}{\hat{\pi}}}{_{\perp\indiq[#1]{2}}}}%
{A2m}{\tensor[^{\text{T}}]{\hat{\pi}}{_{\indiq[#1]{3}}}}%
}[\packageError{cosmicclass}{Unidentified Critical Case: #1}{}]%
}

\newrobustcmd{\PiPu}[2][placeholder]{%
\IfEqCase{#2}{%
{B0p}{\hat{\pi}}%
{B1p}{\tensor{\smash{\overset{\wedge}{\hat{\pi}}}}{^{\indiq[#1]{2}}}}%
{B1m}{\tensor{\smash{\hat{\pi}}}{^{\perp\indiq[#1]{1}}}}%
{B2p}{\tensor{\smash{\overset{\sim}{\hat{\pi}}}}{^{\indiq[#1]{2}}}}%
{A0p}{\tensor{\smash{\hat{\pi}}}{^\perp}}%
{A0m}{\tensor[^{\text{P}}]{\smash{\hat{\pi}}}{}}%
{A1p}{\tensor{\smash{\overset{\wedge}{\hat{\pi}}}}{^{\perp\indiq[#1]{2}}}}%
{A1m}{\tensor{\smash{\overset{\rightharpoonup}{\hat{\pi}}}}{^{\indiq[#1]{1}}}}%
{A2p}{\tensor{\smash{\overset{\sim}{\hat{\pi}}}}{^{\perp\indiq[#1]{2}}}}%
{A2m}{\tensor[^{\text{T}}]{\smash{\hat{\pi}}}{^{\indiq[#1]{3}}}}%
}[\packageError{cosmicclass}{Unidentified Critical Case: #1}{}]%
}

\newrobustcmd{\sicl}[2][placeholder]{%
\IfEqCase{#2}{%
{B0p}{\tensor{\chi}{^{\flat}}}%
{B1p}{\tensor{\smash{\overset{\wedge}{\chi}}}{^{\flat}_{\indiq[#1]{2}}}}%
{B1m}{\tensor{\chi}{^{\flat}_{\perp}_{\indiq[#1]{1}}}}%
{B2p}{\tensor{\smash{\overset{\sim}{\chi}}}{^{\flat}_{\indiq[#1]{2}}}}%
{A0p}{\tensor{\chi}{^{\flat}_\perp}}%
{A0m}{\tensor[^{\text{P}}]{\chi}{^{\flat}}}%
{A1p}{\tensor{\smash{\overset{\wedge}{\chi}}}{^{\flat}_{\perp\indiq[#1]{2}}}}%
{A1m}{\tensor{\smash{\overset{\rightharpoonup}{\chi}}}{^{\flat}_{\indiq[#1]{1}}}}%
{A2p}{\tensor{\smash{\overset{\sim}{\chi}}}{^{\flat}_{\perp\indiq[#1]{2}}}}%
{A2m}{\tensor[^{\text{T}}]{\chi}{^{\flat}_{\indiq[#1]{3}}}}%
}[\packageError{cosmicclass}{Unidentified Critical Case: #1}{}]%
}

\newrobustcmd{\ticl}[2][placeholder]{%
\IfEqCase{#2}{%
{B0p}{\tensor{\zeta}{^{\flat}}}%
{B1p}{\tensor{\smash{\overset{\wedge}{\zeta}}}{^{\flat}_{\indiq[#1]{2}}}}%
{B1m}{\tensor{\zeta}{^{\flat}_{\perp}_{\indiq[#1]{1}}}}%
{B2p}{\tensor{\smash{\overset{\sim}{\zeta}}}{^{\flat}_{\indiq[#1]{2}}}}%
{A0p}{\tensor{\zeta}{^{\flat}_\perp}}%
{A0m}{\tensor[^{\text{P}}]{\zeta}{^{\flat}}}%
{A1p}{\tensor{\smash{\overset{\wedge}{\zeta}}}{^{\flat}_{\perp\indiq[#1]{2}}}}%
{A1m}{\tensor{\smash{\overset{\rightharpoonup}{\zeta}}}{^{\flat}_{\indiq[#1]{1}}}}%
{A2p}{\tensor{\smash{\overset{\sim}{\zeta}}}{^{\flat}_{\perp\indiq[#1]{2}}}}%
{A2m}{\tensor[^{\text{T}}]{\zeta}{^{\flat}_{\indiq[#1]{3}}}}%
}[\packageError{cosmicclass}{Unidentified Critical Case: #1}{}]%
}

\newrobustcmd{\PiPl}[2][placeholder]{%
\IfEqCase{#2}{%
{B0p}{\tensor{\hat{\pi}}{^{\flat}}}%
{B1p}{\tensor{\smash{\overset{\wedge}{\hat{\pi}}}}{^{\flat}_{\indiq[#1]{2}}}}%
{B1m}{\tensor{\hat{\pi}}{^{\flat}_{\perp}_{\indiq[#1]{1}}}}%
{B2p}{\tensor{\smash{\overset{\sim}{\hat{\pi}}}}{^{\flat}_{\indiq[#1]{2}}}}%
{A0p}{\tensor{\hat{\pi}}{_\perp}^{\flat}}%
{A0m}{\tensor[^{\text{P}}]{\hat{\pi}}{^{\flat}}}%
{A1p}{\tensor{\smash{\overset{\wedge}{\hat{\pi}}}}{^{\flat}_{\perp\indiq[#1]{2}}}}%
{A1m}{\tensor{\smash{\overset{\rightharpoonup}{\hat{\pi}}}}{^{\flat}_{\indiq[#1]{1}}}}%
{A2p}{\tensor{\smash{\overset{\sim}{\hat{\pi}}}}{^{\flat}_{\perp\indiq[#1]{2}}}}%
{A2m}{\tensor[^{\text{T}}]{\hat{\pi}}{^{\flat}_{\indiq[#1]{3}}}}%
}[\packageError{cosmicclass}{Unidentified Critical Case: #1}{}]%
}

\newrobustcmd{\sic}[2][placeholder]{%
\IfEqCase{#2}{%
{B0p}{\chi}%
{B1p}{\tensor{\overset{\wedge}{\chi}}{_{\indiq[#1]{2}}}}%
{B1m}{\tensor{\chi}{_{\perp\indiq[#1]{1}}}}%
{B2p}{\tensor{\overset{\sim}{\chi}}{_{\indiq[#1]{2}}}}%
{A0p}{\tensor{\chi}{_\perp}}%
{A0m}{\tensor[^{\text{P}}]{\chi}{}}%
{A1p}{\tensor{\overset{\wedge}{\chi}}{_{\perp\indiq[#1]{2}}}}%
{A1m}{\tensor{\overset{\rightharpoonup}{\chi}}{_{\indiq[#1]{1}}}}%
{A2p}{\tensor{\overset{\sim}{\chi}}{_{\perp\indiq[#1]{2}}}}%
{A2m}{\tensor[^{\text{T}}]{\chi}{_{\indiq[#1]{3}}}}%
}[\packageError{cosmicclass}{Unidentified Critical Case: #1}{}]%
}

\newrobustcmd{\lorsicpar}[2][placeholder]{%
\IfEqCase{#2}{%
{B0m}{\tensor*[^{\text{P}}]{\smash{\underline{\chi}}}{^{\parallel}}}%
{B1p}{\tensor*{\smash{\overset{\wedge}{\chi}}}{^{\parallel}_{\indiq[#1]{2}}}}%
{B1m}{\tensor*{\chi}{^{\parallel}_{\perp\indiq[#1]{1}}}}%
{B2m}{\tensor*[^{\text{T}}]{\smash{\underline{\chi}}}{^{\parallel}_{\indiq[#1]{3}}}}%
{A0p}{\tensor*{\chi}{^{\parallel}_\perp}}%
{A0m}{\tensor*[^{\text{P}}]{\chi}{^{\parallel}}}%
{A1p}{\tensor*{\smash{\overset{\wedge}{\chi}}}{^{\parallel}_{\perp\indiq[#1]{2}}}}%
{A1m}{\tensor*{\smash{\overset{\rightharpoonup}{\chi}}}{^{\parallel}_{\indiq[#1]{1}}}}%
{A2p}{\tensor*{\smash{\overset{\sim}{\chi}}}{^{\parallel}_{\perp\indiq[#1]{2}}}}%
{A2m}{\tensor*[^{\text{T}}]{\chi}{^{\parallel}_{\indiq[#1]{3}}}}%
}[\packageError{cosmicclass}{Unidentified Critical Case: #1}{}]%
}

\newrobustcmd{\lorsicpir}[2][placeholder]{%
\IfEqCase{#2}{%
{B0p}{\tensor*{\chi}{^{\vDash}}}%
{B1p}{\tensor*{\smash{\overset{\wedge}{\chi}}}{^{\vDash}_{\indiq[#1]{2}}}}%
{B1m}{\tensor*{\chi}{^{\vDash}_{\perp\indiq[#1]{1}}}}%
{B2p}{\tensor*{\smash{\overset{\sim}{\chi}}}{^{\vDash}_{\indiq[#1]{2}}}}%
{A0p}{\tensor*{\chi}{^{\vDash}_\perp}}%
{A0m}{\tensor*[^{\text{P}}]{\chi}{^{\vDash}}}%
{A1p}{\tensor*{\smash{\overset{\wedge}{\chi}}}{^{\vDash}_{\perp\indiq[#1]{2}}}}%
{A1m}{\tensor*{\smash{\overset{\rightharpoonup}{\chi}}}{^{\vDash}_{\indiq[#1]{1}}}}%
{A2p}{\tensor*{\smash{\overset{\sim}{\chi}}}{^{\vDash}_{\perp\indiq[#1]{2}}}}%
{A2m}{\tensor*[^{\text{T}}]{\chi}{^{\vDash}_{\indiq[#1]{3}}}}%
}[\packageError{cosmicclass}{Unidentified Critical Case: #1}{}]%
}

\newrobustcmd{\lorsicper}[2][placeholder]{%
\IfEqCase{#2}{%
{B0p}{\tensor*{\chi}{^{\perp}}}%
{B1p}{\tensor*{\smash{\overset{\wedge}{\chi}}}{^{\perp}_{\indiq[#1]{2}}}}%
{B1m}{\tensor*{\chi}{^{\perp}_{\perp\indiq[#1]{1}}}}%
{B2p}{\tensor*{\smash{\overset{\sim}{\chi}}}{^{\perp}_{\indiq[#1]{2}}}}%
{A0p}{\tensor*{\chi}{^{\perp}_\perp}}%
{A0m}{\tensor*[^{\text{P}}]{\chi}{^{\perp}}}%
{A1p}{\tensor*{\smash{\overset{\wedge}{\chi}}}{^{\perp}_{\perp\indiq[#1]{2}}}}%
{A1m}{\tensor*{\smash{\overset{\rightharpoonup}{\chi}}}{^{\perp}_{\indiq[#1]{1}}}}%
{A2p}{\tensor*{\smash{\overset{\sim}{\chi}}}{^{\perp}_{\perp\indiq[#1]{2}}}}%
{A2m}{\tensor*[^{\text{T}}]{\chi}{^{\perp}_{\indiq[#1]{3}}}}%
}[\packageError{cosmicclass}{Unidentified Critical Case: #1}{}]%
}

\newrobustcmd{\Tl}[2][placeholder]{%
\IfEqCase{#2}{%
{B0p}{\tensor{\chi}{^{\flat}}}%
{B1p}{\tensor{\smash{\overset{\wedge}{\chi}}}{^{\flat}_{\indiq[#1]{2}}}}%
{B1m}{\tensor{\chi}{^{\flat}_{\perp}_{\indiq[#1]{1}}}}%
{B2p}{\tensor{\smash{\overset{\sim}{\chi}}}{^{\flat}_{\indiq[#1]{2}}}}%
{A0p}{\tensor{\chi}{^{\flat}_\perp}}%
{A0m}{\tensor[^{\text{P}}]{\mathcal{T}}{^{\flat}}}%
{A1p}{\tensor{\smash{\overset{\wedge}{\chi}}}{^{\flat}_{\perp\indiq[#1]{2}}}}%
{A1m}{\tensor{\smash{\overset{\rightharpoonup}{\mathcal{T}}}}{^{\flat}_{\indiq[#1]{1}}}}%
{A2p}{\tensor{\smash{\overset{\sim}{\chi}}}{^{\flat}_{\perp\indiq[#1]{2}}}}%
{A2m}{\tensor[^{\text{T}}]{\mathcal{T}}{^{\flat}_{\indiq[#1]{3}}}}%
}[\tensor{\mathcal{T}}{^{\flat}_{\indiq[#1]{3}}}]%
}

\newrobustcmd{\cT}[2][placeholder]{%
\IfEqCase{#2}{%
{B1p}{\tensor{\mathcal{T}}{_{\perp\indiq[#1]{2}}}}%
{B1m}{\tensor{\overset{\rightharpoonup}{\mathcal{T}}}{_{\indiq[#1]{1}}}}%
{A0m}{\tensor[^{\text{P}}]{\mathcal{T}}{}}%
{A2m}{\tensor[^{\text{T}}]{\mathcal{T}}{_{\indiq[#1]{3}}}}%
}[\packageError{cosmicclass}{Unidentified Critical Case: #1}{}]%
}

\newrobustcmd{\cTLambda}[2][placeholder]{%
\IfEqCase{#2}{%
{B1p}{\tensor{\zeta}{_{\perp\indiq[#1]{2}}}}%
{B1m}{\tensor{\overset{\rightharpoonup}{\zeta}}{_{\indiq[#1]{1}}}}%
{A0m}{\tensor[^{\text{P}}]{\zeta}{}}%
{A2m}{\tensor[^{\text{T}}]{\zeta}{_{\indiq[#1]{3}}}}%
}[\packageError{cosmicclass}{Unidentified Critical Case: #1}{}]%
}

\newrobustcmd{\cTpic}[2][placeholder]{%
\IfEqCase{#2}{%
{B1p}{\tensor{\phi}{_{\perp\indiq[#1]{2}}}}%
{B1m}{\tensor{\overset{\rightharpoonup}{\phi}}{_{\indiq[#1]{1}}}}%
{A0m}{\tensor[^{\text{P}}]{\phi}{}}%
{A2m}{\tensor[^{\text{T}}]{\phi}{_{\indiq[#1]{3}}}}%
}[\packageError{cosmicclass}{Unidentified Critical Case: #1}{}]%
}

\newrobustcmd{\cTpicl}[2][placeholder]{%
\IfEqCase{#2}{%
{B1p}{\tensor{\phi}{^{\flat}_{\perp\indiq[#1]{2}}}}%
{B1m}{\tensor{\overset{\rightharpoonup}{\phi}}{^{\flat}_{\indiq[#1]{1}}}}%
{A0m}{\tensor[^{\text{P}}]{\phi}{^{\flat}}}%
{A2m}{\tensor[^{\text{T}}]{\phi}{^{\flat}_{\indiq[#1]{3}}}}%
}[\packageError{cosmicclass}{Unidentified Critical Case: #1}{}]%
}

\newrobustcmd{\cTPiP}[2][placeholder]{%
\IfEqCase{#2}{%
{B1p}{\tensor{\varpi}{_{\perp\indiq[#1]{2}}}}%
{B1m}{\tensor{\overset{\rightharpoonup}{\varpi}}{_{\indiq[#1]{1}}}}%
{A0m}{\tensor[^{\text{P}}]{\varpi}{}}%
{A2m}{\tensor[^{\text{T}}]{\varpi}{_{\indiq[#1]{3}}}}%
}[\packageError{cosmicclass}{Unidentified Critical Case: #1}{}]%
}

\newrobustcmd{\cTl}[2][placeholder]{%
\IfEqCase{#2}{%
{B1p}{\tensor{\mathcal{T}}{^{\flat}_{\perp\indiq[#1]{2}}}}%
{B1m}{\tensor{\smash{\overset{\rightharpoonup}{\mathcal{T}}}}{^{\flat}_{\indiq[#1]{1}}}}%
{A0m}{\tensor[^{\text{P}}]{\mathcal{T}}{^{\flat}}}%
{A2m}{\tensor[^{\text{T}}]{\mathcal{T}}{^{\flat}_{\indiq[#1]{3}}}}%
}[\packageError{cosmicclass}{Unidentified Critical Case: #1}{}]%
}

\newrobustcmd{\cTu}[2][placeholder]{%
\IfEqCase{#2}{%
{B1p}{\tensor{\mathcal{T}}{^{\perp\indiq[#1]{2}}}}%
{B1m}{\tensor{\smash{\overset{\rightharpoonup}{\mathcal{T}}}}{^{\indiq[#1]{1}}}}%
{A0m}{\tensor[^{\text{P}}]{\mathcal{T}}{}}%
{A2m}{\tensor[^{\text{T}}]{\mathcal{T}}{^{\indiq[#1]{3}}}}%
}[\packageError{cosmicclass}{Unidentified Critical Case: #1}{}]%
}

\newrobustcmd{\ncTLambda}[2][placeholder]{%
\IfEqCase{#2}{%
{B0p}{\tensor{\zeta}{^{\indiq[#1]{1}}_{\indiq[#1]{1}\perp}}}%
{B1p}{\tensor{\zeta}{_{[\indiq[#1]{2}]\perp}}}%
{B1m}{\tensor{\zeta}{_{\perp\indiq[#1]{1}\perp}}}%
{B2p}{\tensor{\zeta}{_{\langle\indiq[#1]{2}\rangle\perp}}}%
}[\packageError{cosmicclass}{Unidentified Critical Case: #1}{}]%
}

\newrobustcmd{\ncTmul}[2][placeholder]{%
\IfEqCase{#2}{%
{B0p}{\tensor{\upsilon}{^{\indiq[#1]{1}}_{\indiq[#1]{1}\perp}}}%
{B1p}{\tensor{\upsilon}{_{[\indiq[#1]{2}]\perp}}}%
{B1m}{\tensor{\upsilon}{_{\perp\indiq[#1]{1}\perp}}}%
{B2p}{\tensor{\upsilon}{_{\langle\indiq[#1]{2}\rangle\perp}}}%
}[\packageError{cosmicclass}{Unidentified Critical Case: #1}{}]%
}

\newrobustcmd{\ncTpic}[2][placeholder]{%
\IfEqCase{#2}{%
{B0p}{\tensor{\phi}{^{\indiq[#1]{1}}_{\indiq[#1]{1}\perp}}}%
{B1p}{\tensor{\phi}{_{[\indiq[#1]{2}]\perp}}}%
{B1m}{\tensor{\phi}{_{\perp\indiq[#1]{1}\perp}}}%
{B2p}{\tensor{\phi}{_{\langle\indiq[#1]{2}\rangle\perp}}}%
}[\packageError{cosmicclass}{Unidentified Critical Case: #1}{}]%
}

\newrobustcmd{\ncTpicl}[2][placeholder]{%
\IfEqCase{#2}{%
  {B0p}{\tensor{\phi}{^{\flat}^{\indiq[#1]{1}}_{\indiq[#1]{1}\perp}}}%
{B1p}{\tensor{\phi}{^{\flat}_{[\indiq[#1]{2}]\perp}}}%
{B1m}{\tensor{\phi}{^{\flat}_{\perp\indiq[#1]{1}\perp}}}%
{B2p}{\tensor{\phi}{^{\flat}_{\langle\indiq[#1]{2}\rangle\perp}}}%
}[\packageError{cosmicclass}{Unidentified Critical Case: #1}{}]%
}

\newrobustcmd{\ncTPiP}[2][placeholder]{%
\IfEqCase{#2}{%
{B0p}{\tensor{\varpi}{^{\indiq[#1]{1}}_{\indiq[#1]{1}\perp}}}%
{B1p}{\tensor{\varpi}{_{[\indiq[#1]{2}]\perp}}}%
{B1m}{\tensor{\varpi}{_{\perp\indiq[#1]{1}\perp}}}%
{B2p}{\tensor{\varpi}{_{\langle\indiq[#1]{2}\rangle\perp}}}%
}[\packageError{cosmicclass}{Unidentified Critical Case: #1}{}]%
}

\newrobustcmd{\ncT}[2][placeholder]{%
\IfEqCase{#2}{%
{B0p}{\tensor{\mathcal{T}}{^{\indiq[#1]{1}}_{\indiq[#1]{1}\perp}}}%
{B1p}{\tensor{\mathcal{T}}{_{[\indiq[#1]{2}]\perp}}}%
{B1m}{\tensor{\mathcal{T}}{_{\perp\indiq[#1]{1}\perp}}}%
{B2p}{\tensor{\mathcal{T}}{_{\langle\indiq[#1]{2}\rangle\perp}}}%
}[\packageError{cosmicclass}{Unidentified Critical Case: #1}{}]%
}

\newrobustcmd{\cR}[2][placeholder]{%
\IfEqCase{#2}{%
{A0p}{\tensor{\underline{\mathcal{R}}}{}}%
{A0m}{\tensor[^{\text{P}}]{\mathcal{R}}{_{\perp\circ}}}%
{A1p}{\tensor{\underline{\mathcal{R}}}{_{[\indiq[#1]{2}]}}}%
{A1m}{\tensor{\mathcal{R}}{_{\perp\indiq[#1]{1}}}}%
{A2p}{\tensor{\underline{\mathcal{R}}}{_{\langle\indiq[#1]{2}\rangle}}}%
{A2m}{\tensor[^{\text{T}}]{\mathcal{R}}{_{\perp\indiq[#1]{3}}}}%
}[\packageError{cosmicclass}{Unidentified Critical Case: #1}{}]%
}

\newrobustcmd{\cRLambda}[2][placeholder]{%
\IfEqCase{#2}{%
{A0p}{\tensor{\underline{\zeta}}{}}%
{A0m}{\tensor[^{\text{P}}]{\zeta}{_{\perp\circ}}}%
{A1p}{\tensor{\underline{\zeta}}{_{[\indiq[#1]{2}]}}}%
{A1m}{\tensor{\zeta}{_{\perp\indiq[#1]{1}}}}%
{A2p}{\tensor{\underline{\zeta}}{_{\langle\indiq[#1]{2}\rangle}}}%
{A2m}{\tensor[^{\text{T}}]{\zeta}{_{\perp\indiq[#1]{3}}}}%
}[\packageError{cosmicclass}{Unidentified Critical Case: #1}{}]%
}

\newrobustcmd{\cRl}[2][placeholder]{%
\IfEqCase{#2}{%
  {A0p}{\tensor{\underline{\mathcal{R}}}{^{\flat}}}%
{A0m}{\tensor[^{\text{P}}]{\mathcal{R}}{^{\flat}_{\perp\circ}}}%
{A1p}{\tensor{\underline{\mathcal{R}}}{^{\flat}_{[\indiq[#1]{2}]}}}%
{A1m}{\tensor{\mathcal{R}}{^{\flat}_{\perp\indiq[#1]{1}}}}%
{A2p}{\tensor{\underline{\mathcal{R}}}{^{\flat}_{\langle\indiq[#1]{2}\rangle}}}%
{A2m}{\tensor[^{\text{T}}]{\mathcal{R}}{^{\flat}_{\perp\indiq[#1]{3}}}}%
}[\packageError{cosmicclass}{Unidentified Critical Case: #1}{}]%
}

\newrobustcmd{\cRu}[2][placeholder]{%
\IfEqCase{#2}{%
{A0p}{\tensor{\underline{\mathcal{R}}}{}}%
{A0m}{\tensor[^{\text{P}}]{\mathcal{R}}{_{\perp\circ}}}%
{A1p}{\tensor{\underline{\mathcal{R}}}{^{[\indiq[#1]{2}]}}}%
{A1m}{\tensor{\mathcal{R}}{^{\perp\indiq[#1]{1}}}}%
{A2p}{\tensor{\underline{\mathcal{R}}}{^{\langle\indiq[#1]{2}\rangle}}}%
{A2m}{\tensor[^{\text{T}}]{\mathcal{R}}{^{\perp\indiq[#1]{3}}}}%
}[\packageError{cosmicclass}{Unidentified Critical Case: #1}{}]%
}

\newrobustcmd{\ncR}[2][placeholder]{%
\IfEqCase{#2}{%
  {A0p}{\tensor{\mathcal{R}}{_{\perp\perp}}}%
{A0m}{\tensor[^{\text{P}}]{\mathcal{R}}{_{\circ\perp}}}%
{A1p}{\tensor{\mathcal{R}}{_{\perp[\indiq[#1]{2}]\perp}}}%
{A1m}{\tensor{\mathcal{R}}{_{\indiq[#1]{1}\perp}}}%
{A2p}{\tensor{\mathcal{R}}{_{\perp\langle\indiq[#1]{2}\rangle\perp}}}%
{A2m}{\tensor[^{\text{T}}]{\mathcal{R}}{_{\indiq[#1]{3}\perp}}}%
}[\packageError{cosmicclass}{Unidentified Critical Case: #1}{}]%
}

\newrobustcmd{\ncRLambda}[2][placeholder]{%
\IfEqCase{#2}{%
  {A0p}{\tensor{\zeta}{_{\perp\perp}}}%
{A0m}{\tensor[^{\text{P}}]{\zeta}{_{\circ\perp}}}%
{A1p}{\tensor{\zeta}{_{\perp[\indiq[#1]{2}]\perp}}}%
{A1m}{\tensor{\zeta}{_{\indiq[#1]{1}\perp}}}%
{A2p}{\tensor{\zeta}{_{\perp\langle\indiq[#1]{2}\rangle\perp}}}%
{A2m}{\tensor[^{\text{T}}]{\zeta}{_{\indiq[#1]{3}\perp}}}%
}[\packageError{cosmicclass}{Unidentified Critical Case: #1}{}]%
}

\newrobustcmd{\Proj}[2][placeholder]{%
\IfEqCase{#2}{%
  {A2m}{\tensor[^{\text{T}}]{\check{\mathcal{P}}}{#1}}%
}[\packageError{cosmicclass}{Unidentified Critical Case: #1}{}]%
}

\newrobustcmd{\Projl}[2][placeholder]{%
\IfEqCase{#2}{%
  {A2m}{\tensor[^{\text{T}}]{\check{\mathcal{P}}}{^{\flat}#1}}%
}[\packageError{cosmicclass}{Unidentified Critical Case: #1}{}]%
}

\newrobustcmd{\fA}{%
  {\tensor{\mathcal{  A}}{_{\acu{u}}}}%
}
\newrobustcmd{\fB}{%
  {\tensor{\mathcal{  B}}{_{\acu{v}}}}%
}
\newrobustcmd{\fC}{%
  {\tensor{\mathcal{  C}}{^{\acu{v}}}}%
}
\newrobustcmd{\fphi}{%
  {\tensor{\phi}{^{\acu{w}}}}%
}
\newrobustcmd{\fpi}{%
  {\tensor{\pi}{_{\acu{w}}}}%
}

\newrobustcmd{\covard}[2]{%
  {\frac{\bar{\delta}#1}{\bar{\delta}#2}}
}
\newrobustcmd{\copard}[2]{%
  {\frac{\bar{\partial}#1}{\bar{\partial}#2}}
}
\newrobustcmd{\pard}[2]{%
  {\frac{\partial #1}{\partial #2}}
}

\newrobustcmd{\PPM}[1]{%
  {\left[\tensor*{\mathsf{M}}{_{\ }^{\left(\text{#1}\right)}}\right]}%
}

\tikzset{
  good/.style={circle, opacity=0.7, draw=green!60, fill=green!5, line width=.8mm, minimum size=3.5mm},
  bad/.style={circle, opacity=0.7, draw=red!60, fill=red!5, line width=.8mm, minimum size=3.5mm},
  badx/.style={circle, opacity=0.4, draw=red!60, fill=red!5, line width=.8mm, minimum size=3.5mm},
  save/.style={circle, opacity=0.7, draw=green!60, fill=green!60, line width=1.5mm, minimum size=3.5mm},
  kill/.style={circle, opacity=0.7, draw=red!60, fill=red!60, line width=1.5mm, minimum size=3.5mm},
  bind/.style={draw=green!60, opacity=0.7, line width=1mm,},
  wrap/.style={draw=red!60, opacity=0.7, line width=1mm,},
}

\usepackage{hyperref}
\hypersetup{%
     colorlinks = true,%
     linkcolor = blue,%
     citecolor = blue,%
     filecolor = blue,%
     urlcolor = blue%
     }%
\usepackage[capitalize]{cleveref} 

\begin{document}

\title{Einstein-Proca theory from the Einstein-Cartan formulation}

\author{Will Barker}
\email{wb263@cam.ac.uk}
\affiliation{Astrophysics Group, Cavendish Laboratory, JJ Thomson Avenue, Cambridge CB3 0HE, UK}
\affiliation{Kavli Institute for Cosmology, Madingley Road, Cambridge CB3 0HA, UK}
\author{Sebastian Zell}
\email{sebastian.zell@uclouvain.be}
\affiliation{Centre for Cosmology, Particle Physics and Phenomenology -- CP3,
	Universit\'e catholique de Louvain, B-1348 Louvain-la-Neuve, Belgium}


\begin{abstract}
	We construct a theory of gravity in which a propagating massive vector field arises from a quadratic curvature invariant. 
	The Einstein-Cartan formulation and a partial suppression of torsion ensure the absence of ghost and strong-coupling problems, as we prove with nonlinear Lagrangian and Hamiltonian analysis. Augmenting General Relativity with a propagating torsion vector, our theory provides a purely gravitational origin of Einstein-Proca models and constrains their parameter space. As an outlook to phenomenology, we discuss the gravitational production of fermionic dark matter.
\end{abstract}

\maketitle

\section{Open questions in General Relativity}
The theory of General Relativity (GR) provides an outstandingly successful description of gravity and has been confirmed by countless experiments, such as the breakthrough discovery of gravitational waves~\cite{LIGOScientific:2016aoc}. Nevertheless, many far-reaching questions have remained unanswered. Gravitational interactions have long revealed the existence of a non-baryonic form of matter~\cite{Zwicky:1933gu}. It is unknown, however, what this dark matter is composed of and how it was produced in the cosmological history~\cite{Feng:2010gw,Peebles:2013hla,Bertone:2016nfn,Arbey:2021gdg}. Moreover, the very early moments of our Universe -- due to a suspected phase of inflation~\cite{Starobinsky:1980te, Guth:1980zm, Linde:1981mu, Mukhanov:1981xt} -- and its far future -- because of the observed dark energy~\cite{SupernovaSearchTeam:1998fmf, SupernovaCosmologyProject:1998vns} -- appear to have in common an accelerated expansion. Agreement with recent	measurements, in particular of the cosmic microwave background (CMB)~\cite{Planck:2018jri, BICEP:2021xfz}, is excellent but a microscopic origin has still not been determined both for dark energy and for inflation.

\begin{figure}
		\vspace{-0.5\baselineskip}
	\includegraphics[width=1\linewidth]{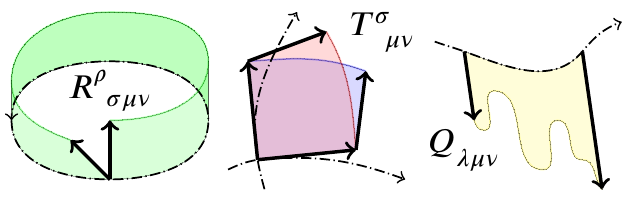}
		\vspace{-2.5\baselineskip}
		\caption{Schematic representation of parallel transport in the presence of curvature $\tensor{R}{^\rho_{\sigma\mu\nu}}$ -- rotation around closed curves -- torsion $\tensor{T}{^\sigma_{\mu\nu}}$ -- non-closure of (infinitesimal) parallelograms -- and nonmetricity $\tensor{Q}{_{\lambda\mu\nu}}$ -- non-conservation of vector norms.}
	\label{fig:schema}
\end{figure}

It is exciting to explore if GR also contains an answer to these problems.  As we shall show, this question is closely connected to the geometry of spacetime, which a priori is characterized by three independent properties curvature $\rcR{}$, torsion $T^\alpha_{~~\beta \gamma}$ and nonmetricity $Q_{\alpha\mu\nu}$ (see~\cref{fig:schema}). The presence or absence of $\rcR{}$, $Q_{\alpha\mu\nu}$ and $T^\alpha_{~~\beta \gamma}$ defines seven formulations of GR~\cite{Einstein:1915,Weyl:1918, Palatini:1919,Weyl:1922,Cartan:1922, Eddington:1923,  Cartan:1923,Eddington:1923, Cartan:1924, Cartan:1925,Einstein:1925, Einstein:1928, Einstein:19282,Schroedinger:1950,Moller:1961, Pellegrini:1963, Hayashi:1967se,Cho:1975dh,Hehl:1976kt, Hehl:1976kv, Hehl:1976my, Hehl:1977fj,Kijowski:1978,Hayashi:1979qx, Nester:1998mp, BeltranJimenez:2019odq} (see~\cite{Heisenberg:2018vsk, BeltranJimenez:2019esp, Rigouzzo:2022yan,Heisenberg:2023lru} for an overview).  
For example, assuming that both $Q_{\alpha\mu\nu}$ and $T^\alpha_{~~\beta \gamma}$ vanish leads to the commonly-used metric variant of GR~\cite{Einstein:1915} corresponding to Riemannian geometry, whereas Einstein-Cartan (EC) gravity is defined by including both $\rcR{}$ and $T^\alpha_{~~\beta \gamma}$ while excluding $Q_{\alpha\mu\nu}$~\cite{Cartan:1922,Cartan:1923,Cartan:1924,Cartan:1925,Einstein:1925, Einstein:1928,Einstein:19282}. The EC version stands out since it can be derived by gauging the Poincar\'e group~\cite{Utiyama:1956sy, Kibble:1961ba, Sciama:1962}, which puts gravity on the same footing as the other fundamental forces.
This fact -- together with its phenomenological virtues -- motivates us to focus on EC gravity in the following.
 
 At first sight, the various versions of GR appear to be very different. In the classical, matter-free limit, however, they are all fully equivalent.
There are two main ways to break this equivalence. First, one can couple matter (non-minimally) to GR. In this case, the spectrum of gravity remains unchanged, \ie no additional propagating \dof emerge apart from the massless spin-2 graviton, but new effects arise at high energies~\cite{Freidel:2005sn,Bauer:2008zj,Poplawski:2011xf,Diakonov:2011fs,Khriplovich:2012xg,Magueijo:2012ug,Khriplovich:2013tqa,Markkanen:2017tun,Carrilho:2018ffi,Enckell:2018hmo,Rasanen:2018fom,Rubio:2019ypq,Shaposhnikov:2020geh,Karananas:2020qkp,Langvik:2020nrs,Shaposhnikov:2020gts,Mikura:2020qhc,Shaposhnikov:2020aen,Kubota:2020ehu,Enckell:2020lvn,Iosifidis:2021iuw,Racioppi:2021ynx,Cheong:2021kyc,Dioguardi:2021fmr,Piani:2022gon,Dux:2022kuk,Rasanen:2022ijc,Gialamas:2023emn,Gialamas:2023flv,Piani:2023aof,Poisson:2023tja,Rigouzzo:2023sbb}. An example, which we shall discuss later, is as a novel mechanism for producing dark matter in EC gravity~\cite{Shaposhnikov:2020aen}.  Second, one can add terms nonlinear in curvature. Almost inevitably, this leads to the existence of new dynamical particles~\cite{Stelle:1977ry,Neville:1978bk,Neville:1979rb,Sezgin:1979zf,Hayashi:1979wj, Hayashi:1980qp}. 
These additional propagating \dof are generically plagued by various types of inconsistencies~\cite{Stelle:1977ry,Neville:1978bk,Neville:1979rb,Sezgin:1979zf,Hayashi:1979wj, Hayashi:1980qp,Hecht:1990wn,Hecht:1991jh,Chen:1998ad}. 
Finding a model which is consistent and weakly coupled has turned out to be challenging and, for torsion, so far was only achieved for a propagating scalar mode~\cite{Hecht:1996np,Chen:1998ad,Yo:1999ex,Yo:2001sy}. 

In this paper, we will use the concept of partial Lagrange multipliers~\cite{Barker:2022jsh} to obtain for the first time a propagating torsion vector from a curvature-squared term in a model for which consistency can be proven. In addition to a proof of concept for a new class of models, our findings can have implications for phenomenology since dynamical (Abelian) massive vector fields have been employed for inflation~\cite{Ford:1989me,Golovnev:2008cf,Chiba:2008eh,Koivisto:2008xf,Kanno:2008gn,Emami:2016ldl,Rodriguez:2017wkg}, dark energy~\cite{Armendariz-Picon:2004say,Koivisto:2007bp,Koivisto:2008xf,Tasinato:2014eka,DeFelice:2016yws,DeFelice:2016uil,BeltranJimenez:2016afo,deFelice:2017paw,Rodriguez:2017wkg,Nakamura:2018oyy,Heisenberg:2020xak,Benisty:2021sul,deRham:2021efp} and dark matter~\cite{Hambye:2008bq,Hambye:2009fg,Arina:2009uq,Hisano:2010yh,Diaz-Cruz:2010czr,Lebedev:2011iq,Farzan:2012hh,Baek:2012se,Belyaev:2016icc,Arcadi:2020jqf,Barman:2021qds}, among others (see also~\cite{Bekenstein:1971hc,Heisenberg:2017hwb,Garcia-Saenz:2021uyv} for implications for black holes). In these works, however, an a priori independent vector field was considered as an ad hoc addition to GR. Our theory endows these Einstein-Proca models with a gravitational origin. This is not only conceptually appealing but also leads to constraints on parameters, which would otherwise seem arbitrary from an effective field theory perspective. Moreover, we show that the massive vector mode can be integrated out in relevant parts of parameter space, thereby establishing a connection to works with non-propagating torsion using the example of~\cite{Shaposhnikov:2020aen}.

\section{Obstructions to dynamical torsion}
The simplest choice of action for EC gravity is (see~\cite{Hehl:1976kj, Shapiro:2001rz, Blagojevic:2002, Blagojevic:2012bc} for reviews, and \cref{Conventions} for details of our conventions)
\begin{equation} \label{EC}
	L_{\text{EC}}  \equiv -\frac{1}{2} \Planck^2 \rcR{} 	+ \TLambda[Q]{_{\mu\nu\sigma}}Q{^{\mu\nu\sigma}}+L_{\text{M}}\left(\Gamma\right) \;,
\end{equation}
where the multiplier field $\TLambda[Q]{_{\mu\nu\sigma}}$ enforces the vanishing of nonmetricity and we included matter $L_{\text{M}}\left(\Gamma\right)$. We can decompose curvature in its Riemannian part $\rR{}$ and contributions from irreducible parts of torsion:
	\begin{align}
		\rcR{}\left(Q_{\alpha\mu\nu}\mapsto 0\right) & = \rR{}
		+
		\frac{8}{9}
		\T[1]{_{\mu\nu\sigma}}
		\T[1]{^{\mu[\nu\sigma]}}
		-
		\frac{2}{3}
		\T[2]{_{\mu}}
		\T[2]{^{\mu}}	
		\nonumber\\
		&\ \ \ 
		+
		\frac{3}{2}
		\T[3]{_{\mu}}
		\T[3]{^{\mu}}	
		-
		2\rD{_\mu}\T[2]{^\mu} \;.
\label{decomposition}
	\end{align}
Varying~\cref{EC} \wrt torsion and without matter, it follows that $\T[1]{^{\mu\nu\sigma}}\approx\T[2]{^{\mu}}\approx\T[3]{^{\mu}}\approx 0$. Plugging this result back in the action shows that the theory \eqref{EC} is equivalent to the metric formulation of GR, $L_{\text{GR}} \equiv -\frac{1}{2} \Planck^2 \rR{}$. 

Adding terms that are quadratic in curvature~\cite{Utiyama:1956sy,Sciama:1964wt,Hayashi:1980ir,Hayashi:1980qp} to~\eqref{EC} generically leads to two types of problems caused by additional propagating d.o.f. The first one consists of ghost and tachyonic instabilities, \ie kinetic or mass terms with the wrong sign~\cite{Stelle:1977ry,Neville:1978bk,Neville:1979rb,Sezgin:1979zf,Hayashi:1979wj, Hayashi:1980qp}. Superficially, one can eliminate these inconsistencies by tuning the kernel of the linearised wave operator, and indeed a `zoo' of such unitary cases is available~\cite{Sezgin:1981xs,Blagojevic:1983zz,Blagojevic:1986dm,Kuhfuss:1986rb,Yo:1999ex,Yo:2001sy,Blagojevic:2002,Puetzfeld:2004yg,Yo:2006qs,Shie:2008ms,Nair:2008yh,Nikiforova:2009qr,Chen:2009at,Ni:2009fg,Baekler:2010fr,Ho:2011qn,Ho:2011xf,Ong:2013qja,Puetzfeld:2014sja,Karananas:2014pxa,Ni:2015poa,Ho:2015ulu,Karananas:2016ltn,Obukhov:2017pxa,Blagojevic:2017ssv,Blagojevic:2018dpz,Tseng:2018feo,Lin:2018awc,BeltranJimenez:2019acz,Zhang:2019mhd,Aoki:2019rvi,Zhang:2019xek,Jimenez:2019qjc,Lin:2019ugq,Percacci:2019hxn,Barker:2020gcp,BeltranJimenez:2020sqf,Barker:2021oez,MaldonadoTorralba:2020mbh,Marzo:2021esg,Marzo:2021iok,delaCruzDombriz:2021nrg,Baldazzi:2021kaf,Annala:2022gtl}. However, a second class of issues, first discussed in the metric-compatible teleparallel formulation of GR~\cite{Einstein:1928, Einstein:19282, Moller:1961, Pellegrini:1963, Hayashi:1967se,Cho:1975dh, Hayashi:1979qx} and its generalizations \cite{Hayashi:1979qx,Dimakis:1989az,Dimakis:1989ba,Lemke:1990su,Hecht:1990wn,Hecht:1991jh,Afshordi:2006ad,Magueijo:2008sx}, is only visible in a nonlinear analysis~\cite{Velo:1969txo,Aragone:1971kh,Cheng:1988zg,Chen:1998ad,Chen:1998ad,Blixt:2018znp,Blixt:2019ene,Blixt:2020ekl,Krasnov:2021zen,Bahamonde:2021gfp,Delhom:2022vae}. Applied to EC gravity, an especially restrictive report~\cite{Yo:2001sy} found that for propagating tensor/vector torsion, weak-field linearisation causes over-counting of the constraints~\cite{Blagojevic:1983zz,Blagojevic:1986dm,Blagojevic:2002,Ong:2013qja,Blagojevic:2018dpz,Barker:2021oez} -- peculiar to torsion theories, the activated fields are ghosts whenever the linear spectrum is unitary~\cite{Hayashi:1980qp,Blagojevic:1983zz,Yo:2001sy,Blagojevic:2013dea,Blagojevic:2013taa,BeltranJimenez:2019hrm,Aoki:2020rae,Barker:2021oez}. In the modern parlance (see~\cref{SCIntro},~\cite{Yo:2001sy,Charmousis:2008ce,Charmousis:2009tc,Papazoglou:2009fj,Baumann:2011dt,Baumann:2011dt,Wang:2017brl,JimenezCano:2021rlu,Barker:2022kdk,Delhom:2022vae,Annala:2022gtl}), this means that these new particles become \emph{strongly coupled} on Minkowski spacetime, which is thereby rendered dynamically unreachable~\cite{DAmico:2011eto,Gumrukcuoglu:2012aa,Mazuet:2017rgq,BeltranJimenez:2020lee,Barker:2022kdk}.\footnote{Such issues of strong coupling have been discovered in a variety of theories~\cite{Vainshtein:1972sx,Deffayet:2001uk,Deffayet:2005ys,Charmousis:2008ce,Charmousis:2009tc,Papazoglou:2009fj,deRham:2014zqa,Deser:2014hga,Wang:2017brl}.} As stated above, models that avoid this problem and remain weakly coupled have so far only been constructed for a propagating \emph{scalar} mode of torsion~\cite{Hecht:1996np,Yo:1999ex,Yo:2001sy} (see e.g.~\cite{Yo:2006qs,Shie:2008ms,Chen:2009at,Baekler:2010fr,Ho:2011qn,Ho:2011xf,Ho:2015ulu,Tseng:2018feo,Zhang:2019mhd,Zhang:2019xek,MaldonadoTorralba:2020mbh,delaCruzDombriz:2021nrg} and~\cite{Puetzfeld:2004yg,Ni:2009fg,Puetzfeld:2014sja,Ni:2015poa,Barker:2020gcp} for applications and reviews, respectively, and~\cite{BeltranJimenez:2019acz,BeltranJimenez:2019esp,Percacci:2020ddy,BeltranJimenez:2020sqf,Marzo:2021esg,Piva:2021nyj,Marzo:2021iok,Iosifidis:2021xdx,Jimenez-Cano:2022sds,Iosifidis:2023pvz} for similar prospects in the nonmetric sector).

In this paper we construct a healthy theory with \emph{propagating vector} (PV) torsion. Extending~\cref{EC}, it is defined
\begin{align}
\LPV{}&\equiv-\frac{1}{2}\Planck^2\rcR{}
+2\alp{5}\rcR{_{[\mu\nu]}}
\rcR{^{[\mu\nu]}}
+\bet{2}\Planck^2\T[2]{_{\mu}}\T[2]{^{\mu}}
\nonumber\\  
	&\hspace{-30pt} 
+\TLambda[1]{_{\mu}^{\nu\sigma}}\T[1]{^{\mu}_{\nu\sigma}}
+\TLambda[3]{^\mu}\T[3]{_\mu}
+\TLambda[Q]{_{\mu\nu\sigma}}Q{^{\mu\nu\sigma}}
+L_{\text{M}}\left(\Gamma\right) \;,\label{DC}
\end{align}
where $\TLambda[1]{_{\mu}^{\nu\sigma}}$ and $\TLambda[3]{^\mu}$ are two Lagrange multipliers and $\alp{5}<0$ throughout.\footnote{One can construct an analogous theory $\LPV{(2)\rightleftarrows(3)}$ with $\alp{5}>0$.} Unlike in recent attempts~\cite{BeltranJimenez:2019acz,BeltranJimenez:2019esp,BeltranJimenez:2020sqf,Jimenez-Cano:2022sds,Iosifidis:2023pvz}, we will demonstrate how consistent dynamics and unitarity are simultaneously upheld by \emph{complete} nonlinear analysis. Our results represent a twofold progress. First, we do not postulate the existence of a kinetic term for $\T[2]{_\mu}$, as is commonly done in EC gravity (see \eg~\cite{Carroll:1994dq,Belyaev:1997zv, Shapiro:2001rz,Vollick:2006uq,Belyaev:2016icc,Barman:2019mlj,Katanaev:2020xdv}). Instead, it arises from the curvature-squared term $\rcR{_{[\mu\nu]}} \rcR{^{[\mu\nu]}}$. Second, it is known that other formulations of GR allow for consistent propagating vectors contained in nonmetricity \cite{BeltranJimenez:2014iie,Iosifidis:2018diy,Iosifidis:2018zwo,Helpin:2019vrv,OrejuelaGarcia:2020viw,Ghilencea:2020piz,BeltranJimenez:2020sih,Xu:2020yeg,Yang:2021fjy,Quiros:2021eju,Quiros:2022uns,Yang:2022icz,Burikham:2023bil,Haghani:2023nrm} (see also~\cite{Aringazin:1991,Vitagliano:2010pq,Iosifidis:2021xdx}) -- indeed our $\rcR{_{[\mu\nu]}}$ in~\cref{DC} is analogous to Weyl's homothetic curvature~\cite{BeltranJimenez:2016wxw,Iosifidis:2018jwu}. Therefore, our theory establishes a direct analogy of \emph{vector torsion} with Weyl's vector nonmetricity.

\section{Nontrivial effect of multipliers} 
Our key innovation in our~\cref{DC} is the \emph{partial} multiplier fields~\cite{Barker:2022jsh}. Without them, the pure $\alp{5}\rcR{_{[\mu\nu]}}\rcR{^{[\mu\nu]}}$ operator propagates $\T[3]{_\mu}$ and $\T[2]{_\mu}$ vectors, both of which are strongly-coupled~\cite{Yo:2001sy,Barker:2022jsh} (absent from the linear spectrum), whilst $\T[3]{_\mu}$ is additionally a ghost (for ${\alpha<0}$). Overlooking strong-coupling, we first naively try to `suppress' the $\T[3]{_\mu}$ ghost solely using the multiplier $\TLambda[3]{^\mu}$ in the theory $L_{\text{PV}}\left(\TLambda[1]{_\mu^{\nu\sigma}},\bet{2}\mapsto 0\right)$. We package six \dof into the 2-form field 
\begin{equation}\label{KalbRamond}
	\begin{aligned}
		\B{_{\mu\nu}}&\equiv 
		\Planck^{-1}\Big(2\rD{_{\sigma}}\T[1]{^{\sigma}_{[\mu\nu]}}
	-\F[2]{_{\mu\nu}}
	+\frac{3}{4}\tensor{\epsilon}{^{\sigma\lambda}_{\mu\nu}}\F[3]{_{\sigma\lambda}}\\
		&\hspace{20pt}
		+2\T[2]{_{\sigma}}\T[1]{^{\sigma}_{[\mu\nu]}}
		+3\tensor{\epsilon}{^{\sigma\lambda\rho}_{[\mu}}\T[3]{_{\sigma}}\T[1]{_{\nu]\lambda\rho}}\Big) \;,
	\end{aligned}
\end{equation}
where the Maxwell field strengths are $\F[2]{_{\mu\nu}}\equiv 2\rD{_{[\mu}}\T[2]{_{\nu]}}$ etc. We introduce~\cref{KalbRamond} just to simplify the $\tensor{\omega}{^{ij}_\mu}$-equations $\delta/\delta \tensor{\omega}{^{ij}_\mu}\int\mathrm{d}^4xeL_{\text{naive}}\approx 0$, which decompose to (see Supplemental Material \cite{FieldEquations} for details about the subsequent computations)
\begin{subequations}
	\begin{align}
		\T[1]{^\sigma_{\mu\nu}}
		&\approx
		\frac{\alp{5}}{\Planck}\left[
		\OmitIndices{\rD{}\B{}}
		+\frac{\OmitIndices{\B{}\rD{}\B{}}}{\Planck}
		+\frac{\OmitIndices{\B{}^2\T[1]{}}}{\Planck}
		\right] \;,\label{TensorEquation}\\
		\T[2]{_\mu}
		&\approx
		\frac{4\alp{5}}{3\Planck}\Big[\rD{_\nu}\B{_\mu^\nu}
		-\B{_{\sigma\lambda}}\T[1]{_{\mu}^{\sigma\lambda}}\Big]\;,\label{VectorEquation}
	\end{align}
\end{subequations}
and the algebraic relation $\TLambda[3]{^\mu}\approx\dots$, with $\T[3]{_\mu}\approx 0$,
where $\OmitIndices{...}$ suppresses contractions, but all parts are simplified by~\eqref{KalbRamond}. With $\alp{5}\mapsto 0$ we recover entirely vanishing vacuum torsion as expected in EC theory, otherwise~\crefrange{TensorEquation}{VectorEquation} should be wavelike for dynamical torsion (if any). It is simplest to notice how all torsion dynamics can be confined to $\B{_{\mu\nu}}$, though that variable eliminates a single derivative in~\cref{KalbRamond}. To extract the propagating (second-derivative) equation in $\B{_{\mu\nu}}$, we take the antisymmetrised divergence of~\eqref{TensorEquation}, next eliminating $\rD{_{\sigma}}\T[1]{^{\sigma}_{[\mu\nu]}}$ for $\B{_{\mu\nu}}$, $\T[1]{^\mu_{\nu\sigma}}$ and $\T[2]{_\mu}$ using~\cref{KalbRamond}, then using~\cref{VectorEquation} to eliminate $\T[2]{_\mu}$ for $\B{_{\mu\nu}}$ and $\T[1]{^\mu_{\nu\sigma}}$, before finally recycling~\cref{TensorEquation} to eliminate all remaining $\T[1]{^\mu_{\nu\sigma}}$ \emph{perturbatively} in terms of $\B{_{\mu\nu}}$. Upon integrating, we find (at least on flat space without matter) that the resulting equation descends from the effective theory 
\begin{align}\label{KalbRamondEffectiveTheory}
	&L_{\text{PV}}\left(\TLambda[1]{_\mu^{\nu\sigma}},\bet{2}\mapsto 0\right)
	\cong 	
	-\frac{\Planck^2}{2}\B{_{\mu\nu}}\B{^{\mu\nu}}
	+2\alp{5}\rD{_{[\mu}}\B{_{\nu\sigma]}}\rD{^{[\mu}}\B{^{\nu\sigma]}}
\nonumber\\
&\hspace{40pt} 
	+\frac{\alp{5}^2}{\Planck}\OmitIndices{\B{}^2\rD{}^2\B{}}
	+\frac{\alp{5}^3}{\Planck^3}\OmitIndices{\B{}^2\rD{}^4\B{}}+\dots
\end{align} 
At~$\mathcal{O}\left(B^2\right)$, \cref{KalbRamondEffectiveTheory} reduces to massive $p$-form electrodynamics~\cite{Henneaux:1986ht} and hence propagates \emph{three} \dof from $\B{_{\mu\nu}}$, while $\rD{^\mu}\big(\delta L_{\text{naive}}/\delta\B{^{\mu\nu}}\big)\propto \rD{_\mu}\B{^\mu_{\nu}}\approx 0$ will constrain the remaining three --- formerly the $\T[3]{_\mu}$ ghost.
Already the fact that the flat-space linear spectrum contains \emph{anything} is remarkable: we accidentally fixed strong coupling of the healthy $\T[2]{_\mu}$ mode by trying to kill the ghost! More strangely still, the ghost is not even dead: the constraint does not survive at~$\mathcal{O}\left(B^3\right)$, so the ghost remains \emph{nonlinearly} active (see~\cref{SCIntro} for a toy model illustrating strong coupling). This conclusion is even true if we ignore the four-derivative term in \cref{KalbRamondEffectiveTheory}, which may lead to additional problems of its own.

\section{Healthy spectrum with multipliers} 
\begin{table}
\begin{tabularx}{\linewidth}{X|c|c|c|c|c|c|c|c}
\hline\hline
	Linear d.o.f & \UnitaryTensor{} & \multicolumn{3}{c|}{\UnitaryTensor{}+\UnitaryVector{}+\GhostVector{}} & \multicolumn{4}{c}{\UnitaryTensor{}+\UnitaryVector{}}\\
\hline
	Nonlinear d.o.f & \UnitaryTensor{}+\UnitaryStrongVector{}+\GhostStrongVector{} & \multicolumn{3}{c|}{\UnitaryTensor{}+\UnitaryVector{}+\GhostVector{}} & \multicolumn{2}{c|}{\UnitaryTensor{}+\UnitaryVector{}+\GhostStrongVector{}} & \multicolumn{2}{c}{\UnitaryTensor{}+\UnitaryVector{}}\\
\hline
	$+\TLambda[3]{^\mu}\T[3]{_\mu}$ &\xmark &\xmark &\xmark &\cmark &\cmark &\xmark &\cmark &\cmark \\
	$+\TLambda[1]{_{\mu}^{\nu\sigma}}\T[1]{^{\mu}_{\nu\sigma}}$ &\xmark &\cmark &\cmark &\xmark &\xmark &\xmark &\cmark &\cmark \\
	$+\bet{2}\Planck^2\T[2]{_{\mu}}\T[2]{^{\mu}}$ &\xmark &\xmark &\cmark &\cmark &\xmark &\cmark &\xmark &\cmark \\
\hline
	Remarks & {\color{red}Failure} & \multicolumn{5}{c|}{{\color{red}Counterintuitive failure}} & \multicolumn{2}{c}{{\color{SeaGreen}Success}}\\
\hline\hline
\end{tabularx}
	\caption{\label{WhacAMole} Additions to $L=-\frac{1}{2}\Planck^2\rcR{}+2\alp{5}\rcR{_{[\mu\nu]}}\rcR{^{[\mu\nu]}}$ needed for a \emph{single} (i.e. non-ghost) $\T[2]{_{\mu}}$ vector in~\cref{DC} with $\alp{5}<0$. Counterintuitively, the pathological $\T[3]{_{\mu}}$ mode \emph{cannot} just trivially be removed via a Lagrange multiplier without activating other ghosts (subscript `g') or strongly-coupled vectors (subscript `s'); see Supplemental Material \cite{NonTrivialMultipliers}.
		All these problems are resolved by our $\TLambda[1]{_{\mu}^{\nu\sigma}}$, which also yields a `conventional' massless limit for $\T[2]{_{\mu}}$ in~\cref{Mass2}.}
\end{table}

Our remarkable experience in~\cref{KalbRamondEffectiveTheory} leads to a highly non-trivial game of `Whac-a-Mole' against strongly-coupled ghosts, whose final score is shown in~\cref{WhacAMole}. 
In another unexpected twist, it turns out that the introduction of the explicit mass parameter $\bet{2}$ has a multiplier-like effect, despite \emph{implicit} $\T[2]{_{\mu}}$ masses being already present within the basic $-\frac{1}{2}\Planck^2\rcR{}$ operator in~\cref{decomposition}. Ultimately, our $\TLambda[1]{_{\mu\nu\sigma}}\T[1]{^{\mu\nu\sigma}}$ term is the cure, even though $\T[1]{^{\mu\nu\sigma}}$ was never implicated in the original pathology~\cite{Yo:2001sy}. For general $\alp{5}$, we restore the whole axial vector sector and, using the spin tensor of matter $e\tensor{S}{^\mu_{ij}}\equiv -2\delta/\delta \tensor{\omega}{^{ij}_\mu}\int\mathrm{d}^4xeL_{\text{M}}$, we obtain the following effective torsion-free theory, to be compared with~\cref{KalbRamondEffectiveTheory}:
\begin{align}\label{MasterLagrangian}
	&\hspace{-15pt}\LPV{\TLambda[3]{^\mu}\mapsto 0}+\bet{3}\Planck^2\T[3]{^\mu}\T[3]{_\mu}
	\nonumber\\
	&
	\cong 
	-\frac{\Planck^2\rR{}}{2}
	+
	\frac{2\alp{5}}{9}
	\F[2]{_{\mu\nu}}
	\F[2]{^{\mu\nu}}
	-
	\frac{\alp{5}}{2}
	\F[3]{_{\mu\nu}}
	\F[3]{^{\mu\nu}}
	\nonumber\\
	&
	+\Planck^2
	\left(
	\tfrac{\left(1+3\bet{2}\right)}{3}
	\T[2]{_{\mu}}
	\T[2]{^{\mu}}	
	-
	\tfrac{\left(3-4\bet{3}\right)}{4}
	\T[3]{_{\mu}}
	\T[3]{^{\mu}}	
	\right)
	\nonumber\\
	&
	-
	\frac{1}{3}
	\T[2]{_{\mu}}
	\Si[2]{^{\mu}}	
	-
	\frac{3}{2}
	\T[3]{_{\mu}}
	\Si[3]{^{\mu}}
	+L_{\text{M}}\left(\rCon{}\right) \;.
\end{align}
In~\cref{MasterLagrangian}, we notice that the residual torsion reduces to the Proca pair, one of which is a ghost, and the full model $\LPV{}$, or $\LPV{(2)\rightleftarrows(3)}$, kills off the ghost in either case. 
In~\cref{DC}, valid for $\alp{5}<0$, the mass of $\T[2]{_\mu}$ is 
\begin{equation}
	\mass[2]{}^2\equiv-3\Planck^2(1+3\bet{2})/4\alp{5} \;.\label{Mass2}
\end{equation}

\begin{figure}[t]
  \center
  \includegraphics[width=\linewidth]{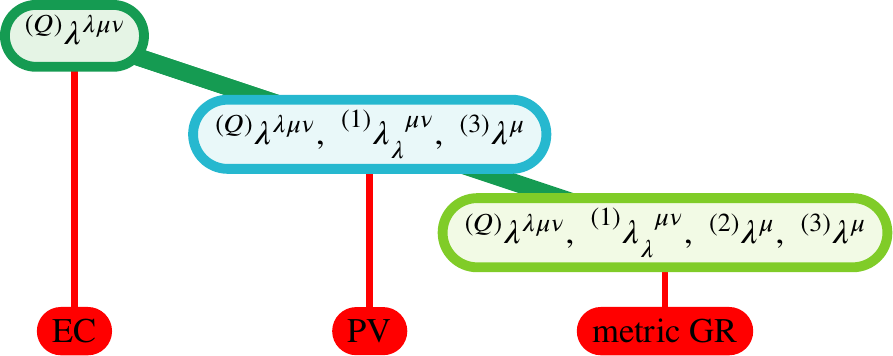}
	\caption{\label{Propaganda} 
	Multipliers are already ubiquitous in gravity. Placement of our PV theory~\cref{DC}, relative to EC~\cite{Cartan:1922,Cartan:1923,Cartan:1924,Cartan:1925,Einstein:1925, Einstein:1928,Einstein:19282} and metric GR~\cite{Einstein:1915}.
	}
\end{figure}

We shall briefly comment on how our model \eqref{DC} relates to the known formulations of GR. Since some of the irreducible representations of torsion vanish because of partial multipliers~\cite{Barker:2022jsh}, one could classify it as interpolating between the metric and EC formulations, as illustrated in~\cref{Propaganda}. However, our theory \eqref{DC} also shares properties with teleparallel versions of GR, in which curvature is excluded~\cite{Einstein:1928, Einstein:19282, Moller:1961, Pellegrini:1963, Hayashi:1967se,Cho:1975dh, Hayashi:1979qx, Nester:1998mp, BeltranJimenez:2019odq}. The reason is that -- as shown -- the multiplier $\TLambda[1]{_{\mu}^{\nu\sigma}}$ changes the spectrum. The same happens in teleparallel theories for the Lagrange multiplier that enforces the vanishing of curvature (see~\cite{BeltranJimenez:2018vdo,Rigouzzo:2022yan}).

\section{Illustration of strong coupling alleviation} 
\begin{figure}[t]
  \center
  \includegraphics[width=\linewidth]{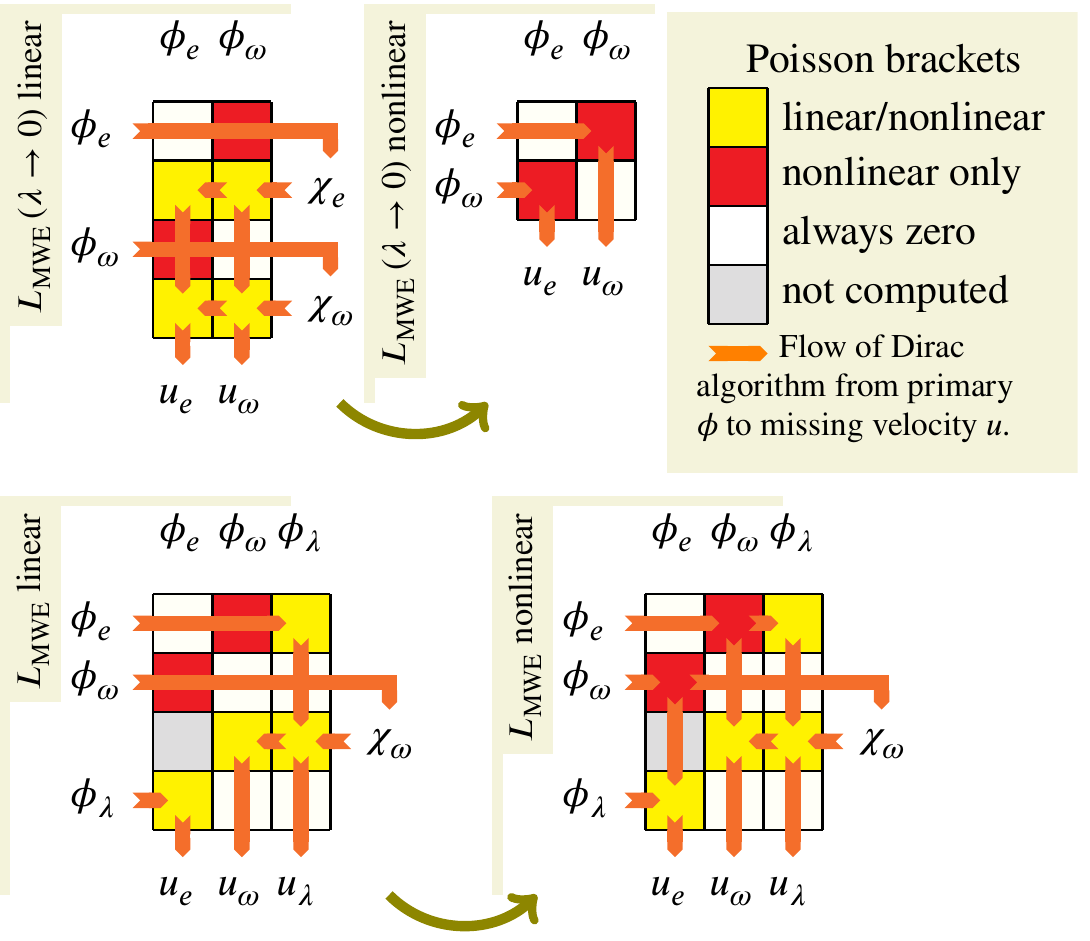}
	\caption{\label{ConstraintAlgebra} 
	Our solution to non-Riemannian strong coupling can easily be understood through the 1D minimal working example in~\cref{MWE}. Without our method, the $\chi_e$, $\chi_\omega$ constraints convert into a nonlinear torsion d.o.f. Accordingly, in actual nonlinear gravity~\cite{Yo:2001sy,Barker:2021oez}, constraints are converted into the $\B{_{\mu\nu}}$ field in~\cref{KalbRamondEffectiveTheory}.
	}
\end{figure}

We shall illustrate the mechanism by which $\TLambda[1]{_{\mu}^{\nu\sigma}}$ alleviates the strong coupling problem in a simple example -- the corresponding analysis for our theory \eqref{DC}, as output of the software HiGGS \cite{Barker:2022kdk}, is presented in~\cref{FullAnalysis}. Assuming familiarity with the Dirac algorithm~\cite{Anderson:1951ta,Bergmann:1954tc,Castellani:1981us,Chen:1998ad,Yo:1999ex,Yo:2001sy} (see good pedagogical introductions~\cite{Henneaux:1992ig,Blagojevic:2002,Golovnev:2022rui}), we consider analogs of `\emph{tetrad}' $\qe{}$, `\emph{spin connection}' $\qw{}$ and `\emph{torsion}' $\dqe{}+\qw{}$ in a minimal working example (MWE) \begin{equation}\label{MWE}
	\begin{aligned}
		L_{\text{MWE}}&\equiv\qw{}\left[\qe{}\left(1+\qw{}\right)-1\right]\left(\dqw{}+1\right)\\
			&\ \ \ \ \ +\left(\qe{}-1\right)^2+\lambda\left(\dqe{}+\qw{}\right) \;,
	\end{aligned}
\end{equation}
with `\emph{Minkowski background}' $\qe{}-1\approx\qw{}\approx0$.
First try omitting the multiplier: the definitions $p_i\equiv\partial/\partial\dot{q}_i L_{\text{MWE}}\left(\lambda\mapsto 0\right)$ engender two \emph{primary} constraints $\phie{}\equiv\pie{}\approx 0$ and $\phiw{}\equiv\piw{}-\qw{}\left[\qe{}\left(1+\qw{}\right)-1\right]\approx 0$, so the Hamiltonian is $H_{\text{MWE}}\left(\lambda\mapsto 0\right)\equiv\sum_i u_i\phi_i-\qw{}\left[\qe{}\left(1+\qw{}\right)-1\right]-\left(\qe{}-1\right)^2$, where $u_i$ replace Dirac's missing/uninvertible velocities $\dot{q}_i$. The Poisson bracket $\left\{\phie{},\phiw{}\right\}\approx \qw{}\left(1+\qw{}\right)$ has the sickly property that it \emph{vanishes} in the background linearisation of~\cref{MWE}. This feature is the popular understanding of strongly coupled torsion~\cite{Cheng:1988zg,Chen:1998ad,Yo:1999ex,Yo:2001sy,Blixt:2020ekl,Barker:2021oez}. For a vanishing bracket, the consistency conditions $\dot{\phi}_i\equiv\left\{\phi_i,H_{\text{MWE}}\right\}\approx0$ engender \emph{secondary} constraints $\chie{}\equiv 2\left(\qe{}-1\right)+\qw{}\left(\qw{}+1\right)\approx 0$ and $\chiw{}\equiv\qe{}\left(1+2\qw{}\right)-1\approx 0$. Since $\{\chie{},\phie{}\}\approx 2$ and $\{\chiw{},\phiw{}\}\approx 2\qe{}$ do not vanish, so $\dot{\chi}_i\approx 0$ solve for the $u_i$ and terminate the algorithm with \emph{zero} dynamical d.o.f. In the nonlinear theory $\dot{\phi}_i\approx 0$ \emph{already} solve for the $u_i$, the $\chi_i$ are not induced, and a single \dof (two unconstrained Cauchy data) is \emph{activated}~\cite{Golovnev:2022rui}. Next, with $\lambda$ we additionally have $\phil{}\equiv\pil{}\approx 0$ and this time it is a trivial exercise to confirm that background linearisation \emph{changes nothing} (see~\cref{ConstraintAlgebra}). In this metaphor, we compare $\lambda$ in~\cref{MWE} with $\TLambda[1]{^\sigma_{\mu\nu}}$ in~\cref{DC}, the effect is to drag the nonlinear \dof down into the linear regime, and thereby solve a major problem in non-Riemannian gravity~\cite{Kopczynski:1982,Cheng:1988zg,Dimakis:1989az,Dimakis:1989ba,Lemke:1990su,Hecht:1990wn,Chen:1998ad,Yo:1999ex,Yo:2001sy,BeltranJimenez:2019hrm,Blixt:2020ekl,Barker:2021oez}.

\section{Application to dark matter}

We can broadly distinguish two situations, according to whether or not the Proca field $\T[2]{_{\mu}}$ propagates at the relevant energy scales. The first case has been considered in~\cite{Bekenstein:1971hc,Ford:1989me,Armendariz-Picon:2004say,Koivisto:2007bp,Koivisto:2008xf,Hambye:2008bq,Golovnev:2008cf,Chiba:2008eh,Koivisto:2008xf,Kanno:2008gn,Hambye:2009fg,Arina:2009uq,Hisano:2010yh,Diaz-Cruz:2010czr,Lebedev:2011iq,Farzan:2012hh,Baek:2012se,Tasinato:2014eka,DeFelice:2016yws,DeFelice:2016uil,BeltranJimenez:2016afo,Belyaev:2016icc,Emami:2016ldl,Heisenberg:2017hwb,Rodriguez:2017wkg,deFelice:2017paw,Nakamura:2018oyy,Arcadi:2020jqf,Heisenberg:2020xak,Benisty:2021sul,deRham:2021efp,Barman:2021qds,Garcia-Saenz:2021uyv}. Whereas all these works used as a starting point an \emph{effective} Lagrangian with a structure similar to \cref{MasterLagrangian}, our theory \eqref{DC} can endow such models with a fundamental origin in EC gravity.

Formula \eqref{Mass2} for the vector mass hints towards the second case, in which torsion does not propagate at the relevant energy scales. Namely, we already know that $\alp{5}<0$ and $\bet{2}>-1/3$. If $\bet{2}$ is not too close to the border value (\eg $\bet{2}\gtrsim0$) and $|\alp{5}|$ is at most of order one, we observe that $\mass[2]{}^2 \gtrsim \Planck^2$. Thus, one is tempted to conclude that parameter choices for which the Proca field does not propagate below the Planck scale are more generic. This is good news since then the energy scale at which scattering amplitudes involving the massive vector field violate perturbative unitarity also lies above the Planck scale. Consequently, introducing propagating torsion with the theory \eqref{DC} generically does not lower the cutoff scale, above which the theory ceases to be predictive. 

Integrating out the field $\T[2]{_{\mu}}$ at energies below $\mass[2]{}$, we get from \eq \eqref{MasterLagrangian} (with  $\T[3]{_{\mu}}\mapsto 0$)
\begin{equation} \label{effectiveLagrangian}
	\mathcal{L}_{\text{T}} = - \frac{\Planck^2}{2}  \rR{} - \frac{1}{18 \mass[2]{}^2} \Si[2]{_{\mu}} \Si[2]{^{\mu}} + 	L_{\text{M}}\left(\rCon{}\right) \;.
\end{equation}
As a concrete example, we shall focus on fermionic fields in $L_{\text{M}}\left(\rCon{}\right)$ and first consider a generic $\Si[2]{_{\mu}} = -3 \sum_j \big(\zeta_V V^{(j)}_\mu + \zeta_A A^{(j)}_\mu\big)$, where $j$ sums species, $\zeta_V$ and $\zeta_A$ are real constants and $V_\mu = \bar{\Psi} \gamma_\mu \Psi$ and $A_\mu = \bar{\Psi} \gamma_5 \gamma_\mu \Psi$ represent the fermionic vector and axial currents, respectively. Following~\cite{Shaposhnikov:2020aen}, we shall take into account the field content of the Standard Model and add to it a singlet fermion $N$ of mass $M_N$. Such a scenario is strongly motivated since a small mixing with active neutrinos can generate neutrino masses via the seesaw mechanism (see~\cite{Boyarsky:2018tvu} for a review). For the present discussion, however, we shall not specify the precise nature of $N$. Independently of a possible connection to neutrinos, a singlet fermion is a dark matter candidate, but this option is only viable if a sufficient production of $N$ takes places in the early Universe. 

This can be achieved both by a propagating torsion field acting as mediator~\cite{Barman:2019mlj} and by the effective 4-fermion interaction arising from \eq \eqref{effectiveLagrangian} ~\cite{Shaposhnikov:2020aen}. We focus on the second case and consider the parameter choice $\zeta_V \gg 1$ and $\zeta_A \gg 1$, which can arise from a non-minimally coupled kinetic term of the fermion, $i/2 \bar{\Psi} (1+ 2 i \zeta_V + 2 i \zeta_A \gamma_5) \gamma^\mu \mathcal{D}_\mu \Psi + \text{h.c.}$ (see~\cite{Shaposhnikov:2020frq,Karananas:2021zkl,Rigouzzo:2023sbb}).  Then the produced abundance of fermions is~\cite{Shaposhnikov:2020aen}
\begin{equation}
	\frac{\Omega_N}{\Omega_{DM}}\simeq 1.7 \, C \, \frac{M_P}{\mass[2]{}} \left( \frac{M_N}{10~\text{keV}}\right) \, \left( \frac{T_\mathrm{prod}}{\mass[2]{}} \right)^3 \;.
	\label{darkMatterAbundance}
\end{equation}
Here $T_\mathrm{prod}$ is the highest temperature at which fermions are generated, \ie generically the temperature of the hot Big Bang, and $C = \left(15(\zeta_V^2 - \zeta_A^2)^2 + 7(\zeta_V + \zeta_A)^4 + 8 (\zeta_V - \zeta_A)^4\right)$ for a Dirac fermion $N$. Choosing $C$ sufficiently large, producing all of dark matter, $\Omega_N=\Omega_{DM}$, can be achieved for a wide range of masses $M_N$ down to $\text{few}~\text{keV}$, where even lighter fermions are excluded due to well-known observational bounds on warm dark matter~\cite{Garzilli:2018jqh, Garzilli:2019qki}. The resulting mass hierarchy typically is $\mass[2]{}\gg M_P \gg T_\mathrm{prod}$. Also $\mass[2]{}< M_P$ is possible, as long as $T_\mathrm{prod}<\mass[2]{}$ so that $\T[2]{_{\mu}}$ does not propagate.

\section{Conclusions} 
Several long-standing problems related to gravity have remained unsolved. Since many of the proposed solutions call for the introduction of new particles, one naturally wonders if GR can provide them and indeed it is easy to generate additional propagating degrees of freedom, \eg from higher powers of curvature terms. However, such models are generically plagued by problems due to ghost instabilities~\cite{Stelle:1977ry,Neville:1978bk,Neville:1979rb,Sezgin:1979zf,Hayashi:1979wj, Hayashi:1980qp} and strong coupling~\cite{Hecht:1996np,Yo:1999ex,Yo:2001sy}. In this paper, we have simultaneously addressed both issues by providing the first example of propagating vector torsion  arising from a curvature-squared invariant in EC gravity, for which consistency can be proven at the full nonlinear level.

Our theory \eqref{DC} endows effective Einstein-Proca models with a fundamental gravitational origin. This leads to significant constraints on the parameter space even in situations in which torsion does not propagate, and one can say that the remaining freedom is `just right'. On the one hand, it is possible to address some of the unresolved puzzles in cosmology with approaches that do not exist in the metric formulation of GR. As an example, we discussed a novel mechanism for producing fermionic dark matter~\cite{Shaposhnikov:2020aen}. On the other hand, theories such as the one that we have constructed generically feature significantly fewer free parameters as compared to effective Einstein-Proca models~\cite{Bekenstein:1971hc,Ford:1989me,Armendariz-Picon:2004say,Koivisto:2007bp,Golovnev:2008cf,Chiba:2008eh,Koivisto:2008xf,Kanno:2008gn,Koivisto:2008xf,Hambye:2008bq,Hambye:2009fg,Arina:2009uq,Hisano:2010yh,Diaz-Cruz:2010czr,Lebedev:2011iq,Farzan:2012hh,Baek:2012se,Tasinato:2014eka,DeFelice:2016yws,Emami:2016ldl,DeFelice:2016uil,BeltranJimenez:2016afo,deFelice:2017paw,Heisenberg:2017hwb,Rodriguez:2017wkg,Nakamura:2018oyy,Heisenberg:2020xak,Arcadi:2020jqf,Benisty:2021sul,deRham:2021efp,Barman:2021qds,Garcia-Saenz:2021uyv} or proposals~\cite{Freidel:2005sn,Bauer:2008zj,Poplawski:2011xf,Diakonov:2011fs,Khriplovich:2012xg,Magueijo:2012ug,Khriplovich:2013tqa,Markkanen:2017tun,Carrilho:2018ffi,Enckell:2018hmo,Rasanen:2018fom,Rubio:2019ypq,Shaposhnikov:2020geh,Karananas:2020qkp,Langvik:2020nrs,Shaposhnikov:2020gts,Mikura:2020qhc,Shaposhnikov:2020aen,Kubota:2020ehu,Enckell:2020lvn,Iosifidis:2021iuw,Racioppi:2021ynx,Cheong:2021kyc,Dioguardi:2021fmr,Piani:2022gon,Dux:2022kuk,Rasanen:2022ijc,Gialamas:2023emn,Gialamas:2023flv,Piani:2023aof,Poisson:2023tja,Rigouzzo:2023sbb} with non-propagating torsion, and hence are more predictive. 

Despite the attractive by-products in this case (\eg dark matter production), it may be argued that there is \emph{not presently a fundamental need} for a new massive vector in cosmology and fundamental physics and many of the open issues, such as inflation and dark energy, can be addressed more easily with scalar fields. It is interesting therefore that the theoretical efforts required to extract such a vector from torsion are quite strenuous. Our paper shows this very clearly (see~\cref{WhacAMole}); the torsion vectors are badly entangled with the tensor mode. So one might even conclude that the conceptual difficulties in constructing a theory with a propagating torsion vector match nicely with the absence of phenomenological indications for the existence of an additional vector field.

Several directions for future research emerge from our findings. First, our method of analysis has the potential to rule out many of the models~\cite{Sezgin:1981xs,Blagojevic:1983zz,Blagojevic:1986dm,Kuhfuss:1986rb,Yo:1999ex,Yo:2001sy,Blagojevic:2002,Puetzfeld:2004yg,Yo:2006qs,Nair:2008yh,Shie:2008ms,Chen:2009at,Nikiforova:2009qr,Ni:2009fg,Baekler:2010fr,Ho:2011qn,Ho:2011xf,Ong:2013qja,Puetzfeld:2014sja,Karananas:2014pxa,Karananas:2014pxa,Ho:2015ulu,Ni:2015poa,Karananas:2016ltn,Obukhov:2017pxa,Blagojevic:2017ssv,Blagojevic:2018dpz,Lin:2018awc,Blagojevic:2018dpz,Tseng:2018feo,Blagojevic:2018dpz,Zhang:2019xek,Zhang:2019mhd,Jimenez:2019qjc,Lin:2019ugq,BeltranJimenez:2019acz,Aoki:2019rvi,Percacci:2019hxn,BeltranJimenez:2020sqf,Barker:2020gcp,MaldonadoTorralba:2020mbh,Marzo:2021esg,delaCruzDombriz:2021nrg,Marzo:2021iok,Baldazzi:2021kaf,Barker:2021oez,Annala:2022gtl} that -- as a result of linear study only -- are seemingly consistent. Second, we have developed an approach to construct consistent theories. In particular, it remains to be determined if our theory is special or if classes of models with analogous properties exist in EC gravity and other formulations of GR. Third, our model can constrain the parameter space in inflationary models that suffer from a loss of predictivity due to numerous unknown coupling constants (see \eg \cite{Bauer:2008zj,Rodriguez:2017wkg,Rasanen:2018ihz, Raatikainen:2019qey, Langvik:2020nrs, Shaposhnikov:2020gts,Rigouzzo:2022yan}) and have further observables consequences such as in gravitational waves~\cite{Battista:2021rlh}. Finally, consistent gravitational theories with curvature-squared terms may provide new approaches for the ultimate challenge of UV-completing GR~\cite{Lin:2019ugq,Melichev:2023lwj}.

\begin{acknowledgments}
This work was performed using resources provided by the Cambridge Service for Data Driven Discovery (CSD3) operated by the University of Cambridge Research Computing Service (\href{www.csd3.cam.ac.uk}{www.csd3.cam.ac.uk}), provided by Dell EMC and Intel using Tier-2 funding from the Engineering and Physical Sciences Research Council (capital grant EP/T022159/1), and DiRAC funding from the Science and Technology Facilities Council (\href{www.dirac.ac.uk}{www.dirac.ac.uk}).

We thank Jaakko Annala, Daniel Blixt, Francisco José Maldonado Torralba, Carlo Marzo, Roberto Percacci, Syksy Räsänen, Claire Rigouzzo, Inar Timiryasov and Tom Złośnik for useful comments. Moreover we are indebted to anonymous referees for insightful feedback. W.B. is grateful for the kind hospitality of Leiden University and the Lorentz Institute, and the support of Girton College, Cambridge. S.Z.~acknowledges support of the Fonds de la Recherche Scientifique -- FNRS.
\end{acknowledgments}

\appendix

\section{Conventions}\label{Conventions}

In this appendix we quote our conventions. The metric $g_{\mu\nu}$ has signature $\left(+,-,-,-\right)$ and the covariant derivative of a covector $A_\nu$ is $\rcD{_\mu} A_\nu = \partial_\mu A_\nu - \Gamma^\alpha_{~~ \mu\nu} A_\alpha$ with an a priori independent affine connection $\Gamma^\alpha_{~~\beta \gamma}$ and torsion $\tensor{T}{^\mu_{\nu\sigma}}\equiv -2\rcCon{^\mu_{[\nu\sigma]}}\equiv\frac{4}{3}\T[1]{^\mu_{[\nu\sigma]}}+\frac{2}{3}\tensor*{\delta}{^\mu_{[\nu}}\T[2]{_{\sigma]}}+\tensor{\epsilon}{^\mu_{\nu\sigma\lambda}}\T[3]{^\lambda}$ decomposed into tensor $\T[1]{^\lambda_{\mu\nu}}\equiv\frac{1}{2}\tensor{T}{^\lambda_{\mu\nu}}+\tfrac{1}{2}\tensor{T}{_\mu^\lambda_\nu}+\tfrac{1}{6}\tensor{g}{_{\mu\nu}}\T[2]{^\lambda}-\tfrac{1}{6}\tensor*{\delta}{^\lambda_\nu}\T[2]{_\mu}+\tfrac{1}{3}\tensor*{\delta}{^\lambda_\mu}\T[2]{_\nu}$, vector $\T[2]{_\mu}\equiv\tensor{T}{^\lambda_{\lambda\mu}}$ and pseudovector $\T[3]{_\mu}\equiv\tfrac{1}{6}\tensor{\epsilon}{_{\mu\nu\sigma\lambda}}\tensor{T}{^{\nu\sigma\lambda}}$ parts, with nonmetricity $Q_{\alpha\mu\nu}\equiv\nabla_\alpha g_{\mu\nu}$, where anti-symmetrization on some tensor $T_{[\mu\nu]}$ is defined as $T_{[\mu\nu]} \equiv \frac{1}{2}(T_{\mu\nu}-T_{\nu\mu})$. The curvature has contractions $\rcR{}\equiv\rcR{^\lambda_\lambda}\equiv\rcR{^{\mu\nu}_{\mu\nu}}$, with $\rcR{_\lambda^\sigma_{\mu\nu}}\equiv 2\PD{_{[\nu}}\rcCon{^\sigma_{\mu]\lambda}}+2\rcCon{^\sigma_{[\nu|\kappa}}\rcCon{^\kappa_{|\mu]\lambda}}$, and the Riemann curvature $\rR{_\lambda^\sigma_{\mu\nu}}$ is analogously defined in terms of the metric-compatible and torsion-free Levi-Civita connection $\rCon{^\sigma_{\mu\nu}}$, with covariant derivative $\rD{_\mu}$. We also use the tetrad $\tensor{e}{_i^\mu}\tensor{e}{_j^\nu}\tensor{\eta}{^{ij}}\equiv\tensor{g}{^{\mu\nu}}$, with $e\equiv\det\tensor{e}{^i_\mu}\equiv\sqrt{-g}\equiv\sqrt{-\det\tensor{g}{_{\mu\nu}}}$, and independent spin connection $\tensor{\omega}{^{ij}_\mu}$. For our gamma matrices, we use the convention $\left\{\gamma_i, \gamma_j \right\} \equiv - 2 \eta_{ij}$, $\gamma_5 \equiv i \gamma_0 \gamma_1 \gamma_2 \gamma_3$, with Greek indices on gamma matrices implying contraction with an (inverse) tetrad field, and in flat space with Cartesian coordinates the Levi-Civita tensor has the same components as the symbol, $\epsilon_{0123}\equiv 1$.

\begin{figure}[t]
  \center
  \includegraphics[width=\linewidth]{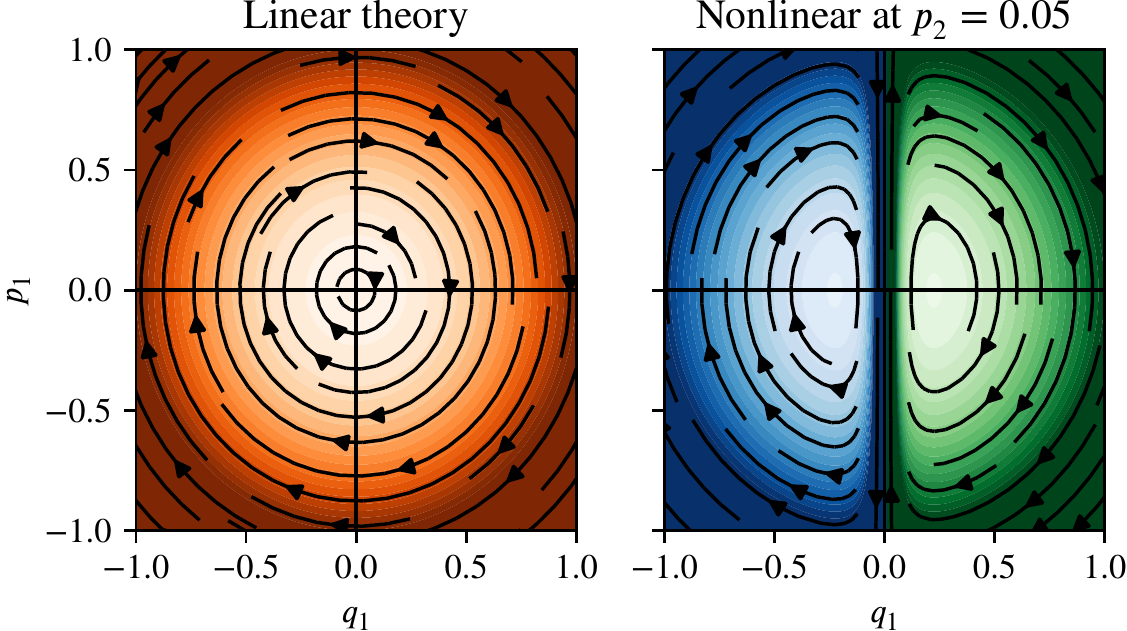}
	\caption{\label{PhaseSpace} 
	Strong coupling splits the perturbative vacuum. Left: the linear approximation to~\cref{ToyModel} erroneously anticipates the stable vacuum. Right: in the bulk of the true nonlinear phase space this vacuum is torn apart by a high-energy separatrix.
	}
\end{figure}

\begin{figure*}[h]
  \center
	\vspace{20pt}
  \includegraphics[width=\linewidth]{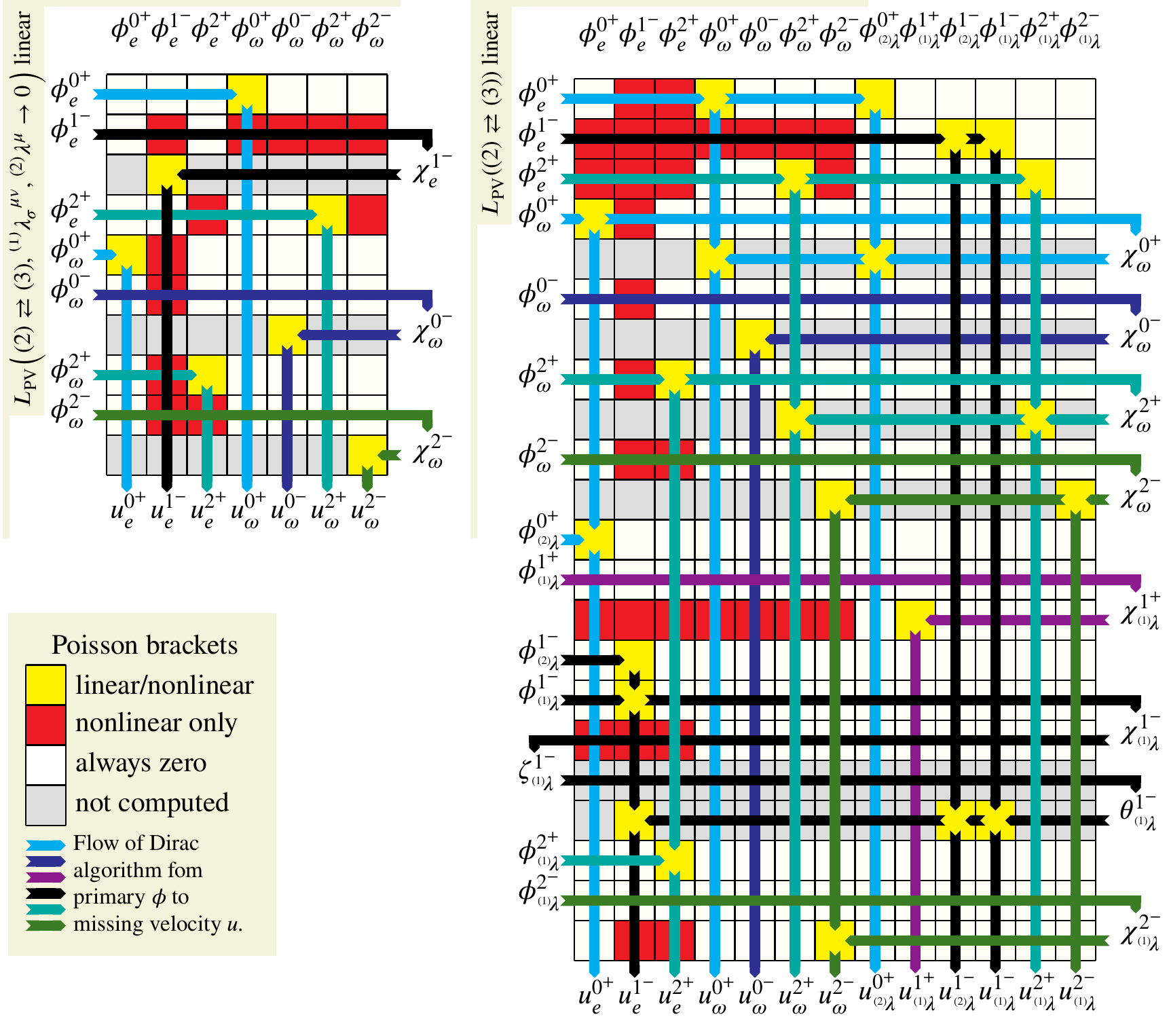}
	\caption{\label{FullAnalysisFigure} 
	On-shell constraint algebra of $\LPV{(2)\rightleftarrows(3)}$ in~\eqref{DC}, with and without our multipliers, with the logical flow of the \emph{linear} Dirac algorithm superimposed. This figure supplements~\cref{ConstraintAlgebra} for completeness. Without our method, the induced secondary constraints $\tensor*{\chi}{_e^{1^-}}$, $\tensor*{\chi}{_\omega^{0^-}}$, $\tensor*{\chi}{_\omega^{2^-}}$ of the theory~\cref{DC} would collectively lose six (canonical) \dof in nonlinear gravity~\cite{Yo:2001sy,Barker:2021oez}, activating half of the $\B{_{\mu\nu}}$ field in~\cref{KalbRamond}. If $\bet{2}\to 0$ had additionally been enforced, the whole of $\B{_{\mu\nu}}$ would have become activated. The reason we do not show the nonlinear algebra in this case is because the loss of secondaries occurs \emph{among} specific components of given spin sectors, so the convenient spin-parity decomposition of the algebra falls apart. The full nonlinear constraint chains are listed in~\cite{HamiltonianAnalysis}.
	}
\end{figure*}

\section{Toy-model of strong coupling}\label{SCIntro}

\vspace{10pt}

In this appendix, we clarify what we mean by \emph{strong coupling}, and why it is commonly diagnosed in the Hamiltonian framework. Consider the double-oscillator toy model (TM) of \dof $q_1$ and $q_2$ in one dimension
\begin{equation}\label{ToyModel}
	L_{\text{TM}}=\frac{1}{2}\dot{q}_1^2-\frac{1}{2}q_1^2+\frac{1}{2}q_1^2\dot{q}_2^2-\frac{1}{2}q_2^2 \;.
\end{equation}
The perturbative `approximation' to~\cref{ToyModel} is $L_{\text{TM}}=\dot{q}_1^2/2-q_1^2/2-q_2^2/2+\mathcal{O}\left(q^4\right)$, for which the second oscillator drops out algebraically as $q_2\approx 0$, leaving the Hamiltonian phase subspace $\left(q_1,p_1\equiv\dot{q}_1\right)$ in which the first oscillator orbits the perfectly healthy vacuum at $q_1\approx p_1\approx 0$ with Hamilton's equations $\left(\dot{q}_1\approx p_1,\dot{p}_1\approx-q_1\right)$. However, the true physics always lives in the space $\left(q_1,p_1,q_1,p_2\right)$, in which the Hamilton equation $\dot{p}_1\approx -q_1+p_2^2/q_1^2$ is \emph{inherently non-perturbative}. As illustrated in~\cref{PhaseSpace}, the effect is absolutely fatal to the perturbative vacuum. Generalising to field theory, strongly-coupled \dof, analogous to $q_2$, reveal themselves via structural changes to the Hamiltonian constraint structure when passing from linearised to nonlinear gravity theories. In such cases it is not clear how the background of the linearisation can then be a viable spacetime, even though it may perfectly well be an exact solution to the nonlinear field equations. The sick kinetic structure of~\eqref{ToyModel} may be compared to that in~\eqref{KalbRamondEffectiveTheory}.

It is important to point out that the splitting of the vacuum in~\cref{PhaseSpace} is an effect which \emph{need not} be associated with ghost modes. Indeed, in~\cref{ToyModel} we are careful to preserve the sign of the kinetic term. In our practical, field theoretic example in~\cref{KalbRamondEffectiveTheory} however, it may turn out to be harder to bound the Hamiltonian due to the multiple components of $\B{_{\mu\nu}}$. 

\section{Full nonlinear Hamiltonian analysis}\label{FullAnalysis}

In this appendix, we quote the full, nonlinear Hamiltonian analysis of $\LPV{(2)\rightleftarrows(3)}$ corresponding to $\alp{5}>0$ w.l.o.g. The full Dirac algorithm is shown schematically in~\cref{FullAnalysisFigure}, using precisely the same notation as in~\cref{ConstraintAlgebra}. Note the appearance of tertiary $\zeta$ and quaternary $\theta$ constraints. The full nonlinear constraint chains are found in~\cite{HamiltonianAnalysis}, these are fully computed by the HiGGS software based only on the definition of the Lagrangian~\cite{Barker:2022kdk}. The key observation is that the size of the algebra (i.e. the number of induced non-primary constraints) does not change in passing from the linear to the nonlinear theories (though we do not exclude this effect at other points which punctuate the bulk of the phase space). This is consistent with the \emph{manifestly} healthy and dynamically equivalent Einstein-Proca model.

\bibliography{Manuscript}

\begin{thebibliography}{252}%
\makeatletter
\providecommand \@ifxundefined [1]{%
 \@ifx{#1\undefined}
}%
\providecommand \@ifnum [1]{%
 \ifnum #1\expandafter \@firstoftwo
 \else \expandafter \@secondoftwo
 \fi
}%
\providecommand \@ifx [1]{%
 \ifx #1\expandafter \@firstoftwo
 \else \expandafter \@secondoftwo
 \fi
}%
\providecommand \natexlab [1]{#1}%
\providecommand \enquote  [1]{``#1''}%
\providecommand \bibnamefont  [1]{#1}%
\providecommand \bibfnamefont [1]{#1}%
\providecommand \citenamefont [1]{#1}%
\providecommand \href@noop [0]{\@secondoftwo}%
\providecommand \href [0]{\begingroup \@sanitize@url \@href}%
\providecommand \@href[1]{\@@startlink{#1}\@@href}%
\providecommand \@@href[1]{\endgroup#1\@@endlink}%
\providecommand \@sanitize@url [0]{\catcode `\\12\catcode `\$12\catcode
  `\&12\catcode `\#12\catcode `\^12\catcode `\_12\catcode `\%12\relax}%
\providecommand \@@startlink[1]{}%
\providecommand \@@endlink[0]{}%
\providecommand \url  [0]{\begingroup\@sanitize@url \@url }%
\providecommand \@url [1]{\endgroup\@href {#1}{\urlprefix }}%
\providecommand \urlprefix  [0]{URL }%
\providecommand \Eprint [0]{\href }%
\providecommand \doibase [0]{http://dx.doi.org/}%
\providecommand \selectlanguage [0]{\@gobble}%
\providecommand \bibinfo  [0]{\@secondoftwo}%
\providecommand \bibfield  [0]{\@secondoftwo}%
\providecommand \translation [1]{[#1]}%
\providecommand \BibitemOpen [0]{}%
\providecommand \bibitemStop [0]{}%
\providecommand \bibitemNoStop [0]{.\EOS\space}%
\providecommand \EOS [0]{\spacefactor3000\relax}%
\providecommand \BibitemShut  [1]{\csname bibitem#1\endcsname}%
\let\auto@bib@innerbib\@empty
\bibitem [{\citenamefont {Abbott}\ \emph {et~al.}(2016)\citenamefont {Abbott}
  \emph {et~al.}}]{LIGOScientific:2016aoc}%
  \BibitemOpen
  \bibfield  {author} {\bibinfo {author} {\bibfnamefont {B.~P.}\ \bibnamefont
  {Abbott}} \emph {et~al.} (\bibinfo {collaboration} {LIGO Scientific,
  Virgo}),\ }\bibfield  {title} {\enquote {\bibinfo {title} {{Observation of
  Gravitational Waves from a Binary Black Hole Merger}},}\ }\href {\doibase
  10.1103/PhysRevLett.116.061102} {\bibfield  {journal} {\bibinfo  {journal}
  {Phys. Rev. Lett.}\ }\textbf {\bibinfo {volume} {116}},\ \bibinfo {pages}
  {061102} (\bibinfo {year} {2016})},\ \Eprint
  {http://arxiv.org/abs/1602.03837} {arXiv:1602.03837 [gr-qc]} \BibitemShut
  {NoStop}%
\bibitem [{\citenamefont {Zwicky}(1933)}]{Zwicky:1933gu}%
  \BibitemOpen
  \bibfield  {author} {\bibinfo {author} {\bibfnamefont {F.}~\bibnamefont
  {Zwicky}},\ }\bibfield  {title} {\enquote {\bibinfo {title} {{Die
  Rotverschiebung von extragalaktischen Nebeln}},}\ }\href {\doibase
  10.1007/s10714-008-0707-4} {\bibfield  {journal} {\bibinfo  {journal} {Helv.
  Phys. Acta}\ }\textbf {\bibinfo {volume} {6}},\ \bibinfo {pages} {110--127}
  (\bibinfo {year} {1933})}\BibitemShut {NoStop}%
\bibitem [{\citenamefont {Feng}(2010)}]{Feng:2010gw}%
  \BibitemOpen
  \bibfield  {author} {\bibinfo {author} {\bibfnamefont {Jonathan~L.}\
  \bibnamefont {Feng}},\ }\bibfield  {title} {\enquote {\bibinfo {title} {{Dark
  Matter Candidates from Particle Physics and Methods of Detection}},}\ }\href
  {\doibase 10.1146/annurev-astro-082708-101659} {\bibfield  {journal}
  {\bibinfo  {journal} {Ann. Rev. Astron. Astrophys.}\ }\textbf {\bibinfo
  {volume} {48}},\ \bibinfo {pages} {495--545} (\bibinfo {year} {2010})},\
  \Eprint {http://arxiv.org/abs/1003.0904} {arXiv:1003.0904 [astro-ph.CO]}
  \BibitemShut {NoStop}%
\bibitem [{\citenamefont {Peebles}(2015)}]{Peebles:2013hla}%
  \BibitemOpen
  \bibfield  {author} {\bibinfo {author} {\bibfnamefont {P.~J.~E.}\
  \bibnamefont {Peebles}},\ }\bibfield  {title} {\enquote {\bibinfo {title}
  {{Dark Matter}},}\ }\href {\doibase 10.1073/pnas.1308786111} {\bibfield
  {journal} {\bibinfo  {journal} {Proc. Nat. Acad. Sci.}\ }\textbf {\bibinfo
  {volume} {112}},\ \bibinfo {pages} {2246} (\bibinfo {year} {2015})},\ \Eprint
  {http://arxiv.org/abs/1305.6859} {arXiv:1305.6859 [astro-ph.CO]} \BibitemShut
  {NoStop}%
\bibitem [{\citenamefont {Bertone}\ and\ \citenamefont
  {Hooper}(2018)}]{Bertone:2016nfn}%
  \BibitemOpen
  \bibfield  {author} {\bibinfo {author} {\bibfnamefont {Gianfranco}\
  \bibnamefont {Bertone}}\ and\ \bibinfo {author} {\bibfnamefont {Dan}\
  \bibnamefont {Hooper}},\ }\bibfield  {title} {\enquote {\bibinfo {title}
  {{History of dark matter}},}\ }\href {\doibase 10.1103/RevModPhys.90.045002}
  {\bibfield  {journal} {\bibinfo  {journal} {Rev. Mod. Phys.}\ }\textbf
  {\bibinfo {volume} {90}},\ \bibinfo {pages} {045002} (\bibinfo {year}
  {2018})},\ \Eprint {http://arxiv.org/abs/1605.04909} {arXiv:1605.04909
  [astro-ph.CO]} \BibitemShut {NoStop}%
\bibitem [{\citenamefont {Arbey}\ and\ \citenamefont
  {Mahmoudi}(2021)}]{Arbey:2021gdg}%
  \BibitemOpen
  \bibfield  {author} {\bibinfo {author} {\bibfnamefont {A.}~\bibnamefont
  {Arbey}}\ and\ \bibinfo {author} {\bibfnamefont {F.}~\bibnamefont
  {Mahmoudi}},\ }\bibfield  {title} {\enquote {\bibinfo {title} {{Dark matter
  and the early Universe: a review}},}\ }\href {\doibase
  10.1016/j.ppnp.2021.103865} {\bibfield  {journal} {\bibinfo  {journal} {Prog.
  Part. Nucl. Phys.}\ }\textbf {\bibinfo {volume} {119}},\ \bibinfo {pages}
  {103865} (\bibinfo {year} {2021})},\ \Eprint
  {http://arxiv.org/abs/2104.11488} {arXiv:2104.11488 [hep-ph]} \BibitemShut
  {NoStop}%
\bibitem [{\citenamefont {Starobinsky}(1980)}]{Starobinsky:1980te}%
  \BibitemOpen
  \bibfield  {author} {\bibinfo {author} {\bibfnamefont {Alexei~A.}\
  \bibnamefont {Starobinsky}},\ }\bibfield  {title} {\enquote {\bibinfo {title}
  {{A New Type of Isotropic Cosmological Models Without Singularity}},}\ }\href
  {\doibase 10.1016/0370-2693(80)90670-X} {\bibfield  {journal} {\bibinfo
  {journal} {Phys. Lett. B}\ }\textbf {\bibinfo {volume} {91}},\ \bibinfo
  {pages} {99--102} (\bibinfo {year} {1980})}\BibitemShut {NoStop}%
\bibitem [{\citenamefont {Guth}(1981)}]{Guth:1980zm}%
  \BibitemOpen
  \bibfield  {author} {\bibinfo {author} {\bibfnamefont {Alan~H.}\ \bibnamefont
  {Guth}},\ }\bibfield  {title} {\enquote {\bibinfo {title} {{The Inflationary
  Universe: A Possible Solution to the Horizon and Flatness Problems}},}\
  }\href {\doibase 10.1103/PhysRevD.23.347} {\bibfield  {journal} {\bibinfo
  {journal} {Phys. Rev. D}\ }\textbf {\bibinfo {volume} {23}},\ \bibinfo
  {pages} {347--356} (\bibinfo {year} {1981})}\BibitemShut {NoStop}%
\bibitem [{\citenamefont {Linde}(1982)}]{Linde:1981mu}%
  \BibitemOpen
  \bibfield  {author} {\bibinfo {author} {\bibfnamefont {Andrei~D.}\
  \bibnamefont {Linde}},\ }\bibfield  {title} {\enquote {\bibinfo {title} {{A
  New Inflationary Universe Scenario: A Possible Solution of the Horizon,
  Flatness, Homogeneity, Isotropy and Primordial Monopole Problems}},}\ }\href
  {\doibase 10.1016/0370-2693(82)91219-9} {\bibfield  {journal} {\bibinfo
  {journal} {Phys. Lett. B}\ }\textbf {\bibinfo {volume} {108}},\ \bibinfo
  {pages} {389--393} (\bibinfo {year} {1982})}\BibitemShut {NoStop}%
\bibitem [{\citenamefont {Mukhanov}\ and\ \citenamefont
  {Chibisov}(1981)}]{Mukhanov:1981xt}%
  \BibitemOpen
  \bibfield  {author} {\bibinfo {author} {\bibfnamefont {Viatcheslav~F.}\
  \bibnamefont {Mukhanov}}\ and\ \bibinfo {author} {\bibfnamefont {G.~V.}\
  \bibnamefont {Chibisov}},\ }\bibfield  {title} {\enquote {\bibinfo {title}
  {{Quantum Fluctuations and a Nonsingular Universe}},}\ }\href
  {http://jetpletters.ru/ps/1510/article_23079.shtml} {\bibfield  {journal}
  {\bibinfo  {journal} {JETP Lett.}\ }\textbf {\bibinfo {volume} {33}},\
  \bibinfo {pages} {532--535} (\bibinfo {year} {1981})}\BibitemShut {NoStop}%
\bibitem [{\citenamefont {Riess}\ \emph {et~al.}(1998)\citenamefont {Riess}
  \emph {et~al.}}]{SupernovaSearchTeam:1998fmf}%
  \BibitemOpen
  \bibfield  {author} {\bibinfo {author} {\bibfnamefont {Adam~G.}\ \bibnamefont
  {Riess}} \emph {et~al.} (\bibinfo {collaboration} {Supernova Search Team}),\
  }\bibfield  {title} {\enquote {\bibinfo {title} {{Observational evidence from
  supernovae for an accelerating universe and a cosmological constant}},}\
  }\href {\doibase 10.1086/300499} {\bibfield  {journal} {\bibinfo  {journal}
  {Astron. J.}\ }\textbf {\bibinfo {volume} {116}},\ \bibinfo {pages}
  {1009--1038} (\bibinfo {year} {1998})},\ \Eprint
  {http://arxiv.org/abs/astro-ph/9805201} {arXiv:astro-ph/9805201} \BibitemShut
  {NoStop}%
\bibitem [{\citenamefont {Perlmutter}\ \emph {et~al.}(1999)\citenamefont
  {Perlmutter} \emph {et~al.}}]{SupernovaCosmologyProject:1998vns}%
  \BibitemOpen
  \bibfield  {author} {\bibinfo {author} {\bibfnamefont {S.}~\bibnamefont
  {Perlmutter}} \emph {et~al.} (\bibinfo {collaboration} {Supernova Cosmology
  Project}),\ }\bibfield  {title} {\enquote {\bibinfo {title} {{Measurements of
  $\Omega$ and $\Lambda$ from 42 high redshift supernovae}},}\ }\href {\doibase
  10.1086/307221} {\bibfield  {journal} {\bibinfo  {journal} {Astrophys. J.}\
  }\textbf {\bibinfo {volume} {517}},\ \bibinfo {pages} {565--586} (\bibinfo
  {year} {1999})},\ \Eprint {http://arxiv.org/abs/astro-ph/9812133}
  {arXiv:astro-ph/9812133} \BibitemShut {NoStop}%
\bibitem [{\citenamefont {Akrami}\ \emph {et~al.}(2020)\citenamefont {Akrami}
  \emph {et~al.}}]{Planck:2018jri}%
  \BibitemOpen
  \bibfield  {author} {\bibinfo {author} {\bibfnamefont {Y.}~\bibnamefont
  {Akrami}} \emph {et~al.} (\bibinfo {collaboration} {Planck}),\ }\bibfield
  {title} {\enquote {\bibinfo {title} {{Planck 2018 results. X. Constraints on
  inflation}},}\ }\href {\doibase 10.1051/0004-6361/201833887} {\bibfield
  {journal} {\bibinfo  {journal} {Astron. Astrophys.}\ }\textbf {\bibinfo
  {volume} {641}},\ \bibinfo {pages} {A10} (\bibinfo {year} {2020})},\ \Eprint
  {http://arxiv.org/abs/1807.06211} {arXiv:1807.06211 [astro-ph.CO]}
  \BibitemShut {NoStop}%
\bibitem [{\citenamefont {Ade}\ \emph {et~al.}(2021)\citenamefont {Ade} \emph
  {et~al.}}]{BICEP:2021xfz}%
  \BibitemOpen
  \bibfield  {author} {\bibinfo {author} {\bibfnamefont {P.~A.~R.}\
  \bibnamefont {Ade}} \emph {et~al.} (\bibinfo {collaboration} {BICEP, Keck}),\
  }\bibfield  {title} {\enquote {\bibinfo {title} {{Improved Constraints on
  Primordial Gravitational Waves using Planck, WMAP, and BICEP/Keck
  Observations through the 2018 Observing Season}},}\ }\href {\doibase
  10.1103/PhysRevLett.127.151301} {\bibfield  {journal} {\bibinfo  {journal}
  {Phys. Rev. Lett.}\ }\textbf {\bibinfo {volume} {127}},\ \bibinfo {pages}
  {151301} (\bibinfo {year} {2021})},\ \Eprint
  {http://arxiv.org/abs/2110.00483} {arXiv:2110.00483 [astro-ph.CO]}
  \BibitemShut {NoStop}%
\bibitem [{\citenamefont {Einstein}(1915)}]{Einstein:1915}%
  \BibitemOpen
  \bibfield  {author} {\bibinfo {author} {\bibfnamefont {A}~\bibnamefont
  {Einstein}},\ }\bibfield  {title} {\enquote {\bibinfo {title} {{Die
  Feldgleichungen der Gravitation}},}\ }\href@noop {} {\bibfield  {journal}
  {\bibinfo  {journal} {Sitzungsber. Preuss. Akad. Wiss}\ }\textbf {\bibinfo
  {volume} {18}},\ \bibinfo {pages} {844} (\bibinfo {year} {1915})}\BibitemShut
  {NoStop}%
\bibitem [{\citenamefont {Weyl}(1918)}]{Weyl:1918}%
  \BibitemOpen
  \bibfield  {author} {\bibinfo {author} {\bibfnamefont {H.}~\bibnamefont
  {Weyl}},\ }\bibfield  {title} {\enquote {\bibinfo {title} {{Gravitation and
  electricity}},}\ }\href@noop {} {\bibfield  {journal} {\bibinfo  {journal}
  {Sitzungsber. Preuss. Akad. Wiss.}\ }\textbf {\bibinfo {volume} {26}},\
  \bibinfo {pages} {465} (\bibinfo {year} {1918})}\BibitemShut {NoStop}%
\bibitem [{\citenamefont {Palatini}(1919)}]{Palatini:1919}%
  \BibitemOpen
  \bibfield  {author} {\bibinfo {author} {\bibfnamefont {Attilio}\ \bibnamefont
  {Palatini}},\ }\bibfield  {title} {\enquote {\bibinfo {title} {{Deduzione
  invariantiva delle equazioni gravitazionali dal principio di Hamilton}},}\
  }\href {\doibase 10.1007/BF03014670} {\bibfield  {journal} {\bibinfo
  {journal} {Rendiconti del Circolo Matematico di Palermo}\ }\textbf {\bibinfo
  {volume} {43}},\ \bibinfo {pages} {203--212} (\bibinfo {year}
  {1919})}\BibitemShut {NoStop}%
\bibitem [{\citenamefont {Weyl}(1922)}]{Weyl:1922}%
  \BibitemOpen
  \bibfield  {author} {\bibinfo {author} {\bibfnamefont {H.}~\bibnamefont
  {Weyl}},\ }\href@noop {} {\emph {\bibinfo {title} {{Space-Time-Matter}}}}\
  (\bibinfo  {publisher} {Dover Publications},\ \bibinfo {year}
  {1922})\BibitemShut {NoStop}%
\bibitem [{\citenamefont {Cartan}(1922)}]{Cartan:1922}%
  \BibitemOpen
  \bibfield  {author} {\bibinfo {author} {\bibfnamefont {{\'E}lie}\
  \bibnamefont {Cartan}},\ }\bibfield  {title} {\enquote {\bibinfo {title}
  {{Sur une g{\'e}n{\'e}ralisation de la notion de courbure de Riemann et les
  espaces {\`a} torsion}},}\ }\href@noop {} {\bibfield  {journal} {\bibinfo
  {journal} {Comptes Rendus, Ac. Sc. Paris}\ }\textbf {\bibinfo {volume}
  {174}},\ \bibinfo {pages} {593--595} (\bibinfo {year} {1922})}\BibitemShut
  {NoStop}%
\bibitem [{\citenamefont {Eddington}(1923)}]{Eddington:1923}%
  \BibitemOpen
  \bibfield  {author} {\bibinfo {author} {\bibfnamefont {Arthur~Stanley}\
  \bibnamefont {Eddington}},\ }\href@noop {} {\emph {\bibinfo {title} {{The
  Mathematical Theory of Relativity}}}}\ (\bibinfo  {publisher} {Cambridge
  University Press},\ \bibinfo {year} {1923})\BibitemShut {NoStop}%
\bibitem [{\citenamefont {Cartan}(1923)}]{Cartan:1923}%
  \BibitemOpen
  \bibfield  {author} {\bibinfo {author} {\bibfnamefont {{\'E}lie}\
  \bibnamefont {Cartan}},\ }\bibfield  {title} {\enquote {\bibinfo {title}
  {{Sur les vari{\'e}t{\'e}s {\`a} connexion affine et la th{\'e}orie de la
  relativit{\'e} g{\'e}n{\'e}ralis{\'e}e (premi{\`e}re partie)}},}\ }in\
  \href@noop {} {\emph {\bibinfo {booktitle} {Annales scientifiques de
  l'{\'E}cole normale sup{\'e}rieure}}},\ Vol.~\bibinfo {volume} {40}\
  (\bibinfo {year} {1923})\ pp.\ \bibinfo {pages} {325--412}\BibitemShut
  {NoStop}%
\bibitem [{\citenamefont {Cartan}(1924)}]{Cartan:1924}%
  \BibitemOpen
  \bibfield  {author} {\bibinfo {author} {\bibfnamefont {{\'E}lie}\
  \bibnamefont {Cartan}},\ }\bibfield  {title} {\enquote {\bibinfo {title}
  {{Sur les vari{\'e}t{\'e}s {\`a} connexion affine, et la th{\'e}orie de la
  relativit{\'e} g{\'e}n{\'e}ralis{\'e}e (premi{\`e}re partie)(suite)}},}\ }in\
  \href@noop {} {\emph {\bibinfo {booktitle} {Annales scientifiques de
  l'{\'E}cole Normale Sup{\'e}rieure}}},\ Vol.~\bibinfo {volume} {41}\
  (\bibinfo {year} {1924})\ pp.\ \bibinfo {pages} {1--25}\BibitemShut {NoStop}%
\bibitem [{\citenamefont {Cartan}(1925)}]{Cartan:1925}%
  \BibitemOpen
  \bibfield  {author} {\bibinfo {author} {\bibfnamefont {{\'E}lie}\
  \bibnamefont {Cartan}},\ }\bibfield  {title} {\enquote {\bibinfo {title}
  {{Sur les vari{\'e}t{\'e}s {\`a} connexion affine, et la th{\'e}orie de la
  relativit{\'e} g{\'e}n{\'e}ralis{\'e}e (deuxi{\`e}me partie)}},}\ }in\
  \href@noop {} {\emph {\bibinfo {booktitle} {Annales scientifiques de
  l'{\'E}cole normale sup{\'e}rieure}}},\ Vol.~\bibinfo {volume} {42}\
  (\bibinfo {year} {1925})\ pp.\ \bibinfo {pages} {17--88}\BibitemShut
  {NoStop}%
\bibitem [{\citenamefont {Einstein}(1925)}]{Einstein:1925}%
  \BibitemOpen
  \bibfield  {author} {\bibinfo {author} {\bibfnamefont {A}~\bibnamefont
  {Einstein}},\ }\bibfield  {title} {\enquote {\bibinfo {title} {{Einheitliche
  Feldtheorie von Gravitation und Elektrizit\"at}},}\ }\href@noop {} {\bibfield
   {journal} {\bibinfo  {journal} {Sitzungsber. Preuss. Akad. Wiss}\ }\textbf
  {\bibinfo {volume} {22}},\ \bibinfo {pages} {414} (\bibinfo {year}
  {1925})}\BibitemShut {NoStop}%
\bibitem [{\citenamefont {Einstein}(1928{\natexlab{a}})}]{Einstein:1928}%
  \BibitemOpen
  \bibfield  {author} {\bibinfo {author} {\bibfnamefont {Albert}\ \bibnamefont
  {Einstein}},\ }\bibfield  {title} {\enquote {\bibinfo {title}
  {{Riemanngeometrie mit Aufrechterhaltung des Begriffes des
  Fern-Parallelismus}},}\ }\href@noop {} {\bibfield  {journal} {\bibinfo
  {journal} {Sitzungsber. Preuss. Akad. Wiss}\ }\textbf {\bibinfo {volume}
  {17}},\ \bibinfo {pages} {217} (\bibinfo {year}
  {1928}{\natexlab{a}})}\BibitemShut {NoStop}%
\bibitem [{\citenamefont {Einstein}(1928{\natexlab{b}})}]{Einstein:19282}%
  \BibitemOpen
  \bibfield  {author} {\bibinfo {author} {\bibfnamefont {Albert}\ \bibnamefont
  {Einstein}},\ }\bibfield  {title} {\enquote {\bibinfo {title} {{Neue
  M{\"o}glichkeit f{\"u}r eine einheitliche Feldtheorie von Gravitation und
  Elektrizit{\"a}t}},}\ }\href@noop {} {\bibfield  {journal} {\bibinfo
  {journal} {Sitzungsber. Preuss. Akad. Wiss}\ }\textbf {\bibinfo {volume}
  {18}},\ \bibinfo {pages} {224} (\bibinfo {year}
  {1928}{\natexlab{b}})}\BibitemShut {NoStop}%
\bibitem [{\citenamefont {Schr\"odinger}(1950)}]{Schroedinger:1950}%
  \BibitemOpen
  \bibfield  {author} {\bibinfo {author} {\bibfnamefont {Erwin}\ \bibnamefont
  {Schr\"odinger}},\ }\href@noop {} {\emph {\bibinfo {title} {{Space-Time
  Structure}}}}\ (\bibinfo  {publisher} {Cambridge University Press},\ \bibinfo
  {year} {1950})\BibitemShut {NoStop}%
\bibitem [{\citenamefont {M{\o}ller}(1961)}]{Moller:1961}%
  \BibitemOpen
  \bibfield  {author} {\bibinfo {author} {\bibfnamefont {C.}~\bibnamefont
  {M{\o}ller}},\ }\bibfield  {title} {\enquote {\bibinfo {title} {{Conservation
  Law and Absolute Parallelism in General Relativity}},}\ }\href@noop {}
  {\bibfield  {journal} {\bibinfo  {journal} {K. Dan. Vidensk. Selsk. Mat. Fys.
  Skr.}\ }\textbf {\bibinfo {volume} {1}},\ \bibinfo {pages} {1} (\bibinfo
  {year} {1961})}\BibitemShut {NoStop}%
\bibitem [{\citenamefont {Pellegrini}\ and\ \citenamefont
  {Plebanski}(1963)}]{Pellegrini:1963}%
  \BibitemOpen
  \bibfield  {author} {\bibinfo {author} {\bibfnamefont {C.}~\bibnamefont
  {Pellegrini}}\ and\ \bibinfo {author} {\bibfnamefont {J.}~\bibnamefont
  {Plebanski}},\ }\bibfield  {title} {\enquote {\bibinfo {title} {{Tetrad
  Fields and Gravitational Fields}},}\ }\href@noop {} {\bibfield  {journal}
  {\bibinfo  {journal} {K. Dan. Vidensk. Selsk. Mat. Fys. Skr.}\ }\textbf
  {\bibinfo {volume} {2}},\ \bibinfo {pages} {1} (\bibinfo {year}
  {1963})}\BibitemShut {NoStop}%
\bibitem [{\citenamefont {Hayashi}\ and\ \citenamefont
  {Nakano}(1967)}]{Hayashi:1967se}%
  \BibitemOpen
  \bibfield  {author} {\bibinfo {author} {\bibfnamefont {K.}~\bibnamefont
  {Hayashi}}\ and\ \bibinfo {author} {\bibfnamefont {T.}~\bibnamefont
  {Nakano}},\ }\bibfield  {title} {\enquote {\bibinfo {title} {{Extended
  translation invariance and associated gauge fields}},}\ }\href {\doibase
  10.1143/PTP.38.491} {\bibfield  {journal} {\bibinfo  {journal} {Prog. Theor.
  Phys.}\ }\textbf {\bibinfo {volume} {38}},\ \bibinfo {pages} {491--507}
  (\bibinfo {year} {1967})}\BibitemShut {NoStop}%
\bibitem [{\citenamefont {Cho}(1976)}]{Cho:1975dh}%
  \BibitemOpen
  \bibfield  {author} {\bibinfo {author} {\bibfnamefont {Y.~M.}\ \bibnamefont
  {Cho}},\ }\bibfield  {title} {\enquote {\bibinfo {title} {{Einstein
  Lagrangian as the Translational Yang-Mills Lagrangian}},}\ }\href {\doibase
  10.1103/PhysRevD.14.2521} {\bibfield  {journal} {\bibinfo  {journal} {Phys.
  Rev. D}\ }\textbf {\bibinfo {volume} {14}},\ \bibinfo {pages} {2521}
  (\bibinfo {year} {1976})}\BibitemShut {NoStop}%
\bibitem [{\citenamefont {Hehl}\ \emph
  {et~al.}(1976{\natexlab{a}})\citenamefont {Hehl}, \citenamefont {Kerlick},\
  and\ \citenamefont {Von Der~Heyde}}]{Hehl:1976kt}%
  \BibitemOpen
  \bibfield  {author} {\bibinfo {author} {\bibfnamefont {F.~W.}\ \bibnamefont
  {Hehl}}, \bibinfo {author} {\bibfnamefont {G.~D.}\ \bibnamefont {Kerlick}}, \
  and\ \bibinfo {author} {\bibfnamefont {P.}~\bibnamefont {Von Der~Heyde}},\
  }\bibfield  {title} {\enquote {\bibinfo {title} {{On Hypermomentum in General
  Relativity. 2. The Geometry of Space-Time}},}\ }\href
  {https://doi.org/10.1515/zna-1976-0602} {\bibfield  {journal} {\bibinfo
  {journal} {Z. Naturforsch. A}\ }\textbf {\bibinfo {volume} {31}},\ \bibinfo
  {pages} {524--527} (\bibinfo {year} {1976}{\natexlab{a}})}\BibitemShut
  {NoStop}%
\bibitem [{\citenamefont {Hehl}\ \emph
  {et~al.}(1976{\natexlab{b}})\citenamefont {Hehl}, \citenamefont {Kerlick},\
  and\ \citenamefont {Von Der~Heyde}}]{Hehl:1976kv}%
  \BibitemOpen
  \bibfield  {author} {\bibinfo {author} {\bibfnamefont {F.~W.}\ \bibnamefont
  {Hehl}}, \bibinfo {author} {\bibfnamefont {G.~D.}\ \bibnamefont {Kerlick}}, \
  and\ \bibinfo {author} {\bibfnamefont {P.}~\bibnamefont {Von Der~Heyde}},\
  }\bibfield  {title} {\enquote {\bibinfo {title} {{On Hypermomentum in General
  Relativity. 3. Coupling Hypermomentum to Geometry}},}\ }\href@noop {}
  {\bibfield  {journal} {\bibinfo  {journal} {Z. Naturforsch. A}\ }\textbf
  {\bibinfo {volume} {31}},\ \bibinfo {pages} {823--827} (\bibinfo {year}
  {1976}{\natexlab{b}})}\BibitemShut {NoStop}%
\bibitem [{\citenamefont {Hehl}\ \emph
  {et~al.}(1976{\natexlab{c}})\citenamefont {Hehl}, \citenamefont {Kerlick},\
  and\ \citenamefont {Von Der~Heyde}}]{Hehl:1976my}%
  \BibitemOpen
  \bibfield  {author} {\bibinfo {author} {\bibfnamefont {F.~W.}\ \bibnamefont
  {Hehl}}, \bibinfo {author} {\bibfnamefont {G.~D.}\ \bibnamefont {Kerlick}}, \
  and\ \bibinfo {author} {\bibfnamefont {P.}~\bibnamefont {Von Der~Heyde}},\
  }\bibfield  {title} {\enquote {\bibinfo {title} {{On a New Metric Affine
  Theory of Gravitation}},}\ }\href {\doibase 10.1016/0370-2693(76)90393-2}
  {\bibfield  {journal} {\bibinfo  {journal} {Phys. Lett. B}\ }\textbf
  {\bibinfo {volume} {63}},\ \bibinfo {pages} {446--448} (\bibinfo {year}
  {1976}{\natexlab{c}})}\BibitemShut {NoStop}%
\bibitem [{\citenamefont {Hehl}\ \emph {et~al.}(1977)\citenamefont {Hehl},
  \citenamefont {Kerlick}, \citenamefont {Lord},\ and\ \citenamefont
  {Smalley}}]{Hehl:1977fj}%
  \BibitemOpen
  \bibfield  {author} {\bibinfo {author} {\bibfnamefont {F.~W.}\ \bibnamefont
  {Hehl}}, \bibinfo {author} {\bibfnamefont {G.~D.}\ \bibnamefont {Kerlick}},
  \bibinfo {author} {\bibfnamefont {E.~A.}\ \bibnamefont {Lord}}, \ and\
  \bibinfo {author} {\bibfnamefont {L.~L}\ \bibnamefont {Smalley}},\ }\bibfield
   {title} {\enquote {\bibinfo {title} {{Hypermomentum and the Microscopic
  Violation of the Riemannian Constraint in General Relativity}},}\ }\href
  {\doibase 10.1016/0370-2693(77)90347-1} {\bibfield  {journal} {\bibinfo
  {journal} {Phys. Lett. B}\ }\textbf {\bibinfo {volume} {70}},\ \bibinfo
  {pages} {70--72} (\bibinfo {year} {1977})}\BibitemShut {NoStop}%
\bibitem [{\citenamefont {Kijowski}(1978)}]{Kijowski:1978}%
  \BibitemOpen
  \bibfield  {author} {\bibinfo {author} {\bibfnamefont {Jerzy}\ \bibnamefont
  {Kijowski}},\ }\bibfield  {title} {\enquote {\bibinfo {title} {On a new
  variational principle in general relativity and the energy of the
  gravitational field},}\ }\href@noop {} {\bibfield  {journal} {\bibinfo
  {journal} {General Relativity and Gravitation}\ }\textbf {\bibinfo {volume}
  {9}},\ \bibinfo {pages} {857--877} (\bibinfo {year} {1978})}\BibitemShut
  {NoStop}%
\bibitem [{\citenamefont {Hayashi}\ and\ \citenamefont
  {Shirafuji}(1979)}]{Hayashi:1979qx}%
  \BibitemOpen
  \bibfield  {author} {\bibinfo {author} {\bibfnamefont {Kenji}\ \bibnamefont
  {Hayashi}}\ and\ \bibinfo {author} {\bibfnamefont {Takeshi}\ \bibnamefont
  {Shirafuji}},\ }\bibfield  {title} {\enquote {\bibinfo {title} {{New General
  Relativity}},}\ }\href {\doibase 10.1103/PhysRevD.19.3524} {\bibfield
  {journal} {\bibinfo  {journal} {Phys. Rev. D}\ }\textbf {\bibinfo {volume}
  {19}},\ \bibinfo {pages} {3524--3553} (\bibinfo {year} {1979})},\ \bibinfo
  {note} {[Addendum: Phys.Rev.D 24, 3312--3314 (1982)]}\BibitemShut {NoStop}%
\bibitem [{\citenamefont {Nester}\ and\ \citenamefont
  {Yo}(1999)}]{Nester:1998mp}%
  \BibitemOpen
  \bibfield  {author} {\bibinfo {author} {\bibfnamefont {James~M.}\
  \bibnamefont {Nester}}\ and\ \bibinfo {author} {\bibfnamefont {Hwei-Jang}\
  \bibnamefont {Yo}},\ }\bibfield  {title} {\enquote {\bibinfo {title}
  {{Symmetric teleparallel general relativity}},}\ }\href
  {https://inspirehep.net/literature/476449} {\bibfield  {journal} {\bibinfo
  {journal} {Chin. J. Phys.}\ }\textbf {\bibinfo {volume} {37}},\ \bibinfo
  {pages} {113} (\bibinfo {year} {1999})},\ \Eprint
  {http://arxiv.org/abs/gr-qc/9809049} {arXiv:gr-qc/9809049} \BibitemShut
  {NoStop}%
\bibitem [{\citenamefont {Beltr\'an~Jim\'enez}\ \emph
  {et~al.}(2020{\natexlab{a}})\citenamefont {Beltr\'an~Jim\'enez},
  \citenamefont {Heisenberg}, \citenamefont {Iosifidis}, \citenamefont
  {Jim\'enez-Cano},\ and\ \citenamefont {Koivisto}}]{BeltranJimenez:2019odq}%
  \BibitemOpen
  \bibfield  {author} {\bibinfo {author} {\bibfnamefont {Jose}\ \bibnamefont
  {Beltr\'an~Jim\'enez}}, \bibinfo {author} {\bibfnamefont {Lavinia}\
  \bibnamefont {Heisenberg}}, \bibinfo {author} {\bibfnamefont {Damianos}\
  \bibnamefont {Iosifidis}}, \bibinfo {author} {\bibfnamefont {Alejandro}\
  \bibnamefont {Jim\'enez-Cano}}, \ and\ \bibinfo {author} {\bibfnamefont
  {Tomi~S.}\ \bibnamefont {Koivisto}},\ }\bibfield  {title} {\enquote {\bibinfo
  {title} {{General teleparallel quadratic gravity}},}\ }\href {\doibase
  10.1016/j.physletb.2020.135422} {\bibfield  {journal} {\bibinfo  {journal}
  {Phys. Lett. B}\ }\textbf {\bibinfo {volume} {805}},\ \bibinfo {pages}
  {135422} (\bibinfo {year} {2020}{\natexlab{a}})},\ \Eprint
  {http://arxiv.org/abs/1909.09045} {arXiv:1909.09045 [gr-qc]} \BibitemShut
  {NoStop}%
\bibitem [{\citenamefont {Heisenberg}(2019)}]{Heisenberg:2018vsk}%
  \BibitemOpen
  \bibfield  {author} {\bibinfo {author} {\bibfnamefont {Lavinia}\ \bibnamefont
  {Heisenberg}},\ }\bibfield  {title} {\enquote {\bibinfo {title} {{A
  systematic approach to generalisations of General Relativity and their
  cosmological implications}},}\ }\href {\doibase
  10.1016/j.physrep.2018.11.006} {\bibfield  {journal} {\bibinfo  {journal}
  {Phys. Rept.}\ }\textbf {\bibinfo {volume} {796}},\ \bibinfo {pages} {1--113}
  (\bibinfo {year} {2019})},\ \Eprint {http://arxiv.org/abs/1807.01725}
  {arXiv:1807.01725 [gr-qc]} \BibitemShut {NoStop}%
\bibitem [{\citenamefont {Beltr\'an~Jim\'enez}\ \emph
  {et~al.}(2019)\citenamefont {Beltr\'an~Jim\'enez}, \citenamefont
  {Heisenberg},\ and\ \citenamefont {Koivisto}}]{BeltranJimenez:2019esp}%
  \BibitemOpen
  \bibfield  {author} {\bibinfo {author} {\bibfnamefont {Jose}\ \bibnamefont
  {Beltr\'an~Jim\'enez}}, \bibinfo {author} {\bibfnamefont {Lavinia}\
  \bibnamefont {Heisenberg}}, \ and\ \bibinfo {author} {\bibfnamefont
  {Tomi~S.}\ \bibnamefont {Koivisto}},\ }\bibfield  {title} {\enquote {\bibinfo
  {title} {{The Geometrical Trinity of Gravity}},}\ }\href {\doibase
  10.3390/universe5070173} {\bibfield  {journal} {\bibinfo  {journal}
  {Universe}\ }\textbf {\bibinfo {volume} {5}},\ \bibinfo {pages} {173}
  (\bibinfo {year} {2019})},\ \Eprint {http://arxiv.org/abs/1903.06830}
  {arXiv:1903.06830 [hep-th]} \BibitemShut {NoStop}%
\bibitem [{\citenamefont {Rigouzzo}\ and\ \citenamefont
  {Zell}(2022)}]{Rigouzzo:2022yan}%
  \BibitemOpen
  \bibfield  {author} {\bibinfo {author} {\bibfnamefont {Claire}\ \bibnamefont
  {Rigouzzo}}\ and\ \bibinfo {author} {\bibfnamefont {Sebastian}\ \bibnamefont
  {Zell}},\ }\bibfield  {title} {\enquote {\bibinfo {title} {{Coupling
  metric-affine gravity to a Higgs-like scalar field}},}\ }\href {\doibase
  10.1103/PhysRevD.106.024015} {\bibfield  {journal} {\bibinfo  {journal}
  {Phys. Rev. D}\ }\textbf {\bibinfo {volume} {106}},\ \bibinfo {pages}
  {024015} (\bibinfo {year} {2022})},\ \Eprint
  {http://arxiv.org/abs/2204.03003} {arXiv:2204.03003 [hep-th]} \BibitemShut
  {NoStop}%
\bibitem [{\citenamefont {Heisenberg}(2023)}]{Heisenberg:2023lru}%
  \BibitemOpen
  \bibfield  {author} {\bibinfo {author} {\bibfnamefont {Lavinia}\ \bibnamefont
  {Heisenberg}},\ }\bibfield  {title} {\enquote {\bibinfo {title} {{Review on
  $f(Q)$ Gravity}},}\ }\href@noop {} {\  (\bibinfo {year} {2023})},\ \Eprint
  {http://arxiv.org/abs/2309.15958} {arXiv:2309.15958 [gr-qc]} \BibitemShut
  {NoStop}%
\bibitem [{\citenamefont {Utiyama}(1956)}]{Utiyama:1956sy}%
  \BibitemOpen
  \bibfield  {author} {\bibinfo {author} {\bibfnamefont {Ryoyu}\ \bibnamefont
  {Utiyama}},\ }\bibfield  {title} {\enquote {\bibinfo {title} {{Invariant
  theoretical interpretation of interaction}},}\ }\href {\doibase
  10.1103/PhysRev.101.1597} {\bibfield  {journal} {\bibinfo  {journal} {Phys.
  Rev.}\ }\textbf {\bibinfo {volume} {101}},\ \bibinfo {pages} {1597--1607}
  (\bibinfo {year} {1956})}\BibitemShut {NoStop}%
\bibitem [{\citenamefont {Kibble}(1961)}]{Kibble:1961ba}%
  \BibitemOpen
  \bibfield  {author} {\bibinfo {author} {\bibfnamefont {T.~W.~B.}\
  \bibnamefont {Kibble}},\ }\bibfield  {title} {\enquote {\bibinfo {title}
  {{Lorentz invariance and the gravitational field}},}\ }\href {\doibase
  10.1063/1.1703702} {\bibfield  {journal} {\bibinfo  {journal} {J. Math.
  Phys.}\ }\textbf {\bibinfo {volume} {2}},\ \bibinfo {pages} {212--221}
  (\bibinfo {year} {1961})}\BibitemShut {NoStop}%
\bibitem [{\citenamefont {Sciama}(1962)}]{Sciama:1962}%
  \BibitemOpen
  \bibfield  {author} {\bibinfo {author} {\bibfnamefont {Dennis~W}\
  \bibnamefont {Sciama}},\ }\bibfield  {title} {\enquote {\bibinfo {title} {On
  the analogy between charge and spin in general relativity},}\ }in\ \href@noop
  {} {\emph {\bibinfo {booktitle} {Recent developments in general
  relativity}}}\ (\bibinfo  {publisher} {Pergamon Press},\ \bibinfo {address}
  {Oxford},\ \bibinfo {year} {1962})\ p.\ \bibinfo {pages} {415}\BibitemShut
  {NoStop}%
\bibitem [{\citenamefont {Freidel}\ \emph {et~al.}(2005)\citenamefont
  {Freidel}, \citenamefont {Minic},\ and\ \citenamefont
  {Takeuchi}}]{Freidel:2005sn}%
  \BibitemOpen
  \bibfield  {author} {\bibinfo {author} {\bibfnamefont {Laurent}\ \bibnamefont
  {Freidel}}, \bibinfo {author} {\bibfnamefont {Djordje}\ \bibnamefont
  {Minic}}, \ and\ \bibinfo {author} {\bibfnamefont {Tatsu}\ \bibnamefont
  {Takeuchi}},\ }\bibfield  {title} {\enquote {\bibinfo {title} {{Quantum
  gravity, torsion, parity violation and all that}},}\ }\href {\doibase
  10.1103/PhysRevD.72.104002} {\bibfield  {journal} {\bibinfo  {journal} {Phys.
  Rev. D}\ }\textbf {\bibinfo {volume} {72}},\ \bibinfo {pages} {104002}
  (\bibinfo {year} {2005})},\ \Eprint {http://arxiv.org/abs/hep-th/0507253}
  {arXiv:hep-th/0507253} \BibitemShut {NoStop}%
\bibitem [{\citenamefont {Bauer}\ and\ \citenamefont
  {Demir}(2008)}]{Bauer:2008zj}%
  \BibitemOpen
  \bibfield  {author} {\bibinfo {author} {\bibfnamefont {Florian}\ \bibnamefont
  {Bauer}}\ and\ \bibinfo {author} {\bibfnamefont {Durmus~A.}\ \bibnamefont
  {Demir}},\ }\bibfield  {title} {\enquote {\bibinfo {title} {{Inflation with
  Non-Minimal Coupling: Metric versus Palatini Formulations}},}\ }\href
  {\doibase 10.1016/j.physletb.2008.06.014} {\bibfield  {journal} {\bibinfo
  {journal} {Phys. Lett. B}\ }\textbf {\bibinfo {volume} {665}},\ \bibinfo
  {pages} {222--226} (\bibinfo {year} {2008})},\ \Eprint
  {http://arxiv.org/abs/0803.2664} {arXiv:0803.2664 [hep-ph]} \BibitemShut
  {NoStop}%
\bibitem [{\citenamefont {Poplawski}(2011)}]{Poplawski:2011xf}%
  \BibitemOpen
  \bibfield  {author} {\bibinfo {author} {\bibfnamefont {Nikodem~J.}\
  \bibnamefont {Poplawski}},\ }\bibfield  {title} {\enquote {\bibinfo {title}
  {{Matter-antimatter asymmetry and dark matter from torsion}},}\ }\href
  {\doibase 10.1103/PhysRevD.83.084033} {\bibfield  {journal} {\bibinfo
  {journal} {Phys. Rev. D}\ }\textbf {\bibinfo {volume} {83}},\ \bibinfo
  {pages} {084033} (\bibinfo {year} {2011})},\ \Eprint
  {http://arxiv.org/abs/1101.4012} {arXiv:1101.4012 [gr-qc]} \BibitemShut
  {NoStop}%
\bibitem [{\citenamefont {Diakonov}\ \emph {et~al.}(2011)\citenamefont
  {Diakonov}, \citenamefont {Tumanov},\ and\ \citenamefont
  {Vladimirov}}]{Diakonov:2011fs}%
  \BibitemOpen
  \bibfield  {author} {\bibinfo {author} {\bibfnamefont {Dmitri}\ \bibnamefont
  {Diakonov}}, \bibinfo {author} {\bibfnamefont {Alexander~G.}\ \bibnamefont
  {Tumanov}}, \ and\ \bibinfo {author} {\bibfnamefont {Alexey~A.}\ \bibnamefont
  {Vladimirov}},\ }\bibfield  {title} {\enquote {\bibinfo {title} {{Low-energy
  General Relativity with torsion: A Systematic derivative expansion}},}\
  }\href {\doibase 10.1103/PhysRevD.84.124042} {\bibfield  {journal} {\bibinfo
  {journal} {Phys. Rev. D}\ }\textbf {\bibinfo {volume} {84}},\ \bibinfo
  {pages} {124042} (\bibinfo {year} {2011})},\ \Eprint
  {http://arxiv.org/abs/1104.2432} {arXiv:1104.2432 [hep-th]} \BibitemShut
  {NoStop}%
\bibitem [{\citenamefont {Khriplovich}(2012)}]{Khriplovich:2012xg}%
  \BibitemOpen
  \bibfield  {author} {\bibinfo {author} {\bibfnamefont {I.~B.}\ \bibnamefont
  {Khriplovich}},\ }\bibfield  {title} {\enquote {\bibinfo {title}
  {{Gravitational four-fermion interaction on the Planck scale}},}\ }\href
  {\doibase 10.1016/j.physletb.2012.01.072} {\bibfield  {journal} {\bibinfo
  {journal} {Phys. Lett. B}\ }\textbf {\bibinfo {volume} {709}},\ \bibinfo
  {pages} {111--113} (\bibinfo {year} {2012})},\ \Eprint
  {http://arxiv.org/abs/1201.4226} {arXiv:1201.4226 [gr-qc]} \BibitemShut
  {NoStop}%
\bibitem [{\citenamefont {Magueijo}\ \emph {et~al.}(2013)\citenamefont
  {Magueijo}, \citenamefont {Zlosnik},\ and\ \citenamefont
  {Kibble}}]{Magueijo:2012ug}%
  \BibitemOpen
  \bibfield  {author} {\bibinfo {author} {\bibfnamefont {Jo\~ao}\ \bibnamefont
  {Magueijo}}, \bibinfo {author} {\bibfnamefont {T.~G.}\ \bibnamefont
  {Zlosnik}}, \ and\ \bibinfo {author} {\bibfnamefont {T.~W.~B.}\ \bibnamefont
  {Kibble}},\ }\bibfield  {title} {\enquote {\bibinfo {title} {{Cosmology with
  a spin}},}\ }\href {\doibase 10.1103/PhysRevD.87.063504} {\bibfield
  {journal} {\bibinfo  {journal} {Phys. Rev. D}\ }\textbf {\bibinfo {volume}
  {87}},\ \bibinfo {pages} {063504} (\bibinfo {year} {2013})},\ \Eprint
  {http://arxiv.org/abs/1212.0585} {arXiv:1212.0585 [astro-ph.CO]} \BibitemShut
  {NoStop}%
\bibitem [{\citenamefont {Khriplovich}\ and\ \citenamefont
  {Rudenko}(2013)}]{Khriplovich:2013tqa}%
  \BibitemOpen
  \bibfield  {author} {\bibinfo {author} {\bibfnamefont {I.~B.}\ \bibnamefont
  {Khriplovich}}\ and\ \bibinfo {author} {\bibfnamefont {A.~S.}\ \bibnamefont
  {Rudenko}},\ }\bibfield  {title} {\enquote {\bibinfo {title} {{Gravitational
  four-fermion interaction and dynamics of the early Universe}},}\ }\href
  {\doibase 10.1007/JHEP11(2013)174} {\bibfield  {journal} {\bibinfo  {journal}
  {JHEP}\ }\textbf {\bibinfo {volume} {11}},\ \bibinfo {pages} {174} (\bibinfo
  {year} {2013})},\ \Eprint {http://arxiv.org/abs/1303.1348} {arXiv:1303.1348
  [astro-ph.CO]} \BibitemShut {NoStop}%
\bibitem [{\citenamefont {Markkanen}\ \emph {et~al.}(2018)\citenamefont
  {Markkanen}, \citenamefont {Tenkanen}, \citenamefont {Vaskonen},\ and\
  \citenamefont {Veerm\"ae}}]{Markkanen:2017tun}%
  \BibitemOpen
  \bibfield  {author} {\bibinfo {author} {\bibfnamefont {Tommi}\ \bibnamefont
  {Markkanen}}, \bibinfo {author} {\bibfnamefont {Tommi}\ \bibnamefont
  {Tenkanen}}, \bibinfo {author} {\bibfnamefont {Ville}\ \bibnamefont
  {Vaskonen}}, \ and\ \bibinfo {author} {\bibfnamefont {Hardi}\ \bibnamefont
  {Veerm\"ae}},\ }\bibfield  {title} {\enquote {\bibinfo {title} {{Quantum
  corrections to quartic inflation with a non-minimal coupling: metric vs.
  Palatini}},}\ }\href {\doibase 10.1088/1475-7516/2018/03/029} {\bibfield
  {journal} {\bibinfo  {journal} {JCAP}\ }\textbf {\bibinfo {volume} {03}},\
  \bibinfo {pages} {029} (\bibinfo {year} {2018})},\ \Eprint
  {http://arxiv.org/abs/1712.04874} {arXiv:1712.04874 [gr-qc]} \BibitemShut
  {NoStop}%
\bibitem [{\citenamefont {Carrilho}\ \emph {et~al.}(2018)\citenamefont
  {Carrilho}, \citenamefont {Mulryne}, \citenamefont {Ronayne},\ and\
  \citenamefont {Tenkanen}}]{Carrilho:2018ffi}%
  \BibitemOpen
  \bibfield  {author} {\bibinfo {author} {\bibfnamefont {Pedro}\ \bibnamefont
  {Carrilho}}, \bibinfo {author} {\bibfnamefont {David}\ \bibnamefont
  {Mulryne}}, \bibinfo {author} {\bibfnamefont {John}\ \bibnamefont {Ronayne}},
  \ and\ \bibinfo {author} {\bibfnamefont {Tommi}\ \bibnamefont {Tenkanen}},\
  }\bibfield  {title} {\enquote {\bibinfo {title} {{Attractor Behaviour in
  Multifield Inflation}},}\ }\href {\doibase 10.1088/1475-7516/2018/06/032}
  {\bibfield  {journal} {\bibinfo  {journal} {JCAP}\ }\textbf {\bibinfo
  {volume} {06}},\ \bibinfo {pages} {032} (\bibinfo {year} {2018})},\ \Eprint
  {http://arxiv.org/abs/1804.10489} {arXiv:1804.10489 [astro-ph.CO]}
  \BibitemShut {NoStop}%
\bibitem [{\citenamefont {Enckell}\ \emph {et~al.}(2019)\citenamefont
  {Enckell}, \citenamefont {Enqvist}, \citenamefont {Rasanen},\ and\
  \citenamefont {Wahlman}}]{Enckell:2018hmo}%
  \BibitemOpen
  \bibfield  {author} {\bibinfo {author} {\bibfnamefont {Vera-Maria}\
  \bibnamefont {Enckell}}, \bibinfo {author} {\bibfnamefont {Kari}\
  \bibnamefont {Enqvist}}, \bibinfo {author} {\bibfnamefont {Syksy}\
  \bibnamefont {Rasanen}}, \ and\ \bibinfo {author} {\bibfnamefont {Lumi-Pyry}\
  \bibnamefont {Wahlman}},\ }\bibfield  {title} {\enquote {\bibinfo {title}
  {{Inflation with $R^2$ term in the Palatini formalism}},}\ }\href {\doibase
  10.1088/1475-7516/2019/02/022} {\bibfield  {journal} {\bibinfo  {journal}
  {JCAP}\ }\textbf {\bibinfo {volume} {02}},\ \bibinfo {pages} {022} (\bibinfo
  {year} {2019})},\ \Eprint {http://arxiv.org/abs/1810.05536} {arXiv:1810.05536
  [gr-qc]} \BibitemShut {NoStop}%
\bibitem [{\citenamefont {Rasanen}\ and\ \citenamefont
  {Tomberg}(2019)}]{Rasanen:2018fom}%
  \BibitemOpen
  \bibfield  {author} {\bibinfo {author} {\bibfnamefont {Syksy}\ \bibnamefont
  {Rasanen}}\ and\ \bibinfo {author} {\bibfnamefont {Eemeli}\ \bibnamefont
  {Tomberg}},\ }\bibfield  {title} {\enquote {\bibinfo {title} {{Planck scale
  black hole dark matter from Higgs inflation}},}\ }\href {\doibase
  10.1088/1475-7516/2019/01/038} {\bibfield  {journal} {\bibinfo  {journal}
  {JCAP}\ }\textbf {\bibinfo {volume} {01}},\ \bibinfo {pages} {038} (\bibinfo
  {year} {2019})},\ \Eprint {http://arxiv.org/abs/1810.12608} {arXiv:1810.12608
  [astro-ph.CO]} \BibitemShut {NoStop}%
\bibitem [{\citenamefont {Rubio}\ and\ \citenamefont
  {Tomberg}(2019)}]{Rubio:2019ypq}%
  \BibitemOpen
  \bibfield  {author} {\bibinfo {author} {\bibfnamefont {Javier}\ \bibnamefont
  {Rubio}}\ and\ \bibinfo {author} {\bibfnamefont {Eemeli~S.}\ \bibnamefont
  {Tomberg}},\ }\bibfield  {title} {\enquote {\bibinfo {title} {{Preheating in
  Palatini Higgs inflation}},}\ }\href {\doibase 10.1088/1475-7516/2019/04/021}
  {\bibfield  {journal} {\bibinfo  {journal} {JCAP}\ }\textbf {\bibinfo
  {volume} {04}},\ \bibinfo {pages} {021} (\bibinfo {year} {2019})},\ \Eprint
  {http://arxiv.org/abs/1902.10148} {arXiv:1902.10148 [hep-ph]} \BibitemShut
  {NoStop}%
\bibitem [{\citenamefont {Shaposhnikov}\ \emph
  {et~al.}(2021{\natexlab{a}})\citenamefont {Shaposhnikov}, \citenamefont
  {Shkerin},\ and\ \citenamefont {Zell}}]{Shaposhnikov:2020geh}%
  \BibitemOpen
  \bibfield  {author} {\bibinfo {author} {\bibfnamefont {Mikhail}\ \bibnamefont
  {Shaposhnikov}}, \bibinfo {author} {\bibfnamefont {Andrey}\ \bibnamefont
  {Shkerin}}, \ and\ \bibinfo {author} {\bibfnamefont {Sebastian}\ \bibnamefont
  {Zell}},\ }\bibfield  {title} {\enquote {\bibinfo {title} {{Standard Model
  Meets Gravity: Electroweak Symmetry Breaking and Inflation}},}\ }\href
  {\doibase 10.1103/PhysRevD.103.033006} {\bibfield  {journal} {\bibinfo
  {journal} {Phys. Rev. D}\ }\textbf {\bibinfo {volume} {103}},\ \bibinfo
  {pages} {033006} (\bibinfo {year} {2021}{\natexlab{a}})},\ \Eprint
  {http://arxiv.org/abs/2001.09088} {arXiv:2001.09088 [hep-th]} \BibitemShut
  {NoStop}%
\bibitem [{\citenamefont {Karananas}\ \emph {et~al.}(2020)\citenamefont
  {Karananas}, \citenamefont {Michel},\ and\ \citenamefont
  {Rubio}}]{Karananas:2020qkp}%
  \BibitemOpen
  \bibfield  {author} {\bibinfo {author} {\bibfnamefont {Georgios~K.}\
  \bibnamefont {Karananas}}, \bibinfo {author} {\bibfnamefont {Marco}\
  \bibnamefont {Michel}}, \ and\ \bibinfo {author} {\bibfnamefont {Javier}\
  \bibnamefont {Rubio}},\ }\bibfield  {title} {\enquote {\bibinfo {title} {{One
  residue to rule them all: Electroweak symmetry breaking, inflation and
  field-space geometry}},}\ }\href {\doibase 10.1016/j.physletb.2020.135876}
  {\bibfield  {journal} {\bibinfo  {journal} {Phys. Lett. B}\ }\textbf
  {\bibinfo {volume} {811}},\ \bibinfo {pages} {135876} (\bibinfo {year}
  {2020})},\ \Eprint {http://arxiv.org/abs/2006.11290} {arXiv:2006.11290
  [hep-th]} \BibitemShut {NoStop}%
\bibitem [{\citenamefont {L\r{a}ngvik}\ \emph {et~al.}(2021)\citenamefont
  {L\r{a}ngvik}, \citenamefont {Ojanper\"a}, \citenamefont {Raatikainen},\ and\
  \citenamefont {Rasanen}}]{Langvik:2020nrs}%
  \BibitemOpen
  \bibfield  {author} {\bibinfo {author} {\bibfnamefont {Miklos}\ \bibnamefont
  {L\r{a}ngvik}}, \bibinfo {author} {\bibfnamefont {Juha-Matti}\ \bibnamefont
  {Ojanper\"a}}, \bibinfo {author} {\bibfnamefont {Sami}\ \bibnamefont
  {Raatikainen}}, \ and\ \bibinfo {author} {\bibfnamefont {Syksy}\ \bibnamefont
  {Rasanen}},\ }\bibfield  {title} {\enquote {\bibinfo {title} {{Higgs
  inflation with the Holst and the Nieh\textendash{}Yan term}},}\ }\href
  {\doibase 10.1103/PhysRevD.103.083514} {\bibfield  {journal} {\bibinfo
  {journal} {Phys. Rev. D}\ }\textbf {\bibinfo {volume} {103}},\ \bibinfo
  {pages} {083514} (\bibinfo {year} {2021})},\ \Eprint
  {http://arxiv.org/abs/2007.12595} {arXiv:2007.12595 [astro-ph.CO]}
  \BibitemShut {NoStop}%
\bibitem [{\citenamefont {Shaposhnikov}\ \emph
  {et~al.}(2021{\natexlab{b}})\citenamefont {Shaposhnikov}, \citenamefont
  {Shkerin}, \citenamefont {Timiryasov},\ and\ \citenamefont
  {Zell}}]{Shaposhnikov:2020gts}%
  \BibitemOpen
  \bibfield  {author} {\bibinfo {author} {\bibfnamefont {Mikhail}\ \bibnamefont
  {Shaposhnikov}}, \bibinfo {author} {\bibfnamefont {Andrey}\ \bibnamefont
  {Shkerin}}, \bibinfo {author} {\bibfnamefont {Inar}\ \bibnamefont
  {Timiryasov}}, \ and\ \bibinfo {author} {\bibfnamefont {Sebastian}\
  \bibnamefont {Zell}},\ }\bibfield  {title} {\enquote {\bibinfo {title}
  {{Higgs inflation in Einstein-Cartan gravity}},}\ }\href {\doibase
  10.1088/1475-7516/2021/10/E01} {\bibfield  {journal} {\bibinfo  {journal}
  {JCAP}\ }\textbf {\bibinfo {volume} {02}},\ \bibinfo {pages} {008} (\bibinfo
  {year} {2021}{\natexlab{b}})},\ \bibinfo {note} {[Erratum: JCAP 10, E01
  (2021)]},\ \Eprint {http://arxiv.org/abs/2007.14978} {arXiv:2007.14978
  [hep-ph]} \BibitemShut {NoStop}%
\bibitem [{\citenamefont {Mikura}\ \emph {et~al.}(2020)\citenamefont {Mikura},
  \citenamefont {Tada},\ and\ \citenamefont {Yokoyama}}]{Mikura:2020qhc}%
  \BibitemOpen
  \bibfield  {author} {\bibinfo {author} {\bibfnamefont {Yusuke}\ \bibnamefont
  {Mikura}}, \bibinfo {author} {\bibfnamefont {Yuichiro}\ \bibnamefont {Tada}},
  \ and\ \bibinfo {author} {\bibfnamefont {Shuichiro}\ \bibnamefont
  {Yokoyama}},\ }\bibfield  {title} {\enquote {\bibinfo {title} {{Conformal
  inflation in the metric-affine geometry}},}\ }\href {\doibase
  10.1209/0295-5075/132/39001} {\bibfield  {journal} {\bibinfo  {journal}
  {EPL}\ }\textbf {\bibinfo {volume} {132}},\ \bibinfo {pages} {39001}
  (\bibinfo {year} {2020})},\ \Eprint {http://arxiv.org/abs/2008.00628}
  {arXiv:2008.00628 [hep-th]} \BibitemShut {NoStop}%
\bibitem [{\citenamefont {Shaposhnikov}\ \emph
  {et~al.}(2021{\natexlab{c}})\citenamefont {Shaposhnikov}, \citenamefont
  {Shkerin}, \citenamefont {Timiryasov},\ and\ \citenamefont
  {Zell}}]{Shaposhnikov:2020aen}%
  \BibitemOpen
  \bibfield  {author} {\bibinfo {author} {\bibfnamefont {Mikhail}\ \bibnamefont
  {Shaposhnikov}}, \bibinfo {author} {\bibfnamefont {Andrey}\ \bibnamefont
  {Shkerin}}, \bibinfo {author} {\bibfnamefont {Inar}\ \bibnamefont
  {Timiryasov}}, \ and\ \bibinfo {author} {\bibfnamefont {Sebastian}\
  \bibnamefont {Zell}},\ }\bibfield  {title} {\enquote {\bibinfo {title}
  {{Einstein-Cartan Portal to Dark Matter}},}\ }\href {\doibase
  10.1103/PhysRevLett.127.169901} {\bibfield  {journal} {\bibinfo  {journal}
  {Phys. Rev. Lett.}\ }\textbf {\bibinfo {volume} {126}},\ \bibinfo {pages}
  {161301} (\bibinfo {year} {2021}{\natexlab{c}})},\ \bibinfo {note} {[Erratum:
  Phys.Rev.Lett. 127, 169901 (2021)]},\ \Eprint
  {http://arxiv.org/abs/2008.11686} {arXiv:2008.11686 [hep-ph]} \BibitemShut
  {NoStop}%
\bibitem [{\citenamefont {Kubota}\ \emph {et~al.}(2021)\citenamefont {Kubota},
  \citenamefont {Oda}, \citenamefont {Shimada},\ and\ \citenamefont
  {Yamaguchi}}]{Kubota:2020ehu}%
  \BibitemOpen
  \bibfield  {author} {\bibinfo {author} {\bibfnamefont {Mio}\ \bibnamefont
  {Kubota}}, \bibinfo {author} {\bibfnamefont {Kin-Ya}\ \bibnamefont {Oda}},
  \bibinfo {author} {\bibfnamefont {Keigo}\ \bibnamefont {Shimada}}, \ and\
  \bibinfo {author} {\bibfnamefont {Masahide}\ \bibnamefont {Yamaguchi}},\
  }\bibfield  {title} {\enquote {\bibinfo {title} {{Cosmological Perturbations
  in Palatini Formalism}},}\ }\href {\doibase 10.1088/1475-7516/2021/03/006}
  {\bibfield  {journal} {\bibinfo  {journal} {JCAP}\ }\textbf {\bibinfo
  {volume} {03}},\ \bibinfo {pages} {006} (\bibinfo {year} {2021})},\ \Eprint
  {http://arxiv.org/abs/2010.07867} {arXiv:2010.07867 [hep-th]} \BibitemShut
  {NoStop}%
\bibitem [{\citenamefont {Enckell}\ \emph {et~al.}(2021)\citenamefont
  {Enckell}, \citenamefont {Nurmi}, \citenamefont {R\"as\"anen},\ and\
  \citenamefont {Tomberg}}]{Enckell:2020lvn}%
  \BibitemOpen
  \bibfield  {author} {\bibinfo {author} {\bibfnamefont {Vera-Maria}\
  \bibnamefont {Enckell}}, \bibinfo {author} {\bibfnamefont {Sami}\
  \bibnamefont {Nurmi}}, \bibinfo {author} {\bibfnamefont {Syksy}\ \bibnamefont
  {R\"as\"anen}}, \ and\ \bibinfo {author} {\bibfnamefont {Eemeli}\
  \bibnamefont {Tomberg}},\ }\bibfield  {title} {\enquote {\bibinfo {title}
  {{Critical point Higgs inflation in the Palatini formulation}},}\ }\href
  {\doibase 10.1007/JHEP04(2021)059} {\bibfield  {journal} {\bibinfo  {journal}
  {JHEP}\ }\textbf {\bibinfo {volume} {04}},\ \bibinfo {pages} {059} (\bibinfo
  {year} {2021})},\ \Eprint {http://arxiv.org/abs/2012.03660} {arXiv:2012.03660
  [astro-ph.CO]} \BibitemShut {NoStop}%
\bibitem [{\citenamefont {Iosifidis}\ and\ \citenamefont
  {Ravera}(2021)}]{Iosifidis:2021iuw}%
  \BibitemOpen
  \bibfield  {author} {\bibinfo {author} {\bibfnamefont {Damianos}\
  \bibnamefont {Iosifidis}}\ and\ \bibinfo {author} {\bibfnamefont {Lucrezia}\
  \bibnamefont {Ravera}},\ }\bibfield  {title} {\enquote {\bibinfo {title}
  {{The cosmology of quadratic torsionful gravity}},}\ }\href {\doibase
  10.1140/epjc/s10052-021-09532-8} {\bibfield  {journal} {\bibinfo  {journal}
  {Eur. Phys. J. C}\ }\textbf {\bibinfo {volume} {81}},\ \bibinfo {pages} {736}
  (\bibinfo {year} {2021})},\ \Eprint {http://arxiv.org/abs/2101.10339}
  {arXiv:2101.10339 [gr-qc]} \BibitemShut {NoStop}%
\bibitem [{\citenamefont {Racioppi}\ \emph {et~al.}(2022)\citenamefont
  {Racioppi}, \citenamefont {Rajasalu},\ and\ \citenamefont
  {Selke}}]{Racioppi:2021ynx}%
  \BibitemOpen
  \bibfield  {author} {\bibinfo {author} {\bibfnamefont {Antonio}\ \bibnamefont
  {Racioppi}}, \bibinfo {author} {\bibfnamefont {J\"urgen}\ \bibnamefont
  {Rajasalu}}, \ and\ \bibinfo {author} {\bibfnamefont {Kaspar}\ \bibnamefont
  {Selke}},\ }\bibfield  {title} {\enquote {\bibinfo {title} {{Multiple point
  criticality principle and Coleman-Weinberg inflation}},}\ }\href {\doibase
  10.1007/JHEP06(2022)107} {\bibfield  {journal} {\bibinfo  {journal} {JHEP}\
  }\textbf {\bibinfo {volume} {06}},\ \bibinfo {pages} {107} (\bibinfo {year}
  {2022})},\ \Eprint {http://arxiv.org/abs/2109.03238} {arXiv:2109.03238
  [astro-ph.CO]} \BibitemShut {NoStop}%
\bibitem [{\citenamefont {Cheong}\ \emph {et~al.}(2022)\citenamefont {Cheong},
  \citenamefont {Lee},\ and\ \citenamefont {Park}}]{Cheong:2021kyc}%
  \BibitemOpen
  \bibfield  {author} {\bibinfo {author} {\bibfnamefont {Dhong~Yeon}\
  \bibnamefont {Cheong}}, \bibinfo {author} {\bibfnamefont {Sung~Mook}\
  \bibnamefont {Lee}}, \ and\ \bibinfo {author} {\bibfnamefont {Seong~Chan}\
  \bibnamefont {Park}},\ }\bibfield  {title} {\enquote {\bibinfo {title}
  {{Reheating in models with non-minimal coupling in metric and~Palatini
  formalisms}},}\ }\href {\doibase 10.1088/1475-7516/2022/02/029} {\bibfield
  {journal} {\bibinfo  {journal} {JCAP}\ }\textbf {\bibinfo {volume} {02}},\
  \bibinfo {pages} {029} (\bibinfo {year} {2022})},\ \Eprint
  {http://arxiv.org/abs/2111.00825} {arXiv:2111.00825 [hep-ph]} \BibitemShut
  {NoStop}%
\bibitem [{\citenamefont {Dioguardi}\ \emph {et~al.}(2022)\citenamefont
  {Dioguardi}, \citenamefont {Racioppi},\ and\ \citenamefont
  {Tomberg}}]{Dioguardi:2021fmr}%
  \BibitemOpen
  \bibfield  {author} {\bibinfo {author} {\bibfnamefont {Christian}\
  \bibnamefont {Dioguardi}}, \bibinfo {author} {\bibfnamefont {Antonio}\
  \bibnamefont {Racioppi}}, \ and\ \bibinfo {author} {\bibfnamefont {Eemeli}\
  \bibnamefont {Tomberg}},\ }\bibfield  {title} {\enquote {\bibinfo {title}
  {{Slow-roll inflation in Palatini F(R) gravity}},}\ }\href {\doibase
  10.1007/JHEP06(2022)106} {\bibfield  {journal} {\bibinfo  {journal} {JHEP}\
  }\textbf {\bibinfo {volume} {06}},\ \bibinfo {pages} {106} (\bibinfo {year}
  {2022})},\ \Eprint {http://arxiv.org/abs/2112.12149} {arXiv:2112.12149
  [gr-qc]} \BibitemShut {NoStop}%
\bibitem [{\citenamefont {Piani}\ and\ \citenamefont
  {Rubio}(2022)}]{Piani:2022gon}%
  \BibitemOpen
  \bibfield  {author} {\bibinfo {author} {\bibfnamefont {Matteo}\ \bibnamefont
  {Piani}}\ and\ \bibinfo {author} {\bibfnamefont {Javier}\ \bibnamefont
  {Rubio}},\ }\bibfield  {title} {\enquote {\bibinfo {title} {{Higgs-Dilaton
  inflation in Einstein-Cartan gravity}},}\ }\href {\doibase
  10.1088/1475-7516/2022/05/009} {\bibfield  {journal} {\bibinfo  {journal}
  {JCAP}\ }\textbf {\bibinfo {volume} {05}},\ \bibinfo {pages} {009} (\bibinfo
  {year} {2022})},\ \Eprint {http://arxiv.org/abs/2202.04665} {arXiv:2202.04665
  [gr-qc]} \BibitemShut {NoStop}%
\bibitem [{\citenamefont {Dux}\ \emph {et~al.}(2022)\citenamefont {Dux},
  \citenamefont {Florio}, \citenamefont {Klari\'c}, \citenamefont {Shkerin},\
  and\ \citenamefont {Timiryasov}}]{Dux:2022kuk}%
  \BibitemOpen
  \bibfield  {author} {\bibinfo {author} {\bibfnamefont {Fr\'ed\'eric}\
  \bibnamefont {Dux}}, \bibinfo {author} {\bibfnamefont {Adrien}\ \bibnamefont
  {Florio}}, \bibinfo {author} {\bibfnamefont {Juraj}\ \bibnamefont
  {Klari\'c}}, \bibinfo {author} {\bibfnamefont {Andrey}\ \bibnamefont
  {Shkerin}}, \ and\ \bibinfo {author} {\bibfnamefont {Inar}\ \bibnamefont
  {Timiryasov}},\ }\bibfield  {title} {\enquote {\bibinfo {title} {{Preheating
  in Palatini Higgs inflation on the lattice}},}\ }\href {\doibase
  10.1088/1475-7516/2022/09/015} {\bibfield  {journal} {\bibinfo  {journal}
  {JCAP}\ }\textbf {\bibinfo {volume} {09}},\ \bibinfo {pages} {015} (\bibinfo
  {year} {2022})},\ \Eprint {http://arxiv.org/abs/2203.13286} {arXiv:2203.13286
  [hep-ph]} \BibitemShut {NoStop}%
\bibitem [{\citenamefont {Rasanen}\ and\ \citenamefont
  {Verbin}(2022)}]{Rasanen:2022ijc}%
  \BibitemOpen
  \bibfield  {author} {\bibinfo {author} {\bibfnamefont {Syksy}\ \bibnamefont
  {Rasanen}}\ and\ \bibinfo {author} {\bibfnamefont {Yosef}\ \bibnamefont
  {Verbin}},\ }\bibfield  {title} {\enquote {\bibinfo {title} {{Palatini
  formulation for gauge theory: implications for slow-roll inflation}},}\
  }\href {\doibase 10.21105/astro.2211.15584} {\  (\bibinfo {year} {2022}),\
  10.21105/astro.2211.15584},\ \Eprint {http://arxiv.org/abs/2211.15584}
  {arXiv:2211.15584 [astro-ph.CO]} \BibitemShut {NoStop}%
\bibitem [{\citenamefont {Gialamas}\ and\ \citenamefont
  {Veerm\"ae}(2023)}]{Gialamas:2023emn}%
  \BibitemOpen
  \bibfield  {author} {\bibinfo {author} {\bibfnamefont {Ioannis~D.}\
  \bibnamefont {Gialamas}}\ and\ \bibinfo {author} {\bibfnamefont {Hardi}\
  \bibnamefont {Veerm\"ae}},\ }\bibfield  {title} {\enquote {\bibinfo {title}
  {{Electroweak vacuum decay in metric-affine gravity}},}\ }\href@noop {} {\
  (\bibinfo {year} {2023})},\ \Eprint {http://arxiv.org/abs/2305.07693}
  {arXiv:2305.07693 [hep-th]} \BibitemShut {NoStop}%
\bibitem [{\citenamefont {Gialamas}\ \emph {et~al.}(2023)\citenamefont
  {Gialamas}, \citenamefont {Karam}, \citenamefont {Pappas},\ and\
  \citenamefont {Tomberg}}]{Gialamas:2023flv}%
  \BibitemOpen
  \bibfield  {author} {\bibinfo {author} {\bibfnamefont {Ioannis~D.}\
  \bibnamefont {Gialamas}}, \bibinfo {author} {\bibfnamefont {Alexandros}\
  \bibnamefont {Karam}}, \bibinfo {author} {\bibfnamefont {Thomas~D.}\
  \bibnamefont {Pappas}}, \ and\ \bibinfo {author} {\bibfnamefont {Eemeli}\
  \bibnamefont {Tomberg}},\ }\bibfield  {title} {\enquote {\bibinfo {title}
  {{Implications of Palatini gravity for inflation and beyond}},}\ }\href@noop
  {} {\  (\bibinfo {year} {2023})},\ \Eprint {http://arxiv.org/abs/2303.14148}
  {arXiv:2303.14148 [gr-qc]} \BibitemShut {NoStop}%
\bibitem [{\citenamefont {Piani}\ and\ \citenamefont
  {Rubio}(2023)}]{Piani:2023aof}%
  \BibitemOpen
  \bibfield  {author} {\bibinfo {author} {\bibfnamefont {Matteo}\ \bibnamefont
  {Piani}}\ and\ \bibinfo {author} {\bibfnamefont {Javier}\ \bibnamefont
  {Rubio}},\ }\bibfield  {title} {\enquote {\bibinfo {title} {{Preheating in
  Einstein-Cartan Higgs Inflation: Oscillon formation}},}\ }\href@noop {} {\
  (\bibinfo {year} {2023})},\ \Eprint {http://arxiv.org/abs/2304.13056}
  {arXiv:2304.13056 [hep-ph]} \BibitemShut {NoStop}%
\bibitem [{\citenamefont {Poisson}\ \emph {et~al.}(2023)\citenamefont
  {Poisson}, \citenamefont {Timiryasov},\ and\ \citenamefont
  {Zell}}]{Poisson:2023tja}%
  \BibitemOpen
  \bibfield  {author} {\bibinfo {author} {\bibfnamefont {Arthur}\ \bibnamefont
  {Poisson}}, \bibinfo {author} {\bibfnamefont {Inar}\ \bibnamefont
  {Timiryasov}}, \ and\ \bibinfo {author} {\bibfnamefont {Sebastian}\
  \bibnamefont {Zell}},\ }\bibfield  {title} {\enquote {\bibinfo {title}
  {{Critical Points in Palatini Higgs Inflation with Small Non-Minimal
  Coupling}},}\ }\href@noop {} {\  (\bibinfo {year} {2023})},\ \Eprint
  {http://arxiv.org/abs/2306.03893} {arXiv:2306.03893 [hep-ph]} \BibitemShut
  {NoStop}%
\bibitem [{\citenamefont {Rigouzzo}\ and\ \citenamefont
  {Zell}(2023)}]{Rigouzzo:2023sbb}%
  \BibitemOpen
  \bibfield  {author} {\bibinfo {author} {\bibfnamefont {Claire}\ \bibnamefont
  {Rigouzzo}}\ and\ \bibinfo {author} {\bibfnamefont {Sebastian}\ \bibnamefont
  {Zell}},\ }\bibfield  {title} {\enquote {\bibinfo {title} {{Coupling
  Metric-Affine Gravity to the Standard Model and Dark Matter Fermions}},}\
  }\href@noop {} {\  (\bibinfo {year} {2023})},\ \Eprint
  {http://arxiv.org/abs/2306.13134} {arXiv:2306.13134 [gr-qc]} \BibitemShut
  {NoStop}%
\bibitem [{\citenamefont {Stelle}(1978)}]{Stelle:1977ry}%
  \BibitemOpen
  \bibfield  {author} {\bibinfo {author} {\bibfnamefont {K.~S.}\ \bibnamefont
  {Stelle}},\ }\bibfield  {title} {\enquote {\bibinfo {title} {{Classical
  Gravity with Higher Derivatives}},}\ }\href {\doibase 10.1007/BF00760427}
  {\bibfield  {journal} {\bibinfo  {journal} {Gen. Rel. Grav.}\ }\textbf
  {\bibinfo {volume} {9}},\ \bibinfo {pages} {353--371} (\bibinfo {year}
  {1978})}\BibitemShut {NoStop}%
\bibitem [{\citenamefont {Neville}(1978)}]{Neville:1978bk}%
  \BibitemOpen
  \bibfield  {author} {\bibinfo {author} {\bibfnamefont {Donald~E.}\
  \bibnamefont {Neville}},\ }\bibfield  {title} {\enquote {\bibinfo {title} {{A
  Gravity Lagrangian With Ghost Free Curvature**2 Terms}},}\ }\href {\doibase
  10.1103/PhysRevD.18.3535} {\bibfield  {journal} {\bibinfo  {journal} {Phys.
  Rev. D}\ }\textbf {\bibinfo {volume} {18}},\ \bibinfo {pages} {3535}
  (\bibinfo {year} {1978})}\BibitemShut {NoStop}%
\bibitem [{\citenamefont {Neville}(1980)}]{Neville:1979rb}%
  \BibitemOpen
  \bibfield  {author} {\bibinfo {author} {\bibfnamefont {Donald~E.}\
  \bibnamefont {Neville}},\ }\bibfield  {title} {\enquote {\bibinfo {title}
  {{Gravity Theories With Propagating Torsion}},}\ }\href {\doibase
  10.1103/PhysRevD.21.867} {\bibfield  {journal} {\bibinfo  {journal} {Phys.
  Rev. D}\ }\textbf {\bibinfo {volume} {21}},\ \bibinfo {pages} {867} (\bibinfo
  {year} {1980})}\BibitemShut {NoStop}%
\bibitem [{\citenamefont {Sezgin}\ and\ \citenamefont {van
  Nieuwenhuizen}(1980)}]{Sezgin:1979zf}%
  \BibitemOpen
  \bibfield  {author} {\bibinfo {author} {\bibfnamefont {E.}~\bibnamefont
  {Sezgin}}\ and\ \bibinfo {author} {\bibfnamefont {P.}~\bibnamefont {van
  Nieuwenhuizen}},\ }\bibfield  {title} {\enquote {\bibinfo {title} {{New Ghost
  Free Gravity Lagrangians with Propagating Torsion}},}\ }\href {\doibase
  10.1103/PhysRevD.21.3269} {\bibfield  {journal} {\bibinfo  {journal} {Phys.
  Rev. D}\ }\textbf {\bibinfo {volume} {21}},\ \bibinfo {pages} {3269}
  (\bibinfo {year} {1980})}\BibitemShut {NoStop}%
\bibitem [{\citenamefont {Hayashi}\ and\ \citenamefont
  {Shirafuji}(1980{\natexlab{a}})}]{Hayashi:1979wj}%
  \BibitemOpen
  \bibfield  {author} {\bibinfo {author} {\bibfnamefont {Kenji}\ \bibnamefont
  {Hayashi}}\ and\ \bibinfo {author} {\bibfnamefont {Takeshi}\ \bibnamefont
  {Shirafuji}},\ }\bibfield  {title} {\enquote {\bibinfo {title} {{Gravity from
  Poincare Gauge Theory of the Fundamental Particles. 1. Linear and Quadratic
  Lagrangians}},}\ }\href {\doibase 10.1143/PTP.64.866} {\bibfield  {journal}
  {\bibinfo  {journal} {Prog. Theor. Phys.}\ }\textbf {\bibinfo {volume}
  {64}},\ \bibinfo {pages} {866} (\bibinfo {year} {1980}{\natexlab{a}})},\
  \bibinfo {note} {[Erratum: Prog.Theor.Phys. 65, 2079 (1981)]}\BibitemShut
  {NoStop}%
\bibitem [{\citenamefont {Hayashi}\ and\ \citenamefont
  {Shirafuji}(1980{\natexlab{b}})}]{Hayashi:1980qp}%
  \BibitemOpen
  \bibfield  {author} {\bibinfo {author} {\bibfnamefont {Kenji}\ \bibnamefont
  {Hayashi}}\ and\ \bibinfo {author} {\bibfnamefont {Takeshi}\ \bibnamefont
  {Shirafuji}},\ }\bibfield  {title} {\enquote {\bibinfo {title} {{Gravity From
  Poincare Gauge Theory of the Fundamental Particles. 4. Mass and Energy of
  Particle Spectrum}},}\ }\href {\doibase 10.1143/PTP.64.2222} {\bibfield
  {journal} {\bibinfo  {journal} {Prog. Theor. Phys.}\ }\textbf {\bibinfo
  {volume} {64}},\ \bibinfo {pages} {2222} (\bibinfo {year}
  {1980}{\natexlab{b}})}\BibitemShut {NoStop}%
\bibitem [{\citenamefont {Hecht}\ \emph {et~al.}(1990)\citenamefont {Hecht},
  \citenamefont {Lemke},\ and\ \citenamefont {Wallner}}]{Hecht:1990wn}%
  \BibitemOpen
  \bibfield  {author} {\bibinfo {author} {\bibfnamefont {R.~D.}\ \bibnamefont
  {Hecht}}, \bibinfo {author} {\bibfnamefont {J.}~\bibnamefont {Lemke}}, \ and\
  \bibinfo {author} {\bibfnamefont {R.~P.}\ \bibnamefont {Wallner}},\
  }\bibfield  {title} {\enquote {\bibinfo {title} {{Tachyonic torsion shock
  waves in Poincare gauge theory}},}\ }\href {\doibase
  10.1016/0375-9601(90)90837-E} {\bibfield  {journal} {\bibinfo  {journal}
  {Phys. Lett. A}\ }\textbf {\bibinfo {volume} {151}},\ \bibinfo {pages}
  {12--14} (\bibinfo {year} {1990})}\BibitemShut {NoStop}%
\bibitem [{\citenamefont {Hecht}\ \emph {et~al.}(1991)\citenamefont {Hecht},
  \citenamefont {Lemke},\ and\ \citenamefont {Wallner}}]{Hecht:1991jh}%
  \BibitemOpen
  \bibfield  {author} {\bibinfo {author} {\bibfnamefont {R.~D.}\ \bibnamefont
  {Hecht}}, \bibinfo {author} {\bibfnamefont {J.}~\bibnamefont {Lemke}}, \ and\
  \bibinfo {author} {\bibfnamefont {R.~P.}\ \bibnamefont {Wallner}},\
  }\bibfield  {title} {\enquote {\bibinfo {title} {{Can Poincare gauge theory
  be saved?}}}\ }\href {\doibase 10.1103/PhysRevD.44.2442} {\bibfield
  {journal} {\bibinfo  {journal} {Phys. Rev. D}\ }\textbf {\bibinfo {volume}
  {44}},\ \bibinfo {pages} {2442--2451} (\bibinfo {year} {1991})}\BibitemShut
  {NoStop}%
\bibitem [{\citenamefont {Chen}\ \emph {et~al.}(1998)\citenamefont {Chen},
  \citenamefont {Nester},\ and\ \citenamefont {Yo}}]{Chen:1998ad}%
  \BibitemOpen
  \bibfield  {author} {\bibinfo {author} {\bibfnamefont {Hsin}\ \bibnamefont
  {Chen}}, \bibinfo {author} {\bibfnamefont {J.~M.}\ \bibnamefont {Nester}}, \
  and\ \bibinfo {author} {\bibfnamefont {Hwei-Jang}\ \bibnamefont {Yo}},\
  }\bibfield  {title} {\enquote {\bibinfo {title} {{Acausal PGT modes and the
  nonlinear constraint effect}},}\ }\href
  {https://www.actaphys.uj.edu.pl/index_n.php?I=R&V=29&N=4#961} {\bibfield
  {journal} {\bibinfo  {journal} {Acta Phys. Polon. B}\ }\textbf {\bibinfo
  {volume} {29}},\ \bibinfo {pages} {961--970} (\bibinfo {year}
  {1998})}\BibitemShut {NoStop}%
\bibitem [{\citenamefont {Hecht}\ \emph {et~al.}(1996)\citenamefont {Hecht},
  \citenamefont {Nester},\ and\ \citenamefont {Zhytnikov}}]{Hecht:1996np}%
  \BibitemOpen
  \bibfield  {author} {\bibinfo {author} {\bibfnamefont {R.~D.}\ \bibnamefont
  {Hecht}}, \bibinfo {author} {\bibfnamefont {J.~M.}\ \bibnamefont {Nester}}, \
  and\ \bibinfo {author} {\bibfnamefont {V.~V.}\ \bibnamefont {Zhytnikov}},\
  }\bibfield  {title} {\enquote {\bibinfo {title} {{Some Poincare gauge theory
  Lagrangians with well posed initial value problems}},}\ }\href {\doibase
  10.1016/0375-9601(96)00622-6} {\bibfield  {journal} {\bibinfo  {journal}
  {Phys. Lett. A}\ }\textbf {\bibinfo {volume} {222}},\ \bibinfo {pages}
  {37--42} (\bibinfo {year} {1996})}\BibitemShut {NoStop}%
\bibitem [{\citenamefont {Yo}\ and\ \citenamefont {Nester}(1999)}]{Yo:1999ex}%
  \BibitemOpen
  \bibfield  {author} {\bibinfo {author} {\bibfnamefont {Hwei-jang}\
  \bibnamefont {Yo}}\ and\ \bibinfo {author} {\bibfnamefont {James~M.}\
  \bibnamefont {Nester}},\ }\bibfield  {title} {\enquote {\bibinfo {title}
  {{Hamiltonian analysis of Poincare gauge theory scalar modes}},}\ }\href
  {\doibase 10.1142/S021827189900033X} {\bibfield  {journal} {\bibinfo
  {journal} {Int. J. Mod. Phys. D}\ }\textbf {\bibinfo {volume} {8}},\ \bibinfo
  {pages} {459--479} (\bibinfo {year} {1999})},\ \Eprint
  {http://arxiv.org/abs/gr-qc/9902032} {arXiv:gr-qc/9902032} \BibitemShut
  {NoStop}%
\bibitem [{\citenamefont {Yo}\ and\ \citenamefont {Nester}(2002)}]{Yo:2001sy}%
  \BibitemOpen
  \bibfield  {author} {\bibinfo {author} {\bibfnamefont {Hwei-Jang}\
  \bibnamefont {Yo}}\ and\ \bibinfo {author} {\bibfnamefont {James~M.}\
  \bibnamefont {Nester}},\ }\bibfield  {title} {\enquote {\bibinfo {title}
  {{Hamiltonian analysis of Poincare gauge theory: Higher spin modes}},}\
  }\href {\doibase 10.1142/S0218271802001998} {\bibfield  {journal} {\bibinfo
  {journal} {Int. J. Mod. Phys. D}\ }\textbf {\bibinfo {volume} {11}},\
  \bibinfo {pages} {747--780} (\bibinfo {year} {2002})},\ \Eprint
  {http://arxiv.org/abs/gr-qc/0112030} {arXiv:gr-qc/0112030} \BibitemShut
  {NoStop}%
\bibitem [{\citenamefont {Barker}(2022)}]{Barker:2022jsh}%
  \BibitemOpen
  \bibfield  {author} {\bibinfo {author} {\bibfnamefont {W.~E.~V.}\
  \bibnamefont {Barker}},\ }\bibfield  {title} {\enquote {\bibinfo {title}
  {{Geometric multipliers and partial teleparallelism in Poincar\'e gauge
  theory}},}\ }\href@noop {} {\  (\bibinfo {year} {2022})},\ \Eprint
  {http://arxiv.org/abs/2205.13534} {arXiv:2205.13534 [gr-qc]} \BibitemShut
  {NoStop}%
\bibitem [{\citenamefont {Ford}(1989)}]{Ford:1989me}%
  \BibitemOpen
  \bibfield  {author} {\bibinfo {author} {\bibfnamefont {L.~H.}\ \bibnamefont
  {Ford}},\ }\bibfield  {title} {\enquote {\bibinfo {title} {{INFLATION DRIVEN
  BY A VECTOR FIELD}},}\ }\href {\doibase 10.1103/PhysRevD.40.967} {\bibfield
  {journal} {\bibinfo  {journal} {Phys. Rev. D}\ }\textbf {\bibinfo {volume}
  {40}},\ \bibinfo {pages} {967} (\bibinfo {year} {1989})}\BibitemShut
  {NoStop}%
\bibitem [{\citenamefont {Golovnev}\ \emph {et~al.}(2008)\citenamefont
  {Golovnev}, \citenamefont {Mukhanov},\ and\ \citenamefont
  {Vanchurin}}]{Golovnev:2008cf}%
  \BibitemOpen
  \bibfield  {author} {\bibinfo {author} {\bibfnamefont {Alexey}\ \bibnamefont
  {Golovnev}}, \bibinfo {author} {\bibfnamefont {Viatcheslav}\ \bibnamefont
  {Mukhanov}}, \ and\ \bibinfo {author} {\bibfnamefont {Vitaly}\ \bibnamefont
  {Vanchurin}},\ }\bibfield  {title} {\enquote {\bibinfo {title} {{Vector
  Inflation}},}\ }\href {\doibase 10.1088/1475-7516/2008/06/009} {\bibfield
  {journal} {\bibinfo  {journal} {JCAP}\ }\textbf {\bibinfo {volume} {06}},\
  \bibinfo {pages} {009} (\bibinfo {year} {2008})},\ \Eprint
  {http://arxiv.org/abs/0802.2068} {arXiv:0802.2068 [astro-ph]} \BibitemShut
  {NoStop}%
\bibitem [{\citenamefont {Chiba}(2008)}]{Chiba:2008eh}%
  \BibitemOpen
  \bibfield  {author} {\bibinfo {author} {\bibfnamefont {Takeshi}\ \bibnamefont
  {Chiba}},\ }\bibfield  {title} {\enquote {\bibinfo {title} {{Initial
  Conditions for Vector Inflation}},}\ }\href {\doibase
  10.1088/1475-7516/2008/08/004} {\bibfield  {journal} {\bibinfo  {journal}
  {JCAP}\ }\textbf {\bibinfo {volume} {08}},\ \bibinfo {pages} {004} (\bibinfo
  {year} {2008})},\ \Eprint {http://arxiv.org/abs/0805.4660} {arXiv:0805.4660
  [gr-qc]} \BibitemShut {NoStop}%
\bibitem [{\citenamefont {Koivisto}\ and\ \citenamefont
  {Mota}(2008{\natexlab{a}})}]{Koivisto:2008xf}%
  \BibitemOpen
  \bibfield  {author} {\bibinfo {author} {\bibfnamefont {Tomi}\ \bibnamefont
  {Koivisto}}\ and\ \bibinfo {author} {\bibfnamefont {David~F.}\ \bibnamefont
  {Mota}},\ }\bibfield  {title} {\enquote {\bibinfo {title} {{Vector Field
  Models of Inflation and Dark Energy}},}\ }\href {\doibase
  10.1088/1475-7516/2008/08/021} {\bibfield  {journal} {\bibinfo  {journal}
  {JCAP}\ }\textbf {\bibinfo {volume} {08}},\ \bibinfo {pages} {021} (\bibinfo
  {year} {2008}{\natexlab{a}})},\ \Eprint {http://arxiv.org/abs/0805.4229}
  {arXiv:0805.4229 [astro-ph]} \BibitemShut {NoStop}%
\bibitem [{\citenamefont {Kanno}\ \emph {et~al.}(2008)\citenamefont {Kanno},
  \citenamefont {Kimura}, \citenamefont {Soda},\ and\ \citenamefont
  {Yokoyama}}]{Kanno:2008gn}%
  \BibitemOpen
  \bibfield  {author} {\bibinfo {author} {\bibfnamefont {Sugumi}\ \bibnamefont
  {Kanno}}, \bibinfo {author} {\bibfnamefont {Masashi}\ \bibnamefont {Kimura}},
  \bibinfo {author} {\bibfnamefont {Jiro}\ \bibnamefont {Soda}}, \ and\
  \bibinfo {author} {\bibfnamefont {Shuichiro}\ \bibnamefont {Yokoyama}},\
  }\bibfield  {title} {\enquote {\bibinfo {title} {{Anisotropic Inflation from
  Vector Impurity}},}\ }\href {\doibase 10.1088/1475-7516/2008/08/034}
  {\bibfield  {journal} {\bibinfo  {journal} {JCAP}\ }\textbf {\bibinfo
  {volume} {08}},\ \bibinfo {pages} {034} (\bibinfo {year} {2008})},\ \Eprint
  {http://arxiv.org/abs/0806.2422} {arXiv:0806.2422 [hep-ph]} \BibitemShut
  {NoStop}%
\bibitem [{\citenamefont {Emami}\ \emph {et~al.}(2017)\citenamefont {Emami},
  \citenamefont {Mukohyama}, \citenamefont {Namba},\ and\ \citenamefont
  {Zhang}}]{Emami:2016ldl}%
  \BibitemOpen
  \bibfield  {author} {\bibinfo {author} {\bibfnamefont {Razieh}\ \bibnamefont
  {Emami}}, \bibinfo {author} {\bibfnamefont {Shinji}\ \bibnamefont
  {Mukohyama}}, \bibinfo {author} {\bibfnamefont {Ryo}\ \bibnamefont {Namba}},
  \ and\ \bibinfo {author} {\bibfnamefont {Ying-li}\ \bibnamefont {Zhang}},\
  }\bibfield  {title} {\enquote {\bibinfo {title} {{Stable solutions of
  inflation driven by vector fields}},}\ }\href {\doibase
  10.1088/1475-7516/2017/03/058} {\bibfield  {journal} {\bibinfo  {journal}
  {JCAP}\ }\textbf {\bibinfo {volume} {03}},\ \bibinfo {pages} {058} (\bibinfo
  {year} {2017})},\ \Eprint {http://arxiv.org/abs/1612.09581} {arXiv:1612.09581
  [hep-th]} \BibitemShut {NoStop}%
\bibitem [{\citenamefont {Rodr\'\i{}guez}\ and\ \citenamefont
  {Navarro}(2018)}]{Rodriguez:2017wkg}%
  \BibitemOpen
  \bibfield  {author} {\bibinfo {author} {\bibfnamefont {Yeinzon}\ \bibnamefont
  {Rodr\'\i{}guez}}\ and\ \bibinfo {author} {\bibfnamefont {Andr\'es~A.}\
  \bibnamefont {Navarro}},\ }\bibfield  {title} {\enquote {\bibinfo {title}
  {{Non-Abelian $S$-term dark energy and inflation}},}\ }\href {\doibase
  10.1016/j.dark.2018.01.003} {\bibfield  {journal} {\bibinfo  {journal} {Phys.
  Dark Univ.}\ }\textbf {\bibinfo {volume} {19}},\ \bibinfo {pages} {129--136}
  (\bibinfo {year} {2018})},\ \Eprint {http://arxiv.org/abs/1711.01935}
  {arXiv:1711.01935 [gr-qc]} \BibitemShut {NoStop}%
\bibitem [{\citenamefont {Armendariz-Picon}(2004)}]{Armendariz-Picon:2004say}%
  \BibitemOpen
  \bibfield  {author} {\bibinfo {author} {\bibfnamefont {Christian}\
  \bibnamefont {Armendariz-Picon}},\ }\bibfield  {title} {\enquote {\bibinfo
  {title} {{Could dark energy be vector-like?}}}\ }\href {\doibase
  10.1088/1475-7516/2004/07/007} {\bibfield  {journal} {\bibinfo  {journal}
  {JCAP}\ }\textbf {\bibinfo {volume} {07}},\ \bibinfo {pages} {007} (\bibinfo
  {year} {2004})},\ \Eprint {http://arxiv.org/abs/astro-ph/0405267}
  {arXiv:astro-ph/0405267} \BibitemShut {NoStop}%
\bibitem [{\citenamefont {Koivisto}\ and\ \citenamefont
  {Mota}(2008{\natexlab{b}})}]{Koivisto:2007bp}%
  \BibitemOpen
  \bibfield  {author} {\bibinfo {author} {\bibfnamefont {Tomi}\ \bibnamefont
  {Koivisto}}\ and\ \bibinfo {author} {\bibfnamefont {David~F.}\ \bibnamefont
  {Mota}},\ }\bibfield  {title} {\enquote {\bibinfo {title} {{Accelerating
  Cosmologies with an Anisotropic Equation of State}},}\ }\href {\doibase
  10.1086/587451} {\bibfield  {journal} {\bibinfo  {journal} {Astrophys. J.}\
  }\textbf {\bibinfo {volume} {679}},\ \bibinfo {pages} {1--5} (\bibinfo {year}
  {2008}{\natexlab{b}})},\ \Eprint {http://arxiv.org/abs/0707.0279}
  {arXiv:0707.0279 [astro-ph]} \BibitemShut {NoStop}%
\bibitem [{\citenamefont {Tasinato}(2014)}]{Tasinato:2014eka}%
  \BibitemOpen
  \bibfield  {author} {\bibinfo {author} {\bibfnamefont {Gianmassimo}\
  \bibnamefont {Tasinato}},\ }\bibfield  {title} {\enquote {\bibinfo {title}
  {{Cosmic Acceleration from Abelian Symmetry Breaking}},}\ }\href {\doibase
  10.1007/JHEP04(2014)067} {\bibfield  {journal} {\bibinfo  {journal} {JHEP}\
  }\textbf {\bibinfo {volume} {04}},\ \bibinfo {pages} {067} (\bibinfo {year}
  {2014})},\ \Eprint {http://arxiv.org/abs/1402.6450} {arXiv:1402.6450
  [hep-th]} \BibitemShut {NoStop}%
\bibitem [{\citenamefont {De~Felice}\ \emph
  {et~al.}(2016{\natexlab{a}})\citenamefont {De~Felice}, \citenamefont
  {Heisenberg}, \citenamefont {Kase}, \citenamefont {Mukohyama}, \citenamefont
  {Tsujikawa},\ and\ \citenamefont {Zhang}}]{DeFelice:2016yws}%
  \BibitemOpen
  \bibfield  {author} {\bibinfo {author} {\bibfnamefont {Antonio}\ \bibnamefont
  {De~Felice}}, \bibinfo {author} {\bibfnamefont {Lavinia}\ \bibnamefont
  {Heisenberg}}, \bibinfo {author} {\bibfnamefont {Ryotaro}\ \bibnamefont
  {Kase}}, \bibinfo {author} {\bibfnamefont {Shinji}\ \bibnamefont
  {Mukohyama}}, \bibinfo {author} {\bibfnamefont {Shinji}\ \bibnamefont
  {Tsujikawa}}, \ and\ \bibinfo {author} {\bibfnamefont {Ying-li}\ \bibnamefont
  {Zhang}},\ }\bibfield  {title} {\enquote {\bibinfo {title} {{Cosmology in
  generalized Proca theories}},}\ }\href {\doibase
  10.1088/1475-7516/2016/06/048} {\bibfield  {journal} {\bibinfo  {journal}
  {JCAP}\ }\textbf {\bibinfo {volume} {06}},\ \bibinfo {pages} {048} (\bibinfo
  {year} {2016}{\natexlab{a}})},\ \Eprint {http://arxiv.org/abs/1603.05806}
  {arXiv:1603.05806 [gr-qc]} \BibitemShut {NoStop}%
\bibitem [{\citenamefont {De~Felice}\ \emph
  {et~al.}(2016{\natexlab{b}})\citenamefont {De~Felice}, \citenamefont
  {Heisenberg}, \citenamefont {Kase}, \citenamefont {Mukohyama}, \citenamefont
  {Tsujikawa},\ and\ \citenamefont {Zhang}}]{DeFelice:2016uil}%
  \BibitemOpen
  \bibfield  {author} {\bibinfo {author} {\bibfnamefont {Antonio}\ \bibnamefont
  {De~Felice}}, \bibinfo {author} {\bibfnamefont {Lavinia}\ \bibnamefont
  {Heisenberg}}, \bibinfo {author} {\bibfnamefont {Ryotaro}\ \bibnamefont
  {Kase}}, \bibinfo {author} {\bibfnamefont {Shinji}\ \bibnamefont
  {Mukohyama}}, \bibinfo {author} {\bibfnamefont {Shinji}\ \bibnamefont
  {Tsujikawa}}, \ and\ \bibinfo {author} {\bibfnamefont {Ying-li}\ \bibnamefont
  {Zhang}},\ }\bibfield  {title} {\enquote {\bibinfo {title} {{Effective
  gravitational couplings for cosmological perturbations in generalized Proca
  theories}},}\ }\href {\doibase 10.1103/PhysRevD.94.044024} {\bibfield
  {journal} {\bibinfo  {journal} {Phys. Rev. D}\ }\textbf {\bibinfo {volume}
  {94}},\ \bibinfo {pages} {044024} (\bibinfo {year} {2016}{\natexlab{b}})},\
  \Eprint {http://arxiv.org/abs/1605.05066} {arXiv:1605.05066 [gr-qc]}
  \BibitemShut {NoStop}%
\bibitem [{\citenamefont {Beltran~Jimenez}\ and\ \citenamefont
  {Heisenberg}(2017)}]{BeltranJimenez:2016afo}%
  \BibitemOpen
  \bibfield  {author} {\bibinfo {author} {\bibfnamefont {Jose}\ \bibnamefont
  {Beltran~Jimenez}}\ and\ \bibinfo {author} {\bibfnamefont {Lavinia}\
  \bibnamefont {Heisenberg}},\ }\bibfield  {title} {\enquote {\bibinfo {title}
  {{Generalized multi-Proca fields}},}\ }\href {\doibase
  10.1016/j.physletb.2017.03.002} {\bibfield  {journal} {\bibinfo  {journal}
  {Phys. Lett. B}\ }\textbf {\bibinfo {volume} {770}},\ \bibinfo {pages}
  {16--26} (\bibinfo {year} {2017})},\ \Eprint
  {http://arxiv.org/abs/1610.08960} {arXiv:1610.08960 [hep-th]} \BibitemShut
  {NoStop}%
\bibitem [{\citenamefont {de~Felice}\ \emph {et~al.}(2017)\citenamefont
  {de~Felice}, \citenamefont {Heisenberg},\ and\ \citenamefont
  {Tsujikawa}}]{deFelice:2017paw}%
  \BibitemOpen
  \bibfield  {author} {\bibinfo {author} {\bibfnamefont {Antonio}\ \bibnamefont
  {de~Felice}}, \bibinfo {author} {\bibfnamefont {Lavinia}\ \bibnamefont
  {Heisenberg}}, \ and\ \bibinfo {author} {\bibfnamefont {Shinji}\ \bibnamefont
  {Tsujikawa}},\ }\bibfield  {title} {\enquote {\bibinfo {title}
  {{Observational constraints on generalized Proca theories}},}\ }\href
  {\doibase 10.1103/PhysRevD.95.123540} {\bibfield  {journal} {\bibinfo
  {journal} {Phys. Rev. D}\ }\textbf {\bibinfo {volume} {95}},\ \bibinfo
  {pages} {123540} (\bibinfo {year} {2017})},\ \Eprint
  {http://arxiv.org/abs/1703.09573} {arXiv:1703.09573 [astro-ph.CO]}
  \BibitemShut {NoStop}%
\bibitem [{\citenamefont {Nakamura}\ \emph {et~al.}(2019)\citenamefont
  {Nakamura}, \citenamefont {De~Felice}, \citenamefont {Kase},\ and\
  \citenamefont {Tsujikawa}}]{Nakamura:2018oyy}%
  \BibitemOpen
  \bibfield  {author} {\bibinfo {author} {\bibfnamefont {Shintaro}\
  \bibnamefont {Nakamura}}, \bibinfo {author} {\bibfnamefont {Antonio}\
  \bibnamefont {De~Felice}}, \bibinfo {author} {\bibfnamefont {Ryotaro}\
  \bibnamefont {Kase}}, \ and\ \bibinfo {author} {\bibfnamefont {Shinji}\
  \bibnamefont {Tsujikawa}},\ }\bibfield  {title} {\enquote {\bibinfo {title}
  {{Constraints on massive vector dark energy models from integrated
  Sachs-Wolfe-galaxy cross-correlations}},}\ }\href {\doibase
  10.1103/PhysRevD.99.063533} {\bibfield  {journal} {\bibinfo  {journal} {Phys.
  Rev. D}\ }\textbf {\bibinfo {volume} {99}},\ \bibinfo {pages} {063533}
  (\bibinfo {year} {2019})},\ \Eprint {http://arxiv.org/abs/1811.07541}
  {arXiv:1811.07541 [astro-ph.CO]} \BibitemShut {NoStop}%
\bibitem [{\citenamefont {Heisenberg}\ and\ \citenamefont
  {Villarrubia-Rojo}(2021)}]{Heisenberg:2020xak}%
  \BibitemOpen
  \bibfield  {author} {\bibinfo {author} {\bibfnamefont {Lavinia}\ \bibnamefont
  {Heisenberg}}\ and\ \bibinfo {author} {\bibfnamefont {Hector}\ \bibnamefont
  {Villarrubia-Rojo}},\ }\bibfield  {title} {\enquote {\bibinfo {title} {{Proca
  in the sky}},}\ }\href {\doibase 10.1088/1475-7516/2021/03/032} {\bibfield
  {journal} {\bibinfo  {journal} {JCAP}\ }\textbf {\bibinfo {volume} {03}},\
  \bibinfo {pages} {032} (\bibinfo {year} {2021})},\ \Eprint
  {http://arxiv.org/abs/2010.00513} {arXiv:2010.00513 [astro-ph.CO]}
  \BibitemShut {NoStop}%
\bibitem [{\citenamefont {Benisty}\ \emph {et~al.}(2022)\citenamefont
  {Benisty}, \citenamefont {Guendelman}, \citenamefont {van~de Venn},
  \citenamefont {Vasak}, \citenamefont {Struckmeier},\ and\ \citenamefont
  {Stoecker}}]{Benisty:2021sul}%
  \BibitemOpen
  \bibfield  {author} {\bibinfo {author} {\bibfnamefont {David}\ \bibnamefont
  {Benisty}}, \bibinfo {author} {\bibfnamefont {Eduardo~I.}\ \bibnamefont
  {Guendelman}}, \bibinfo {author} {\bibfnamefont {Armin}\ \bibnamefont {van~de
  Venn}}, \bibinfo {author} {\bibfnamefont {David}\ \bibnamefont {Vasak}},
  \bibinfo {author} {\bibfnamefont {J\"urgen}\ \bibnamefont {Struckmeier}}, \
  and\ \bibinfo {author} {\bibfnamefont {Horst}\ \bibnamefont {Stoecker}},\
  }\bibfield  {title} {\enquote {\bibinfo {title} {{The dark side of the
  torsion: dark energy from propagating torsion}},}\ }\href {\doibase
  10.1140/epjc/s10052-022-10187-2} {\bibfield  {journal} {\bibinfo  {journal}
  {Eur. Phys. J. C}\ }\textbf {\bibinfo {volume} {82}},\ \bibinfo {pages} {264}
  (\bibinfo {year} {2022})},\ \Eprint {http://arxiv.org/abs/2109.01052}
  {arXiv:2109.01052 [astro-ph.CO]} \BibitemShut {NoStop}%
\bibitem [{\citenamefont {de~Rham}\ \emph {et~al.}(2022)\citenamefont
  {de~Rham}, \citenamefont {Garcia-Saenz}, \citenamefont {Heisenberg},\ and\
  \citenamefont {Pozsgay}}]{deRham:2021efp}%
  \BibitemOpen
  \bibfield  {author} {\bibinfo {author} {\bibfnamefont {Claudia}\ \bibnamefont
  {de~Rham}}, \bibinfo {author} {\bibfnamefont {Sebastian}\ \bibnamefont
  {Garcia-Saenz}}, \bibinfo {author} {\bibfnamefont {Lavinia}\ \bibnamefont
  {Heisenberg}}, \ and\ \bibinfo {author} {\bibfnamefont {Victor}\ \bibnamefont
  {Pozsgay}},\ }\bibfield  {title} {\enquote {\bibinfo {title} {{Cosmology of
  Extended Proca-Nuevo}},}\ }\href {\doibase 10.1088/1475-7516/2022/03/053}
  {\bibfield  {journal} {\bibinfo  {journal} {JCAP}\ }\textbf {\bibinfo
  {volume} {03}},\ \bibinfo {pages} {053} (\bibinfo {year} {2022})},\ \Eprint
  {http://arxiv.org/abs/2110.14327} {arXiv:2110.14327 [hep-th]} \BibitemShut
  {NoStop}%
\bibitem [{\citenamefont {Hambye}(2009)}]{Hambye:2008bq}%
  \BibitemOpen
  \bibfield  {author} {\bibinfo {author} {\bibfnamefont {Thomas}\ \bibnamefont
  {Hambye}},\ }\bibfield  {title} {\enquote {\bibinfo {title} {{Hidden vector
  dark matter}},}\ }\href {\doibase 10.1088/1126-6708/2009/01/028} {\bibfield
  {journal} {\bibinfo  {journal} {JHEP}\ }\textbf {\bibinfo {volume} {01}},\
  \bibinfo {pages} {028} (\bibinfo {year} {2009})},\ \Eprint
  {http://arxiv.org/abs/0811.0172} {arXiv:0811.0172 [hep-ph]} \BibitemShut
  {NoStop}%
\bibitem [{\citenamefont {Hambye}\ and\ \citenamefont
  {Tytgat}(2010)}]{Hambye:2009fg}%
  \BibitemOpen
  \bibfield  {author} {\bibinfo {author} {\bibfnamefont {Thomas}\ \bibnamefont
  {Hambye}}\ and\ \bibinfo {author} {\bibfnamefont {Michel H.~G.}\ \bibnamefont
  {Tytgat}},\ }\bibfield  {title} {\enquote {\bibinfo {title} {{Confined hidden
  vector dark matter}},}\ }\href {\doibase 10.1016/j.physletb.2009.11.050}
  {\bibfield  {journal} {\bibinfo  {journal} {Phys. Lett. B}\ }\textbf
  {\bibinfo {volume} {683}},\ \bibinfo {pages} {39--41} (\bibinfo {year}
  {2010})},\ \Eprint {http://arxiv.org/abs/0907.1007} {arXiv:0907.1007
  [hep-ph]} \BibitemShut {NoStop}%
\bibitem [{\citenamefont {Arina}\ \emph {et~al.}(2010)\citenamefont {Arina},
  \citenamefont {Hambye}, \citenamefont {Ibarra},\ and\ \citenamefont
  {Weniger}}]{Arina:2009uq}%
  \BibitemOpen
  \bibfield  {author} {\bibinfo {author} {\bibfnamefont {Chiara}\ \bibnamefont
  {Arina}}, \bibinfo {author} {\bibfnamefont {Thomas}\ \bibnamefont {Hambye}},
  \bibinfo {author} {\bibfnamefont {Alejandro}\ \bibnamefont {Ibarra}}, \ and\
  \bibinfo {author} {\bibfnamefont {Christoph}\ \bibnamefont {Weniger}},\
  }\bibfield  {title} {\enquote {\bibinfo {title} {{Intense Gamma-Ray Lines
  from Hidden Vector Dark Matter Decay}},}\ }\href {\doibase
  10.1088/1475-7516/2010/03/024} {\bibfield  {journal} {\bibinfo  {journal}
  {JCAP}\ }\textbf {\bibinfo {volume} {03}},\ \bibinfo {pages} {024} (\bibinfo
  {year} {2010})},\ \Eprint {http://arxiv.org/abs/0912.4496} {arXiv:0912.4496
  [hep-ph]} \BibitemShut {NoStop}%
\bibitem [{\citenamefont {Hisano}\ \emph {et~al.}(2011)\citenamefont {Hisano},
  \citenamefont {Ishiwata}, \citenamefont {Nagata},\ and\ \citenamefont
  {Yamanaka}}]{Hisano:2010yh}%
  \BibitemOpen
  \bibfield  {author} {\bibinfo {author} {\bibfnamefont {Junji}\ \bibnamefont
  {Hisano}}, \bibinfo {author} {\bibfnamefont {Koji}\ \bibnamefont {Ishiwata}},
  \bibinfo {author} {\bibfnamefont {Natsumi}\ \bibnamefont {Nagata}}, \ and\
  \bibinfo {author} {\bibfnamefont {Masato}\ \bibnamefont {Yamanaka}},\
  }\bibfield  {title} {\enquote {\bibinfo {title} {{Direct Detection of Vector
  Dark Matter}},}\ }\href {\doibase 10.1143/PTP.126.435} {\bibfield  {journal}
  {\bibinfo  {journal} {Prog. Theor. Phys.}\ }\textbf {\bibinfo {volume}
  {126}},\ \bibinfo {pages} {435--456} (\bibinfo {year} {2011})},\ \Eprint
  {http://arxiv.org/abs/1012.5455} {arXiv:1012.5455 [hep-ph]} \BibitemShut
  {NoStop}%
\bibitem [{\citenamefont {Diaz-Cruz}\ and\ \citenamefont
  {Ma}(2011)}]{Diaz-Cruz:2010czr}%
  \BibitemOpen
  \bibfield  {author} {\bibinfo {author} {\bibfnamefont {J.~Lorenzo}\
  \bibnamefont {Diaz-Cruz}}\ and\ \bibinfo {author} {\bibfnamefont {Ernest}\
  \bibnamefont {Ma}},\ }\bibfield  {title} {\enquote {\bibinfo {title}
  {{Neutral SU(2) Gauge Extension of the Standard Model and a Vector-Boson
  Dark-Matter Candidate}},}\ }\href {\doibase 10.1016/j.physletb.2010.11.039}
  {\bibfield  {journal} {\bibinfo  {journal} {Phys. Lett. B}\ }\textbf
  {\bibinfo {volume} {695}},\ \bibinfo {pages} {264--267} (\bibinfo {year}
  {2011})},\ \Eprint {http://arxiv.org/abs/1007.2631} {arXiv:1007.2631
  [hep-ph]} \BibitemShut {NoStop}%
\bibitem [{\citenamefont {Lebedev}\ \emph {et~al.}(2012)\citenamefont
  {Lebedev}, \citenamefont {Lee},\ and\ \citenamefont
  {Mambrini}}]{Lebedev:2011iq}%
  \BibitemOpen
  \bibfield  {author} {\bibinfo {author} {\bibfnamefont {Oleg}\ \bibnamefont
  {Lebedev}}, \bibinfo {author} {\bibfnamefont {Hyun~Min}\ \bibnamefont {Lee}},
  \ and\ \bibinfo {author} {\bibfnamefont {Yann}\ \bibnamefont {Mambrini}},\
  }\bibfield  {title} {\enquote {\bibinfo {title} {{Vector Higgs-portal dark
  matter and the invisible Higgs}},}\ }\href {\doibase
  10.1016/j.physletb.2012.01.029} {\bibfield  {journal} {\bibinfo  {journal}
  {Phys. Lett. B}\ }\textbf {\bibinfo {volume} {707}},\ \bibinfo {pages}
  {570--576} (\bibinfo {year} {2012})},\ \Eprint
  {http://arxiv.org/abs/1111.4482} {arXiv:1111.4482 [hep-ph]} \BibitemShut
  {NoStop}%
\bibitem [{\citenamefont {Farzan}\ and\ \citenamefont
  {Akbarieh}(2012)}]{Farzan:2012hh}%
  \BibitemOpen
  \bibfield  {author} {\bibinfo {author} {\bibfnamefont {Yasaman}\ \bibnamefont
  {Farzan}}\ and\ \bibinfo {author} {\bibfnamefont {Amin~Rezaei}\ \bibnamefont
  {Akbarieh}},\ }\bibfield  {title} {\enquote {\bibinfo {title} {{VDM: A model
  for Vector Dark Matter}},}\ }\href {\doibase 10.1088/1475-7516/2012/10/026}
  {\bibfield  {journal} {\bibinfo  {journal} {JCAP}\ }\textbf {\bibinfo
  {volume} {10}},\ \bibinfo {pages} {026} (\bibinfo {year} {2012})},\ \Eprint
  {http://arxiv.org/abs/1207.4272} {arXiv:1207.4272 [hep-ph]} \BibitemShut
  {NoStop}%
\bibitem [{\citenamefont {Baek}\ \emph {et~al.}(2013)\citenamefont {Baek},
  \citenamefont {Ko}, \citenamefont {Park},\ and\ \citenamefont
  {Senaha}}]{Baek:2012se}%
  \BibitemOpen
  \bibfield  {author} {\bibinfo {author} {\bibfnamefont {Seungwon}\
  \bibnamefont {Baek}}, \bibinfo {author} {\bibfnamefont {P.}~\bibnamefont
  {Ko}}, \bibinfo {author} {\bibfnamefont {Wan-Il}\ \bibnamefont {Park}}, \
  and\ \bibinfo {author} {\bibfnamefont {Eibun}\ \bibnamefont {Senaha}},\
  }\bibfield  {title} {\enquote {\bibinfo {title} {{Higgs Portal Vector Dark
  Matter : Revisited}},}\ }\href {\doibase 10.1007/JHEP05(2013)036} {\bibfield
  {journal} {\bibinfo  {journal} {JHEP}\ }\textbf {\bibinfo {volume} {05}},\
  \bibinfo {pages} {036} (\bibinfo {year} {2013})},\ \Eprint
  {http://arxiv.org/abs/1212.2131} {arXiv:1212.2131 [hep-ph]} \BibitemShut
  {NoStop}%
\bibitem [{\citenamefont {Belyaev}\ \emph {et~al.}(2017)\citenamefont
  {Belyaev}, \citenamefont {Thomas},\ and\ \citenamefont
  {Shapiro}}]{Belyaev:2016icc}%
  \BibitemOpen
  \bibfield  {author} {\bibinfo {author} {\bibfnamefont {Alexander~S.}\
  \bibnamefont {Belyaev}}, \bibinfo {author} {\bibfnamefont {Marc~C.}\
  \bibnamefont {Thomas}}, \ and\ \bibinfo {author} {\bibfnamefont {Ilya~L.}\
  \bibnamefont {Shapiro}},\ }\bibfield  {title} {\enquote {\bibinfo {title}
  {{Torsion as a Dark Matter Candidate from the Higgs Portal}},}\ }\href
  {\doibase 10.1103/PhysRevD.95.095033} {\bibfield  {journal} {\bibinfo
  {journal} {Phys. Rev. D}\ }\textbf {\bibinfo {volume} {95}},\ \bibinfo
  {pages} {095033} (\bibinfo {year} {2017})},\ \Eprint
  {http://arxiv.org/abs/1611.03651} {arXiv:1611.03651 [hep-ph]} \BibitemShut
  {NoStop}%
\bibitem [{\citenamefont {Arcadi}\ \emph {et~al.}(2020)\citenamefont {Arcadi},
  \citenamefont {Djouadi},\ and\ \citenamefont {Kado}}]{Arcadi:2020jqf}%
  \BibitemOpen
  \bibfield  {author} {\bibinfo {author} {\bibfnamefont {Giorgio}\ \bibnamefont
  {Arcadi}}, \bibinfo {author} {\bibfnamefont {Abdelhak}\ \bibnamefont
  {Djouadi}}, \ and\ \bibinfo {author} {\bibfnamefont {Marumi}\ \bibnamefont
  {Kado}},\ }\bibfield  {title} {\enquote {\bibinfo {title} {{The Higgs-portal
  for vector dark matter and the effective field theory approach: A
  reappraisal}},}\ }\href {\doibase 10.1016/j.physletb.2020.135427} {\bibfield
  {journal} {\bibinfo  {journal} {Phys. Lett. B}\ }\textbf {\bibinfo {volume}
  {805}},\ \bibinfo {pages} {135427} (\bibinfo {year} {2020})},\ \Eprint
  {http://arxiv.org/abs/2001.10750} {arXiv:2001.10750 [hep-ph]} \BibitemShut
  {NoStop}%
\bibitem [{\citenamefont {Barman}\ \emph {et~al.}(2022)\citenamefont {Barman},
  \citenamefont {Bernal}, \citenamefont {Das},\ and\ \citenamefont
  {Roshan}}]{Barman:2021qds}%
  \BibitemOpen
  \bibfield  {author} {\bibinfo {author} {\bibfnamefont {Basabendu}\
  \bibnamefont {Barman}}, \bibinfo {author} {\bibfnamefont {Nicol\'as}\
  \bibnamefont {Bernal}}, \bibinfo {author} {\bibfnamefont {Ashmita}\
  \bibnamefont {Das}}, \ and\ \bibinfo {author} {\bibfnamefont {Rishav}\
  \bibnamefont {Roshan}},\ }\bibfield  {title} {\enquote {\bibinfo {title}
  {{Non-minimally coupled vector boson dark matter}},}\ }\href {\doibase
  10.1088/1475-7516/2022/01/047} {\bibfield  {journal} {\bibinfo  {journal}
  {JCAP}\ }\textbf {\bibinfo {volume} {01}},\ \bibinfo {pages} {047} (\bibinfo
  {year} {2022})},\ \Eprint {http://arxiv.org/abs/2108.13447} {arXiv:2108.13447
  [hep-ph]} \BibitemShut {NoStop}%
\bibitem [{\citenamefont {Bekenstein}(1972)}]{Bekenstein:1971hc}%
  \BibitemOpen
  \bibfield  {author} {\bibinfo {author} {\bibfnamefont {Jacob~D.}\
  \bibnamefont {Bekenstein}},\ }\bibfield  {title} {\enquote {\bibinfo {title}
  {{Nonexistence of baryon number for static black holes}},}\ }\href {\doibase
  10.1103/PhysRevD.5.1239} {\bibfield  {journal} {\bibinfo  {journal} {Phys.
  Rev. D}\ }\textbf {\bibinfo {volume} {5}},\ \bibinfo {pages} {1239--1246}
  (\bibinfo {year} {1972})}\BibitemShut {NoStop}%
\bibitem [{\citenamefont {Heisenberg}\ \emph {et~al.}(2017)\citenamefont
  {Heisenberg}, \citenamefont {Kase}, \citenamefont {Minamitsuji},\ and\
  \citenamefont {Tsujikawa}}]{Heisenberg:2017hwb}%
  \BibitemOpen
  \bibfield  {author} {\bibinfo {author} {\bibfnamefont {Lavinia}\ \bibnamefont
  {Heisenberg}}, \bibinfo {author} {\bibfnamefont {Ryotaro}\ \bibnamefont
  {Kase}}, \bibinfo {author} {\bibfnamefont {Masato}\ \bibnamefont
  {Minamitsuji}}, \ and\ \bibinfo {author} {\bibfnamefont {Shinji}\
  \bibnamefont {Tsujikawa}},\ }\bibfield  {title} {\enquote {\bibinfo {title}
  {{Black holes in vector-tensor theories}},}\ }\href {\doibase
  10.1088/1475-7516/2017/08/024} {\bibfield  {journal} {\bibinfo  {journal}
  {JCAP}\ }\textbf {\bibinfo {volume} {08}},\ \bibinfo {pages} {024} (\bibinfo
  {year} {2017})},\ \Eprint {http://arxiv.org/abs/1706.05115} {arXiv:1706.05115
  [gr-qc]} \BibitemShut {NoStop}%
\bibitem [{\citenamefont {Garcia-Saenz}\ \emph {et~al.}(2021)\citenamefont
  {Garcia-Saenz}, \citenamefont {Held},\ and\ \citenamefont
  {Zhang}}]{Garcia-Saenz:2021uyv}%
  \BibitemOpen
  \bibfield  {author} {\bibinfo {author} {\bibfnamefont {Sebastian}\
  \bibnamefont {Garcia-Saenz}}, \bibinfo {author} {\bibfnamefont {Aaron}\
  \bibnamefont {Held}}, \ and\ \bibinfo {author} {\bibfnamefont {Jun}\
  \bibnamefont {Zhang}},\ }\bibfield  {title} {\enquote {\bibinfo {title}
  {{Destabilization of Black Holes and Stars by Generalized Proca Fields}},}\
  }\href {\doibase 10.1103/PhysRevLett.127.131104} {\bibfield  {journal}
  {\bibinfo  {journal} {Phys. Rev. Lett.}\ }\textbf {\bibinfo {volume} {127}},\
  \bibinfo {pages} {131104} (\bibinfo {year} {2021})},\ \Eprint
  {http://arxiv.org/abs/2104.08049} {arXiv:2104.08049 [gr-qc]} \BibitemShut
  {NoStop}%
\bibitem [{\citenamefont {Hehl}\ \emph
  {et~al.}(1976{\natexlab{d}})\citenamefont {Hehl}, \citenamefont {Von
  Der~Heyde}, \citenamefont {Kerlick},\ and\ \citenamefont
  {Nester}}]{Hehl:1976kj}%
  \BibitemOpen
  \bibfield  {author} {\bibinfo {author} {\bibfnamefont {F.~W.}\ \bibnamefont
  {Hehl}}, \bibinfo {author} {\bibfnamefont {P.}~\bibnamefont {Von Der~Heyde}},
  \bibinfo {author} {\bibfnamefont {G.~D.}\ \bibnamefont {Kerlick}}, \ and\
  \bibinfo {author} {\bibfnamefont {J.~M.}\ \bibnamefont {Nester}},\ }\bibfield
   {title} {\enquote {\bibinfo {title} {{General Relativity with Spin and
  Torsion: Foundations and Prospects}},}\ }\href {\doibase
  10.1103/RevModPhys.48.393} {\bibfield  {journal} {\bibinfo  {journal} {Rev.
  Mod. Phys.}\ }\textbf {\bibinfo {volume} {48}},\ \bibinfo {pages} {393--416}
  (\bibinfo {year} {1976}{\natexlab{d}})}\BibitemShut {NoStop}%
\bibitem [{\citenamefont {Shapiro}(2002)}]{Shapiro:2001rz}%
  \BibitemOpen
  \bibfield  {author} {\bibinfo {author} {\bibfnamefont {I.~L.}\ \bibnamefont
  {Shapiro}},\ }\bibfield  {title} {\enquote {\bibinfo {title} {{Physical
  aspects of the space-time torsion}},}\ }\href {\doibase
  10.1016/S0370-1573(01)00030-8} {\bibfield  {journal} {\bibinfo  {journal}
  {Phys. Rept.}\ }\textbf {\bibinfo {volume} {357}},\ \bibinfo {pages} {113}
  (\bibinfo {year} {2002})},\ \Eprint {http://arxiv.org/abs/hep-th/0103093}
  {arXiv:hep-th/0103093} \BibitemShut {NoStop}%
\bibitem [{\citenamefont {Blagojevi{\'c}}(2002)}]{Blagojevic:2002}%
  \BibitemOpen
  \bibfield  {author} {\bibinfo {author} {\bibfnamefont {M.}~\bibnamefont
  {Blagojevi{\'c}}},\ }\href {https://books.google.co.uk/books?id=Bmm2ugEACAAJ}
  {\emph {\bibinfo {title} {Gravitation and Gauge Symmetries}}},\ Series in
  high energy physics, cosmology, and gravitation\ (\bibinfo  {publisher}
  {Institute of Physics Publishing},\ \bibinfo {address} {Bristol, UK},\
  \bibinfo {year} {2002})\BibitemShut {NoStop}%
\bibitem [{\citenamefont {Blagojevic}\ and\ \citenamefont
  {Hehl}(2012)}]{Blagojevic:2012bc}%
  \BibitemOpen
  \bibfield  {author} {\bibinfo {author} {\bibfnamefont {Milutin}\ \bibnamefont
  {Blagojevic}}\ and\ \bibinfo {author} {\bibfnamefont {Friedrich~W.}\
  \bibnamefont {Hehl}},\ }\bibfield  {title} {\enquote {\bibinfo {title}
  {{Gauge Theories of Gravitation}},}\ }\href@noop {} {\  (\bibinfo {year}
  {2012})},\ \Eprint {http://arxiv.org/abs/1210.3775} {arXiv:1210.3775 [gr-qc]}
  \BibitemShut {NoStop}%
\bibitem [{\citenamefont {Sciama}(1964)}]{Sciama:1964wt}%
  \BibitemOpen
  \bibfield  {author} {\bibinfo {author} {\bibfnamefont {Dennis~W.}\
  \bibnamefont {Sciama}},\ }\bibfield  {title} {\enquote {\bibinfo {title}
  {{The Physical structure of general relativity}},}\ }\href {\doibase
  10.1103/RevModPhys.36.1103} {\bibfield  {journal} {\bibinfo  {journal} {Rev.
  Mod. Phys.}\ }\textbf {\bibinfo {volume} {36}},\ \bibinfo {pages} {463--469}
  (\bibinfo {year} {1964})},\ \bibinfo {note} {[Erratum: Rev.Mod.Phys. 36,
  1103--1103 (1964)]}\BibitemShut {NoStop}%
\bibitem [{\citenamefont {Hayashi}\ and\ \citenamefont
  {Shirafuji}(1980{\natexlab{c}})}]{Hayashi:1980ir}%
  \BibitemOpen
  \bibfield  {author} {\bibinfo {author} {\bibfnamefont {Kenji}\ \bibnamefont
  {Hayashi}}\ and\ \bibinfo {author} {\bibfnamefont {Takeshi}\ \bibnamefont
  {Shirafuji}},\ }\bibfield  {title} {\enquote {\bibinfo {title} {{Gravity From
  Poincare Gauge Theory of the Fundamental Particles. 3. Weak Field
  Approximation}},}\ }\href {\doibase 10.1143/PTP.64.1435} {\bibfield
  {journal} {\bibinfo  {journal} {Prog. Theor. Phys.}\ }\textbf {\bibinfo
  {volume} {64}},\ \bibinfo {pages} {1435} (\bibinfo {year}
  {1980}{\natexlab{c}})},\ \bibinfo {note} {[Erratum: Prog.Theor.Phys. 66, 741
  (1981)]}\BibitemShut {NoStop}%
\bibitem [{\citenamefont {Sezgin}(1981)}]{Sezgin:1981xs}%
  \BibitemOpen
  \bibfield  {author} {\bibinfo {author} {\bibfnamefont {E.}~\bibnamefont
  {Sezgin}},\ }\bibfield  {title} {\enquote {\bibinfo {title} {{Class of Ghost
  Free Gravity Lagrangians With Massive or Massless Propagating Torsion}},}\
  }\href {\doibase 10.1103/PhysRevD.24.1677} {\bibfield  {journal} {\bibinfo
  {journal} {Phys. Rev. D}\ }\textbf {\bibinfo {volume} {24}},\ \bibinfo
  {pages} {1677--1680} (\bibinfo {year} {1981})}\BibitemShut {NoStop}%
\bibitem [{\citenamefont {Blagojevic}\ and\ \citenamefont
  {Nikolic}(1983)}]{Blagojevic:1983zz}%
  \BibitemOpen
  \bibfield  {author} {\bibinfo {author} {\bibfnamefont {M.}~\bibnamefont
  {Blagojevic}}\ and\ \bibinfo {author} {\bibfnamefont {I.~A.}\ \bibnamefont
  {Nikolic}},\ }\bibfield  {title} {\enquote {\bibinfo {title} {{Hamiltonian
  dynamics of Poincare gauge theory: General structure in the time gauge}},}\
  }\href {\doibase 10.1103/PhysRevD.28.2455} {\bibfield  {journal} {\bibinfo
  {journal} {Phys. Rev. D}\ }\textbf {\bibinfo {volume} {28}},\ \bibinfo
  {pages} {2455--2463} (\bibinfo {year} {1983})}\BibitemShut {NoStop}%
\bibitem [{\citenamefont {Blagojevic}\ and\ \citenamefont
  {Vasilic}(1987)}]{Blagojevic:1986dm}%
  \BibitemOpen
  \bibfield  {author} {\bibinfo {author} {\bibfnamefont {M.}~\bibnamefont
  {Blagojevic}}\ and\ \bibinfo {author} {\bibfnamefont {M.}~\bibnamefont
  {Vasilic}},\ }\bibfield  {title} {\enquote {\bibinfo {title} {{EXTRA GAUGE
  SYMMETRIES IN A WEAK FIELD APPROXIMATION OF AN R + T**2 + R**2 THEORY OF
  GRAVITY}},}\ }\href {\doibase 10.1103/PhysRevD.35.3748} {\bibfield  {journal}
  {\bibinfo  {journal} {Phys. Rev. D}\ }\textbf {\bibinfo {volume} {35}},\
  \bibinfo {pages} {3748} (\bibinfo {year} {1987})}\BibitemShut {NoStop}%
\bibitem [{\citenamefont {Kuhfuss}\ and\ \citenamefont
  {Nitsch}(1986)}]{Kuhfuss:1986rb}%
  \BibitemOpen
  \bibfield  {author} {\bibinfo {author} {\bibfnamefont {R.}~\bibnamefont
  {Kuhfuss}}\ and\ \bibinfo {author} {\bibfnamefont {J.}~\bibnamefont
  {Nitsch}},\ }\bibfield  {title} {\enquote {\bibinfo {title} {{Propagating
  Modes in Gauge Field Theories of Gravity}},}\ }\href {\doibase
  10.1007/BF00763447} {\bibfield  {journal} {\bibinfo  {journal} {Gen. Rel.
  Grav.}\ }\textbf {\bibinfo {volume} {18}},\ \bibinfo {pages} {1207} (\bibinfo
  {year} {1986})}\BibitemShut {NoStop}%
\bibitem [{\citenamefont {Puetzfeld}(2005)}]{Puetzfeld:2004yg}%
  \BibitemOpen
  \bibfield  {author} {\bibinfo {author} {\bibfnamefont {Dirk}\ \bibnamefont
  {Puetzfeld}},\ }\bibfield  {title} {\enquote {\bibinfo {title} {{Status of
  non-Riemannian cosmology}},}\ }\href {\doibase 10.1016/j.newar.2005.01.022}
  {\bibfield  {journal} {\bibinfo  {journal} {New Astron. Rev.}\ }\textbf
  {\bibinfo {volume} {49}},\ \bibinfo {pages} {59--64} (\bibinfo {year}
  {2005})},\ \Eprint {http://arxiv.org/abs/gr-qc/0404119} {arXiv:gr-qc/0404119}
  \BibitemShut {NoStop}%
\bibitem [{\citenamefont {Yo}\ and\ \citenamefont {Nester}(2007)}]{Yo:2006qs}%
  \BibitemOpen
  \bibfield  {author} {\bibinfo {author} {\bibfnamefont {Hwei-Jang}\
  \bibnamefont {Yo}}\ and\ \bibinfo {author} {\bibfnamefont {James~M.}\
  \bibnamefont {Nester}},\ }\bibfield  {title} {\enquote {\bibinfo {title}
  {{Dynamic Scalar Torsion and an Oscillating Universe}},}\ }\href {\doibase
  10.1142/S0217732307025303} {\bibfield  {journal} {\bibinfo  {journal} {Mod.
  Phys. Lett. A}\ }\textbf {\bibinfo {volume} {22}},\ \bibinfo {pages}
  {2057--2069} (\bibinfo {year} {2007})},\ \Eprint
  {http://arxiv.org/abs/astro-ph/0612738} {arXiv:astro-ph/0612738} \BibitemShut
  {NoStop}%
\bibitem [{\citenamefont {Shie}\ \emph {et~al.}(2008)\citenamefont {Shie},
  \citenamefont {Nester},\ and\ \citenamefont {Yo}}]{Shie:2008ms}%
  \BibitemOpen
  \bibfield  {author} {\bibinfo {author} {\bibfnamefont {Kun-Feng}\
  \bibnamefont {Shie}}, \bibinfo {author} {\bibfnamefont {James~M.}\
  \bibnamefont {Nester}}, \ and\ \bibinfo {author} {\bibfnamefont {Hwei-Jang}\
  \bibnamefont {Yo}},\ }\bibfield  {title} {\enquote {\bibinfo {title}
  {{Torsion Cosmology and the Accelerating Universe}},}\ }\href {\doibase
  10.1103/PhysRevD.78.023522} {\bibfield  {journal} {\bibinfo  {journal} {Phys.
  Rev. D}\ }\textbf {\bibinfo {volume} {78}},\ \bibinfo {pages} {023522}
  (\bibinfo {year} {2008})},\ \Eprint {http://arxiv.org/abs/0805.3834}
  {arXiv:0805.3834 [gr-qc]} \BibitemShut {NoStop}%
\bibitem [{\citenamefont {Nair}\ \emph {et~al.}(2009)\citenamefont {Nair},
  \citenamefont {Randjbar-Daemi},\ and\ \citenamefont {Rubakov}}]{Nair:2008yh}%
  \BibitemOpen
  \bibfield  {author} {\bibinfo {author} {\bibfnamefont {V.~P.}\ \bibnamefont
  {Nair}}, \bibinfo {author} {\bibfnamefont {S.}~\bibnamefont
  {Randjbar-Daemi}}, \ and\ \bibinfo {author} {\bibfnamefont {V.}~\bibnamefont
  {Rubakov}},\ }\bibfield  {title} {\enquote {\bibinfo {title} {{Massive Spin-2
  fields of Geometric Origin in Curved Spacetimes}},}\ }\href {\doibase
  10.1103/PhysRevD.80.104031} {\bibfield  {journal} {\bibinfo  {journal} {Phys.
  Rev. D}\ }\textbf {\bibinfo {volume} {80}},\ \bibinfo {pages} {104031}
  (\bibinfo {year} {2009})},\ \Eprint {http://arxiv.org/abs/0811.3781}
  {arXiv:0811.3781 [hep-th]} \BibitemShut {NoStop}%
\bibitem [{\citenamefont {Nikiforova}\ \emph {et~al.}(2009)\citenamefont
  {Nikiforova}, \citenamefont {Randjbar-Daemi},\ and\ \citenamefont
  {Rubakov}}]{Nikiforova:2009qr}%
  \BibitemOpen
  \bibfield  {author} {\bibinfo {author} {\bibfnamefont {V.}~\bibnamefont
  {Nikiforova}}, \bibinfo {author} {\bibfnamefont {S.}~\bibnamefont
  {Randjbar-Daemi}}, \ and\ \bibinfo {author} {\bibfnamefont {V.}~\bibnamefont
  {Rubakov}},\ }\bibfield  {title} {\enquote {\bibinfo {title} {{Infrared
  Modified Gravity with Dynamical Torsion}},}\ }\href {\doibase
  10.1103/PhysRevD.80.124050} {\bibfield  {journal} {\bibinfo  {journal} {Phys.
  Rev. D}\ }\textbf {\bibinfo {volume} {80}},\ \bibinfo {pages} {124050}
  (\bibinfo {year} {2009})},\ \Eprint {http://arxiv.org/abs/0905.3732}
  {arXiv:0905.3732 [hep-th]} \BibitemShut {NoStop}%
\bibitem [{\citenamefont {Chen}\ \emph {et~al.}(2009)\citenamefont {Chen},
  \citenamefont {Ho}, \citenamefont {Nester}, \citenamefont {Wang},\ and\
  \citenamefont {Yo}}]{Chen:2009at}%
  \BibitemOpen
  \bibfield  {author} {\bibinfo {author} {\bibfnamefont {Hsin}\ \bibnamefont
  {Chen}}, \bibinfo {author} {\bibfnamefont {Fei-Hung}\ \bibnamefont {Ho}},
  \bibinfo {author} {\bibfnamefont {James~M.}\ \bibnamefont {Nester}}, \bibinfo
  {author} {\bibfnamefont {Chih-Hung}\ \bibnamefont {Wang}}, \ and\ \bibinfo
  {author} {\bibfnamefont {Hwei-Jang}\ \bibnamefont {Yo}},\ }\bibfield  {title}
  {\enquote {\bibinfo {title} {{Cosmological dynamics with propagating Lorentz
  connection modes of spin zero}},}\ }\href {\doibase
  10.1088/1475-7516/2009/10/027} {\bibfield  {journal} {\bibinfo  {journal}
  {JCAP}\ }\textbf {\bibinfo {volume} {10}},\ \bibinfo {pages} {027} (\bibinfo
  {year} {2009})},\ \Eprint {http://arxiv.org/abs/0908.3323} {arXiv:0908.3323
  [gr-qc]} \BibitemShut {NoStop}%
\bibitem [{\citenamefont {Ni}(2010)}]{Ni:2009fg}%
  \BibitemOpen
  \bibfield  {author} {\bibinfo {author} {\bibfnamefont {Wei-Tou}\ \bibnamefont
  {Ni}},\ }\bibfield  {title} {\enquote {\bibinfo {title} {{Searches for the
  role of spin and polarization in gravity}},}\ }\href {\doibase
  10.1088/0034-4885/73/5/056901} {\bibfield  {journal} {\bibinfo  {journal}
  {Rept. Prog. Phys.}\ }\textbf {\bibinfo {volume} {73}},\ \bibinfo {pages}
  {056901} (\bibinfo {year} {2010})},\ \Eprint {http://arxiv.org/abs/0912.5057}
  {arXiv:0912.5057 [gr-qc]} \BibitemShut {NoStop}%
\bibitem [{\citenamefont {Baekler}\ \emph {et~al.}(2011)\citenamefont
  {Baekler}, \citenamefont {Hehl},\ and\ \citenamefont
  {Nester}}]{Baekler:2010fr}%
  \BibitemOpen
  \bibfield  {author} {\bibinfo {author} {\bibfnamefont {Peter}\ \bibnamefont
  {Baekler}}, \bibinfo {author} {\bibfnamefont {Friedrich~W.}\ \bibnamefont
  {Hehl}}, \ and\ \bibinfo {author} {\bibfnamefont {James~M.}\ \bibnamefont
  {Nester}},\ }\bibfield  {title} {\enquote {\bibinfo {title} {{Poincare gauge
  theory of gravity: Friedman cosmology with even and odd parity modes.
  Analytic part}},}\ }\href {\doibase 10.1103/PhysRevD.83.024001} {\bibfield
  {journal} {\bibinfo  {journal} {Phys. Rev. D}\ }\textbf {\bibinfo {volume}
  {83}},\ \bibinfo {pages} {024001} (\bibinfo {year} {2011})},\ \Eprint
  {http://arxiv.org/abs/1009.5112} {arXiv:1009.5112 [gr-qc]} \BibitemShut
  {NoStop}%
\bibitem [{\citenamefont {Ho}\ and\ \citenamefont {Nester}(2011)}]{Ho:2011qn}%
  \BibitemOpen
  \bibfield  {author} {\bibinfo {author} {\bibfnamefont {Fei-Hung}\
  \bibnamefont {Ho}}\ and\ \bibinfo {author} {\bibfnamefont {James~M.}\
  \bibnamefont {Nester}},\ }\bibfield  {title} {\enquote {\bibinfo {title}
  {{Poincar\'e gauge theory with even and odd parity dynamic connection modes:
  isotropic Bianchi cosmological models}},}\ }\href {\doibase
  10.1088/1742-6596/330/1/012005} {\bibfield  {journal} {\bibinfo  {journal}
  {J. Phys. Conf. Ser.}\ }\textbf {\bibinfo {volume} {330}},\ \bibinfo {pages}
  {012005} (\bibinfo {year} {2011})},\ \Eprint {http://arxiv.org/abs/1105.5001}
  {arXiv:1105.5001 [gr-qc]} \BibitemShut {NoStop}%
\bibitem [{\citenamefont {Ho}\ and\ \citenamefont {Nester}(2012)}]{Ho:2011xf}%
  \BibitemOpen
  \bibfield  {author} {\bibinfo {author} {\bibfnamefont {Fei-Hung}\
  \bibnamefont {Ho}}\ and\ \bibinfo {author} {\bibfnamefont {James~M.}\
  \bibnamefont {Nester}},\ }\bibfield  {title} {\enquote {\bibinfo {title}
  {{Poincar\'e Gauge Theory With Coupled Even And Odd Parity Dynamic Spin-0
  Modes: Dynamic Equations For Isotropic Bianchi Cosmologies}},}\ }\href
  {\doibase 10.1002/andp.201100101} {\bibfield  {journal} {\bibinfo  {journal}
  {Annalen Phys.}\ }\textbf {\bibinfo {volume} {524}},\ \bibinfo {pages}
  {97--106} (\bibinfo {year} {2012})},\ \Eprint
  {http://arxiv.org/abs/1106.0711} {arXiv:1106.0711 [gr-qc]} \BibitemShut
  {NoStop}%
\bibitem [{\citenamefont {Ong}\ \emph {et~al.}(2013)\citenamefont {Ong},
  \citenamefont {Izumi}, \citenamefont {Nester},\ and\ \citenamefont
  {Chen}}]{Ong:2013qja}%
  \BibitemOpen
  \bibfield  {author} {\bibinfo {author} {\bibfnamefont {Yen~Chin}\
  \bibnamefont {Ong}}, \bibinfo {author} {\bibfnamefont {Keisuke}\ \bibnamefont
  {Izumi}}, \bibinfo {author} {\bibfnamefont {James~M.}\ \bibnamefont
  {Nester}}, \ and\ \bibinfo {author} {\bibfnamefont {Pisin}\ \bibnamefont
  {Chen}},\ }\bibfield  {title} {\enquote {\bibinfo {title} {{Problems with
  Propagation and Time Evolution in f(T) Gravity}},}\ }\href {\doibase
  10.1103/PhysRevD.88.024019} {\bibfield  {journal} {\bibinfo  {journal} {Phys.
  Rev. D}\ }\textbf {\bibinfo {volume} {88}},\ \bibinfo {pages} {024019}
  (\bibinfo {year} {2013})},\ \Eprint {http://arxiv.org/abs/1303.0993}
  {arXiv:1303.0993 [gr-qc]} \BibitemShut {NoStop}%
\bibitem [{\citenamefont {Puetzfeld}\ and\ \citenamefont
  {Obukhov}(2014)}]{Puetzfeld:2014sja}%
  \BibitemOpen
  \bibfield  {author} {\bibinfo {author} {\bibfnamefont {Dirk}\ \bibnamefont
  {Puetzfeld}}\ and\ \bibinfo {author} {\bibfnamefont {Yuri~N.}\ \bibnamefont
  {Obukhov}},\ }\bibfield  {title} {\enquote {\bibinfo {title} {{Prospects of
  detecting spacetime torsion}},}\ }\href {\doibase 10.1142/S0218271814420048}
  {\bibfield  {journal} {\bibinfo  {journal} {Int. J. Mod. Phys. D}\ }\textbf
  {\bibinfo {volume} {23}},\ \bibinfo {pages} {1442004} (\bibinfo {year}
  {2014})},\ \Eprint {http://arxiv.org/abs/1405.4137} {arXiv:1405.4137 [gr-qc]}
  \BibitemShut {NoStop}%
\bibitem [{\citenamefont {Karananas}(2015)}]{Karananas:2014pxa}%
  \BibitemOpen
  \bibfield  {author} {\bibinfo {author} {\bibfnamefont {Georgios~K.}\
  \bibnamefont {Karananas}},\ }\bibfield  {title} {\enquote {\bibinfo {title}
  {{The particle spectrum of parity-violating Poincar\'e gravitational
  theory}},}\ }\href {\doibase 10.1088/0264-9381/32/5/055012} {\bibfield
  {journal} {\bibinfo  {journal} {Class. Quant. Grav.}\ }\textbf {\bibinfo
  {volume} {32}},\ \bibinfo {pages} {055012} (\bibinfo {year} {2015})},\
  \Eprint {http://arxiv.org/abs/1411.5613} {arXiv:1411.5613 [gr-qc]}
  \BibitemShut {NoStop}%
\bibitem [{\citenamefont {Ni}(2016)}]{Ni:2015poa}%
  \BibitemOpen
  \bibfield  {author} {\bibinfo {author} {\bibfnamefont {Wei-Tou}\ \bibnamefont
  {Ni}},\ }\bibfield  {title} {\enquote {\bibinfo {title} {{Searches for the
  role of spin and polarization in gravity: a five-year update}},}\ }\href
  {\doibase 10.1142/S2010194516600107} {\bibfield  {journal} {\bibinfo
  {journal} {Int. J. Mod. Phys. Conf. Ser.}\ }\textbf {\bibinfo {volume}
  {40}},\ \bibinfo {pages} {1660010} (\bibinfo {year} {2016})},\ \Eprint
  {http://arxiv.org/abs/1501.07696} {arXiv:1501.07696 [hep-ph]} \BibitemShut
  {NoStop}%
\bibitem [{\citenamefont {Ho}\ \emph {et~al.}(2015)\citenamefont {Ho},
  \citenamefont {Chen}, \citenamefont {Nester},\ and\ \citenamefont
  {Yo}}]{Ho:2015ulu}%
  \BibitemOpen
  \bibfield  {author} {\bibinfo {author} {\bibfnamefont {Fei-Hung}\
  \bibnamefont {Ho}}, \bibinfo {author} {\bibfnamefont {Hsin}\ \bibnamefont
  {Chen}}, \bibinfo {author} {\bibfnamefont {James~M.}\ \bibnamefont {Nester}},
  \ and\ \bibinfo {author} {\bibfnamefont {Hwei-Jang}\ \bibnamefont {Yo}},\
  }\bibfield  {title} {\enquote {\bibinfo {title} {{General Poincar\'e Gauge
  Theory Cosmology}},}\ }\href {\doibase 10.6122/CJP.20151014} {\bibfield
  {journal} {\bibinfo  {journal} {Chin. J. Phys.}\ }\textbf {\bibinfo {volume}
  {53}},\ \bibinfo {pages} {110109} (\bibinfo {year} {2015})},\ \Eprint
  {http://arxiv.org/abs/1512.01202} {arXiv:1512.01202 [gr-qc]} \BibitemShut
  {NoStop}%
\bibitem [{\citenamefont {Karananas}(2016)}]{Karananas:2016ltn}%
  \BibitemOpen
  \bibfield  {author} {\bibinfo {author} {\bibfnamefont {Georgios~K.}\
  \bibnamefont {Karananas}},\ }\emph {\bibinfo {title} {{Poincar\'e, Scale and
  Conformal Symmetries Gauge Perspective and Cosmological Ramifications}}},\
  \href {\doibase 10.5075/epfl-thesis-7173} {Ph.D. thesis},\ \bibinfo  {school}
  {Ecole Polytechnique, Lausanne} (\bibinfo {year} {2016}),\ \Eprint
  {http://arxiv.org/abs/1608.08451} {arXiv:1608.08451 [hep-th]} \BibitemShut
  {NoStop}%
\bibitem [{\citenamefont {Obukhov}(2017)}]{Obukhov:2017pxa}%
  \BibitemOpen
  \bibfield  {author} {\bibinfo {author} {\bibfnamefont {Yuri~N.}\ \bibnamefont
  {Obukhov}},\ }\bibfield  {title} {\enquote {\bibinfo {title} {{Gravitational
  waves in Poincar\'e gauge gravity theory}},}\ }\href {\doibase
  10.1103/PhysRevD.95.084028} {\bibfield  {journal} {\bibinfo  {journal} {Phys.
  Rev. D}\ }\textbf {\bibinfo {volume} {95}},\ \bibinfo {pages} {084028}
  (\bibinfo {year} {2017})},\ \Eprint {http://arxiv.org/abs/1702.05185}
  {arXiv:1702.05185 [gr-qc]} \BibitemShut {NoStop}%
\bibitem [{\citenamefont {Blagojevi\'c}\ \emph {et~al.}(2017)\citenamefont
  {Blagojevi\'c}, \citenamefont {Cvetkovi\'c},\ and\ \citenamefont
  {Obukhov}}]{Blagojevic:2017ssv}%
  \BibitemOpen
  \bibfield  {author} {\bibinfo {author} {\bibfnamefont {Milutin}\ \bibnamefont
  {Blagojevi\'c}}, \bibinfo {author} {\bibfnamefont {Branislav}\ \bibnamefont
  {Cvetkovi\'c}}, \ and\ \bibinfo {author} {\bibfnamefont {Yuri~N.}\
  \bibnamefont {Obukhov}},\ }\bibfield  {title} {\enquote {\bibinfo {title}
  {{Generalized plane waves in Poincar\'e gauge theory of gravity}},}\ }\href
  {\doibase 10.1103/PhysRevD.96.064031} {\bibfield  {journal} {\bibinfo
  {journal} {Phys. Rev. D}\ }\textbf {\bibinfo {volume} {96}},\ \bibinfo
  {pages} {064031} (\bibinfo {year} {2017})},\ \Eprint
  {http://arxiv.org/abs/1708.08766} {arXiv:1708.08766 [gr-qc]} \BibitemShut
  {NoStop}%
\bibitem [{\citenamefont {Blagojevi\'c}\ and\ \citenamefont
  {Cvetkovi\'c}(2018)}]{Blagojevic:2018dpz}%
  \BibitemOpen
  \bibfield  {author} {\bibinfo {author} {\bibfnamefont {Milutin}\ \bibnamefont
  {Blagojevi\'c}}\ and\ \bibinfo {author} {\bibfnamefont {Branislav}\
  \bibnamefont {Cvetkovi\'c}},\ }\bibfield  {title} {\enquote {\bibinfo {title}
  {{General Poincar\'e gauge theory: Hamiltonian structure and particle
  spectrum}},}\ }\href {\doibase 10.1103/PhysRevD.98.024014} {\bibfield
  {journal} {\bibinfo  {journal} {Phys. Rev. D}\ }\textbf {\bibinfo {volume}
  {98}},\ \bibinfo {pages} {024014} (\bibinfo {year} {2018})},\ \Eprint
  {http://arxiv.org/abs/1804.05556} {arXiv:1804.05556 [gr-qc]} \BibitemShut
  {NoStop}%
\bibitem [{\citenamefont {Tseng}(2018)}]{Tseng:2018feo}%
  \BibitemOpen
  \bibfield  {author} {\bibinfo {author} {\bibfnamefont {Huan-Hsin}\
  \bibnamefont {Tseng}},\ }\emph {\bibinfo {title} {{Gravitational Theories
  with Torsion}}},\ \href@noop {} {Ph.D. thesis},\ \bibinfo  {school} {Taiwan,
  Natl. Tsing Hua U.} (\bibinfo {year} {2018}),\ \Eprint
  {http://arxiv.org/abs/1812.00314} {arXiv:1812.00314 [gr-qc]} \BibitemShut
  {NoStop}%
\bibitem [{\citenamefont {Lin}\ \emph {et~al.}(2019)\citenamefont {Lin},
  \citenamefont {Hobson},\ and\ \citenamefont {Lasenby}}]{Lin:2018awc}%
  \BibitemOpen
  \bibfield  {author} {\bibinfo {author} {\bibfnamefont {Yun-Cherng}\
  \bibnamefont {Lin}}, \bibinfo {author} {\bibfnamefont {Michael~P.}\
  \bibnamefont {Hobson}}, \ and\ \bibinfo {author} {\bibfnamefont {Anthony~N.}\
  \bibnamefont {Lasenby}},\ }\bibfield  {title} {\enquote {\bibinfo {title}
  {{Ghost and tachyon free Poincar\'e gauge theories: A systematic
  approach}},}\ }\href {\doibase 10.1103/PhysRevD.99.064001} {\bibfield
  {journal} {\bibinfo  {journal} {Phys. Rev. D}\ }\textbf {\bibinfo {volume}
  {99}},\ \bibinfo {pages} {064001} (\bibinfo {year} {2019})},\ \Eprint
  {http://arxiv.org/abs/1812.02675} {arXiv:1812.02675 [gr-qc]} \BibitemShut
  {NoStop}%
\bibitem [{\citenamefont {Beltr\'an~Jim\'enez}\ and\ \citenamefont
  {Delhom}(2019)}]{BeltranJimenez:2019acz}%
  \BibitemOpen
  \bibfield  {author} {\bibinfo {author} {\bibfnamefont {Jose}\ \bibnamefont
  {Beltr\'an~Jim\'enez}}\ and\ \bibinfo {author} {\bibfnamefont {Adria}\
  \bibnamefont {Delhom}},\ }\bibfield  {title} {\enquote {\bibinfo {title}
  {{Ghosts in metric-affine higher order curvature gravity}},}\ }\href
  {\doibase 10.1140/epjc/s10052-019-7149-x} {\bibfield  {journal} {\bibinfo
  {journal} {Eur. Phys. J. C}\ }\textbf {\bibinfo {volume} {79}},\ \bibinfo
  {pages} {656} (\bibinfo {year} {2019})},\ \Eprint
  {http://arxiv.org/abs/1901.08988} {arXiv:1901.08988 [gr-qc]} \BibitemShut
  {NoStop}%
\bibitem [{\citenamefont {Zhang}\ and\ \citenamefont
  {Xu}(2019)}]{Zhang:2019mhd}%
  \BibitemOpen
  \bibfield  {author} {\bibinfo {author} {\bibfnamefont {Hongchao}\
  \bibnamefont {Zhang}}\ and\ \bibinfo {author} {\bibfnamefont {Lixin}\
  \bibnamefont {Xu}},\ }\bibfield  {title} {\enquote {\bibinfo {title}
  {{Late-time acceleration and inflation in a Poincar\'e gauge cosmological
  model}},}\ }\href {\doibase 10.1088/1475-7516/2019/09/050} {\bibfield
  {journal} {\bibinfo  {journal} {JCAP}\ }\textbf {\bibinfo {volume} {09}},\
  \bibinfo {pages} {050} (\bibinfo {year} {2019})},\ \Eprint
  {http://arxiv.org/abs/1904.03545} {arXiv:1904.03545 [gr-qc]} \BibitemShut
  {NoStop}%
\bibitem [{\citenamefont {Aoki}\ and\ \citenamefont
  {Shimada}(2019)}]{Aoki:2019rvi}%
  \BibitemOpen
  \bibfield  {author} {\bibinfo {author} {\bibfnamefont {Katsuki}\ \bibnamefont
  {Aoki}}\ and\ \bibinfo {author} {\bibfnamefont {Keigo}\ \bibnamefont
  {Shimada}},\ }\bibfield  {title} {\enquote {\bibinfo {title}
  {{Scalar-metric-affine theories: Can we get ghost-free theories from
  symmetry?}}}\ }\href {\doibase 10.1103/PhysRevD.100.044037} {\bibfield
  {journal} {\bibinfo  {journal} {Phys. Rev. D}\ }\textbf {\bibinfo {volume}
  {100}},\ \bibinfo {pages} {044037} (\bibinfo {year} {2019})},\ \Eprint
  {http://arxiv.org/abs/1904.10175} {arXiv:1904.10175 [hep-th]} \BibitemShut
  {NoStop}%
\bibitem [{\citenamefont {Zhang}\ and\ \citenamefont
  {Xu}(2020)}]{Zhang:2019xek}%
  \BibitemOpen
  \bibfield  {author} {\bibinfo {author} {\bibfnamefont {Hongchao}\
  \bibnamefont {Zhang}}\ and\ \bibinfo {author} {\bibfnamefont {Lixin}\
  \bibnamefont {Xu}},\ }\bibfield  {title} {\enquote {\bibinfo {title}
  {{Inflation in the parity-conserving Poincar\'e gauge cosmology}},}\ }\href
  {\doibase 10.1088/1475-7516/2020/10/003} {\bibfield  {journal} {\bibinfo
  {journal} {JCAP}\ }\textbf {\bibinfo {volume} {10}},\ \bibinfo {pages} {003}
  (\bibinfo {year} {2020})},\ \Eprint {http://arxiv.org/abs/1906.04340}
  {arXiv:1906.04340 [gr-qc]} \BibitemShut {NoStop}%
\bibitem [{\citenamefont {Beltr\'an~Jim\'enez}\ and\ \citenamefont
  {Maldonado~Torralba}(2020{\natexlab{a}})}]{Jimenez:2019qjc}%
  \BibitemOpen
  \bibfield  {author} {\bibinfo {author} {\bibfnamefont {Jose}\ \bibnamefont
  {Beltr\'an~Jim\'enez}}\ and\ \bibinfo {author} {\bibfnamefont
  {Francisco~Jos\'e}\ \bibnamefont {Maldonado~Torralba}},\ }\bibfield  {title}
  {\enquote {\bibinfo {title} {{Revisiting the stability of quadratic
  Poincar\'e gauge gravity}},}\ }\href {\doibase
  10.1140/epjc/s10052-020-8163-8} {\bibfield  {journal} {\bibinfo  {journal}
  {Eur. Phys. J. C}\ }\textbf {\bibinfo {volume} {80}},\ \bibinfo {pages} {611}
  (\bibinfo {year} {2020}{\natexlab{a}})},\ \Eprint
  {http://arxiv.org/abs/1910.07506} {arXiv:1910.07506 [gr-qc]} \BibitemShut
  {NoStop}%
\bibitem [{\citenamefont {Lin}\ \emph {et~al.}(2020)\citenamefont {Lin},
  \citenamefont {Hobson},\ and\ \citenamefont {Lasenby}}]{Lin:2019ugq}%
  \BibitemOpen
  \bibfield  {author} {\bibinfo {author} {\bibfnamefont {Yun-Cherng}\
  \bibnamefont {Lin}}, \bibinfo {author} {\bibfnamefont {Michael~P.}\
  \bibnamefont {Hobson}}, \ and\ \bibinfo {author} {\bibfnamefont {Anthony~N.}\
  \bibnamefont {Lasenby}},\ }\bibfield  {title} {\enquote {\bibinfo {title}
  {{Power-counting renormalizable, ghost-and-tachyon-free Poincar\'e gauge
  theories}},}\ }\href {\doibase 10.1103/PhysRevD.101.064038} {\bibfield
  {journal} {\bibinfo  {journal} {Phys. Rev. D}\ }\textbf {\bibinfo {volume}
  {101}},\ \bibinfo {pages} {064038} (\bibinfo {year} {2020})},\ \Eprint
  {http://arxiv.org/abs/1910.14197} {arXiv:1910.14197 [gr-qc]} \BibitemShut
  {NoStop}%
\bibitem [{\citenamefont {Percacci}\ and\ \citenamefont
  {Sezgin}(2020{\natexlab{a}})}]{Percacci:2019hxn}%
  \BibitemOpen
  \bibfield  {author} {\bibinfo {author} {\bibfnamefont {R.}~\bibnamefont
  {Percacci}}\ and\ \bibinfo {author} {\bibfnamefont {E.}~\bibnamefont
  {Sezgin}},\ }\bibfield  {title} {\enquote {\bibinfo {title} {{New class of
  ghost- and tachyon-free metric affine gravities}},}\ }\href {\doibase
  10.1103/PhysRevD.101.084040} {\bibfield  {journal} {\bibinfo  {journal}
  {Phys. Rev. D}\ }\textbf {\bibinfo {volume} {101}},\ \bibinfo {pages}
  {084040} (\bibinfo {year} {2020}{\natexlab{a}})},\ \Eprint
  {http://arxiv.org/abs/1912.01023} {arXiv:1912.01023 [hep-th]} \BibitemShut
  {NoStop}%
\bibitem [{\citenamefont {Barker}\ \emph {et~al.}(2020)\citenamefont {Barker},
  \citenamefont {Lasenby}, \citenamefont {Hobson},\ and\ \citenamefont
  {Handley}}]{Barker:2020gcp}%
  \BibitemOpen
  \bibfield  {author} {\bibinfo {author} {\bibfnamefont {W.~E.~V.}\
  \bibnamefont {Barker}}, \bibinfo {author} {\bibfnamefont {A.~N.}\
  \bibnamefont {Lasenby}}, \bibinfo {author} {\bibfnamefont {M.~P.}\
  \bibnamefont {Hobson}}, \ and\ \bibinfo {author} {\bibfnamefont {W.~J.}\
  \bibnamefont {Handley}},\ }\bibfield  {title} {\enquote {\bibinfo {title}
  {{Systematic study of background cosmology in unitary Poincar\'e gauge
  theories with application to emergent dark radiation and $H_0$ tension}},}\
  }\href {\doibase 10.1103/PhysRevD.102.024048} {\bibfield  {journal} {\bibinfo
   {journal} {Phys. Rev. D}\ }\textbf {\bibinfo {volume} {102}},\ \bibinfo
  {pages} {024048} (\bibinfo {year} {2020})},\ \Eprint
  {http://arxiv.org/abs/2003.02690} {arXiv:2003.02690 [gr-qc]} \BibitemShut
  {NoStop}%
\bibitem [{\citenamefont {Beltr\'an~Jim\'enez}\ and\ \citenamefont
  {Delhom}(2020)}]{BeltranJimenez:2020sqf}%
  \BibitemOpen
  \bibfield  {author} {\bibinfo {author} {\bibfnamefont {Jose}\ \bibnamefont
  {Beltr\'an~Jim\'enez}}\ and\ \bibinfo {author} {\bibfnamefont {Adri\`a}\
  \bibnamefont {Delhom}},\ }\bibfield  {title} {\enquote {\bibinfo {title}
  {{Instabilities in metric-affine theories of gravity with higher order
  curvature terms}},}\ }\href {\doibase 10.1140/epjc/s10052-020-8143-z}
  {\bibfield  {journal} {\bibinfo  {journal} {Eur. Phys. J. C}\ }\textbf
  {\bibinfo {volume} {80}},\ \bibinfo {pages} {585} (\bibinfo {year} {2020})},\
  \Eprint {http://arxiv.org/abs/2004.11357} {arXiv:2004.11357 [gr-qc]}
  \BibitemShut {NoStop}%
\bibitem [{\citenamefont {Barker}\ \emph {et~al.}(2021)\citenamefont {Barker},
  \citenamefont {Lasenby}, \citenamefont {Hobson},\ and\ \citenamefont
  {Handley}}]{Barker:2021oez}%
  \BibitemOpen
  \bibfield  {author} {\bibinfo {author} {\bibfnamefont {W.~E.~V.}\
  \bibnamefont {Barker}}, \bibinfo {author} {\bibfnamefont {A.~N.}\
  \bibnamefont {Lasenby}}, \bibinfo {author} {\bibfnamefont {M.~P.}\
  \bibnamefont {Hobson}}, \ and\ \bibinfo {author} {\bibfnamefont {W.~J.}\
  \bibnamefont {Handley}},\ }\bibfield  {title} {\enquote {\bibinfo {title}
  {{Nonlinear Hamiltonian analysis of new quadratic torsion theories: Cases
  with curvature-free constraints}},}\ }\href {\doibase
  10.1103/PhysRevD.104.084036} {\bibfield  {journal} {\bibinfo  {journal}
  {Phys. Rev. D}\ }\textbf {\bibinfo {volume} {104}},\ \bibinfo {pages}
  {084036} (\bibinfo {year} {2021})},\ \Eprint
  {http://arxiv.org/abs/2101.02645} {arXiv:2101.02645 [gr-qc]} \BibitemShut
  {NoStop}%
\bibitem [{\citenamefont
  {Maldonado~Torralba}(2020)}]{MaldonadoTorralba:2020mbh}%
  \BibitemOpen
  \bibfield  {author} {\bibinfo {author} {\bibfnamefont {Francisco~Jos\'e}\
  \bibnamefont {Maldonado~Torralba}},\ }\emph {\bibinfo {title} {{New effective
  theories of gravitation and their phenomenological consequences}}},\ \href
  {\doibase 10.33612/diss.143961423} {Ph.D. thesis},\ \bibinfo  {school} {Cape
  Town U., Dept. Math.} (\bibinfo {year} {2020}),\ \Eprint
  {http://arxiv.org/abs/2101.11523} {arXiv:2101.11523 [gr-qc]} \BibitemShut
  {NoStop}%
\bibitem [{\citenamefont {Marzo}(2022{\natexlab{a}})}]{Marzo:2021esg}%
  \BibitemOpen
  \bibfield  {author} {\bibinfo {author} {\bibfnamefont {Carlo}\ \bibnamefont
  {Marzo}},\ }\bibfield  {title} {\enquote {\bibinfo {title} {{Ghost and
  tachyon free propagation up to spin 3 in Lorentz invariant field
  theories}},}\ }\href {\doibase 10.1103/PhysRevD.105.065017} {\bibfield
  {journal} {\bibinfo  {journal} {Phys. Rev. D}\ }\textbf {\bibinfo {volume}
  {105}},\ \bibinfo {pages} {065017} (\bibinfo {year} {2022}{\natexlab{a}})},\
  \Eprint {http://arxiv.org/abs/2108.11982} {arXiv:2108.11982 [hep-ph]}
  \BibitemShut {NoStop}%
\bibitem [{\citenamefont {Marzo}(2022{\natexlab{b}})}]{Marzo:2021iok}%
  \BibitemOpen
  \bibfield  {author} {\bibinfo {author} {\bibfnamefont {Carlo}\ \bibnamefont
  {Marzo}},\ }\bibfield  {title} {\enquote {\bibinfo {title} {{Radiatively
  stable ghost and tachyon freedom in metric affine gravity}},}\ }\href
  {\doibase 10.1103/PhysRevD.106.024045} {\bibfield  {journal} {\bibinfo
  {journal} {Phys. Rev. D}\ }\textbf {\bibinfo {volume} {106}},\ \bibinfo
  {pages} {024045} (\bibinfo {year} {2022}{\natexlab{b}})},\ \Eprint
  {http://arxiv.org/abs/2110.14788} {arXiv:2110.14788 [hep-th]} \BibitemShut
  {NoStop}%
\bibitem [{\citenamefont {de~la Cruz~Dombriz}\ \emph
  {et~al.}(2022)\citenamefont {de~la Cruz~Dombriz}, \citenamefont
  {Maldonado~Torralba},\ and\ \citenamefont {Mota}}]{delaCruzDombriz:2021nrg}%
  \BibitemOpen
  \bibfield  {author} {\bibinfo {author} {\bibfnamefont {\'Alvaro}\
  \bibnamefont {de~la Cruz~Dombriz}}, \bibinfo {author} {\bibfnamefont
  {Francisco~Jos\'e}\ \bibnamefont {Maldonado~Torralba}}, \ and\ \bibinfo
  {author} {\bibfnamefont {David~F.}\ \bibnamefont {Mota}},\ }\bibfield
  {title} {\enquote {\bibinfo {title} {{Dark matter candidate from torsion}},}\
  }\href {\doibase 10.1016/j.physletb.2022.137488} {\bibfield  {journal}
  {\bibinfo  {journal} {Phys. Lett. B}\ }\textbf {\bibinfo {volume} {834}},\
  \bibinfo {pages} {137488} (\bibinfo {year} {2022})},\ \Eprint
  {http://arxiv.org/abs/2112.03957} {arXiv:2112.03957 [gr-qc]} \BibitemShut
  {NoStop}%
\bibitem [{\citenamefont {Baldazzi}\ \emph {et~al.}(2022)\citenamefont
  {Baldazzi}, \citenamefont {Melichev},\ and\ \citenamefont
  {Percacci}}]{Baldazzi:2021kaf}%
  \BibitemOpen
  \bibfield  {author} {\bibinfo {author} {\bibfnamefont {A.}~\bibnamefont
  {Baldazzi}}, \bibinfo {author} {\bibfnamefont {O.}~\bibnamefont {Melichev}},
  \ and\ \bibinfo {author} {\bibfnamefont {R.}~\bibnamefont {Percacci}},\
  }\bibfield  {title} {\enquote {\bibinfo {title} {{Metric-Affine Gravity as an
  effective field theory}},}\ }\href {\doibase 10.1016/j.aop.2022.168757}
  {\bibfield  {journal} {\bibinfo  {journal} {Annals Phys.}\ }\textbf {\bibinfo
  {volume} {438}},\ \bibinfo {pages} {168757} (\bibinfo {year} {2022})},\
  \Eprint {http://arxiv.org/abs/2112.10193} {arXiv:2112.10193 [gr-qc]}
  \BibitemShut {NoStop}%
\bibitem [{\citenamefont {Annala}\ and\ \citenamefont
  {Rasanen}(2023)}]{Annala:2022gtl}%
  \BibitemOpen
  \bibfield  {author} {\bibinfo {author} {\bibfnamefont {Jaakko}\ \bibnamefont
  {Annala}}\ and\ \bibinfo {author} {\bibfnamefont {Syksy}\ \bibnamefont
  {Rasanen}},\ }\bibfield  {title} {\enquote {\bibinfo {title} {{Stability of
  non-degenerate Ricci-type Palatini theories}},}\ }\href {\doibase
  10.1088/1475-7516/2023/04/014} {\bibfield  {journal} {\bibinfo  {journal}
  {JCAP}\ }\textbf {\bibinfo {volume} {04}},\ \bibinfo {pages} {014} (\bibinfo
  {year} {2023})},\ \Eprint {http://arxiv.org/abs/2212.09820} {arXiv:2212.09820
  [gr-qc]} \BibitemShut {NoStop}%
\bibitem [{\citenamefont {Dimakis}(1989{\natexlab{a}})}]{Dimakis:1989az}%
  \BibitemOpen
  \bibfield  {author} {\bibinfo {author} {\bibfnamefont {A.}~\bibnamefont
  {Dimakis}},\ }\bibfield  {title} {\enquote {\bibinfo {title} {{The Initial
  Value Problem of the Poincare Gauge Theory in Vacuum. 1: Second Order
  Formalism}},}\ }\href {https://eudml.org/doc/76473} {\bibfield  {journal}
  {\bibinfo  {journal} {Ann. Inst. H. Poincare Phys. Theor.}\ }\textbf
  {\bibinfo {volume} {51}},\ \bibinfo {pages} {371--388} (\bibinfo {year}
  {1989}{\natexlab{a}})}\BibitemShut {NoStop}%
\bibitem [{\citenamefont {Dimakis}(1989{\natexlab{b}})}]{Dimakis:1989ba}%
  \BibitemOpen
  \bibfield  {author} {\bibinfo {author} {\bibfnamefont {A.}~\bibnamefont
  {Dimakis}},\ }\bibfield  {title} {\enquote {\bibinfo {title} {{THE INITIAL
  VALUE PROBLEM OF THE POINCARE GAUGE THEORY IN VACUUM. 1: FIRST ORDER
  FORMALISM}},}\ }\href {https://eudml.org/doc/76474} {\bibfield  {journal}
  {\bibinfo  {journal} {Ann. Inst. H. Poincare Phys. Theor.}\ }\textbf
  {\bibinfo {volume} {51}},\ \bibinfo {pages} {389--417} (\bibinfo {year}
  {1989}{\natexlab{b}})}\BibitemShut {NoStop}%
\bibitem [{\citenamefont {Lemke}(1990)}]{Lemke:1990su}%
  \BibitemOpen
  \bibfield  {author} {\bibinfo {author} {\bibfnamefont {J.}~\bibnamefont
  {Lemke}},\ }\bibfield  {title} {\enquote {\bibinfo {title} {{Shock waves in
  the Poincare gauge theory of gravitation}},}\ }\href {\doibase
  10.1016/0375-9601(90)90789-Q} {\bibfield  {journal} {\bibinfo  {journal}
  {Phys. Lett. A}\ }\textbf {\bibinfo {volume} {143}},\ \bibinfo {pages}
  {13--16} (\bibinfo {year} {1990})}\BibitemShut {NoStop}%
\bibitem [{\citenamefont {Afshordi}\ \emph {et~al.}(2007)\citenamefont
  {Afshordi}, \citenamefont {Chung},\ and\ \citenamefont
  {Geshnizjani}}]{Afshordi:2006ad}%
  \BibitemOpen
  \bibfield  {author} {\bibinfo {author} {\bibfnamefont {Niayesh}\ \bibnamefont
  {Afshordi}}, \bibinfo {author} {\bibfnamefont {Daniel J.~H.}\ \bibnamefont
  {Chung}}, \ and\ \bibinfo {author} {\bibfnamefont {Ghazal}\ \bibnamefont
  {Geshnizjani}},\ }\bibfield  {title} {\enquote {\bibinfo {title} {{Cuscuton:
  A Causal Field Theory with an Infinite Speed of Sound}},}\ }\href {\doibase
  10.1103/PhysRevD.75.083513} {\bibfield  {journal} {\bibinfo  {journal} {Phys.
  Rev. D}\ }\textbf {\bibinfo {volume} {75}},\ \bibinfo {pages} {083513}
  (\bibinfo {year} {2007})},\ \Eprint {http://arxiv.org/abs/hep-th/0609150}
  {arXiv:hep-th/0609150} \BibitemShut {NoStop}%
\bibitem [{\citenamefont {Magueijo}(2009)}]{Magueijo:2008sx}%
  \BibitemOpen
  \bibfield  {author} {\bibinfo {author} {\bibfnamefont {Joao}\ \bibnamefont
  {Magueijo}},\ }\bibfield  {title} {\enquote {\bibinfo {title} {{Bimetric
  varying speed of light theories and primordial fluctuations}},}\ }\href
  {\doibase 10.1103/PhysRevD.79.043525} {\bibfield  {journal} {\bibinfo
  {journal} {Phys. Rev. D}\ }\textbf {\bibinfo {volume} {79}},\ \bibinfo
  {pages} {043525} (\bibinfo {year} {2009})},\ \Eprint
  {http://arxiv.org/abs/0807.1689} {arXiv:0807.1689 [gr-qc]} \BibitemShut
  {NoStop}%
\bibitem [{\citenamefont {Velo}\ and\ \citenamefont
  {Zwanziger}(1969)}]{Velo:1969txo}%
  \BibitemOpen
  \bibfield  {author} {\bibinfo {author} {\bibfnamefont {G.}~\bibnamefont
  {Velo}}\ and\ \bibinfo {author} {\bibfnamefont {D.}~\bibnamefont
  {Zwanziger}},\ }\bibfield  {title} {\enquote {\bibinfo {title} {{Noncausality
  and other defects of interaction lagrangians for particles with spin one and
  higher}},}\ }\href {\doibase 10.1103/PhysRev.188.2218} {\bibfield  {journal}
  {\bibinfo  {journal} {Phys. Rev.}\ }\textbf {\bibinfo {volume} {188}},\
  \bibinfo {pages} {2218--2222} (\bibinfo {year} {1969})}\BibitemShut {NoStop}%
\bibitem [{\citenamefont {Aragone}\ and\ \citenamefont
  {Deser}(1971)}]{Aragone:1971kh}%
  \BibitemOpen
  \bibfield  {author} {\bibinfo {author} {\bibfnamefont {C.}~\bibnamefont
  {Aragone}}\ and\ \bibinfo {author} {\bibfnamefont {Stanley}\ \bibnamefont
  {Deser}},\ }\bibfield  {title} {\enquote {\bibinfo {title} {{Constraints on
  gravitationally coupled tensor fields}},}\ }\href {\doibase
  10.1007/BF02813572} {\bibfield  {journal} {\bibinfo  {journal} {Nuovo Cim.
  A}\ }\textbf {\bibinfo {volume} {3}},\ \bibinfo {pages} {709--720} (\bibinfo
  {year} {1971})}\BibitemShut {NoStop}%
\bibitem [{\citenamefont {Cheng}\ \emph {et~al.}(1988)\citenamefont {Cheng},
  \citenamefont {Chern},\ and\ \citenamefont {Nester}}]{Cheng:1988zg}%
  \BibitemOpen
  \bibfield  {author} {\bibinfo {author} {\bibfnamefont {Wen-Hao}\ \bibnamefont
  {Cheng}}, \bibinfo {author} {\bibfnamefont {De-Ching}\ \bibnamefont {Chern}},
  \ and\ \bibinfo {author} {\bibfnamefont {J.~M.}\ \bibnamefont {Nester}},\
  }\bibfield  {title} {\enquote {\bibinfo {title} {{Canonical Analysis of the
  One Parameter Teleparallel Theory}},}\ }\href {\doibase
  10.1103/PhysRevD.38.2656} {\bibfield  {journal} {\bibinfo  {journal} {Phys.
  Rev. D}\ }\textbf {\bibinfo {volume} {38}},\ \bibinfo {pages} {2656--2658}
  (\bibinfo {year} {1988})}\BibitemShut {NoStop}%
\bibitem [{\citenamefont {Blixt}\ \emph
  {et~al.}(2019{\natexlab{a}})\citenamefont {Blixt}, \citenamefont {Hohmann},\
  and\ \citenamefont {Pfeifer}}]{Blixt:2018znp}%
  \BibitemOpen
  \bibfield  {author} {\bibinfo {author} {\bibfnamefont {Daniel}\ \bibnamefont
  {Blixt}}, \bibinfo {author} {\bibfnamefont {Manuel}\ \bibnamefont {Hohmann}},
  \ and\ \bibinfo {author} {\bibfnamefont {Christian}\ \bibnamefont
  {Pfeifer}},\ }\bibfield  {title} {\enquote {\bibinfo {title} {{Hamiltonian
  and primary constraints of new general relativity}},}\ }\href {\doibase
  10.1103/PhysRevD.99.084025} {\bibfield  {journal} {\bibinfo  {journal} {Phys.
  Rev. D}\ }\textbf {\bibinfo {volume} {99}},\ \bibinfo {pages} {084025}
  (\bibinfo {year} {2019}{\natexlab{a}})},\ \Eprint
  {http://arxiv.org/abs/1811.11137} {arXiv:1811.11137 [gr-qc]} \BibitemShut
  {NoStop}%
\bibitem [{\citenamefont {Blixt}\ \emph
  {et~al.}(2019{\natexlab{b}})\citenamefont {Blixt}, \citenamefont {Hohmann},
  \citenamefont {Kr\v{s}\v{s}\'ak},\ and\ \citenamefont
  {Pfeifer}}]{Blixt:2019ene}%
  \BibitemOpen
  \bibfield  {author} {\bibinfo {author} {\bibfnamefont {Daniel}\ \bibnamefont
  {Blixt}}, \bibinfo {author} {\bibfnamefont {Manuel}\ \bibnamefont {Hohmann}},
  \bibinfo {author} {\bibfnamefont {Martin}\ \bibnamefont {Kr\v{s}\v{s}\'ak}},
  \ and\ \bibinfo {author} {\bibfnamefont {Christian}\ \bibnamefont
  {Pfeifer}},\ }\bibfield  {title} {\enquote {\bibinfo {title} {{Hamiltonian
  Analysis In New General Relativity}},}\ }in\ \href {\doibase
  10.1142/9789811258251_0038} {\emph {\bibinfo {booktitle} {{15th Marcel
  Grossmann Meeting on Recent Developments in Theoretical and Experimental
  General Relativity, Astrophysics, and Relativistic Field Theories}}}}\
  (\bibinfo  {publisher} {World Scientific Publishing Co Pte Ltd},\ \bibinfo
  {address} {Singapore},\ \bibinfo {year} {2019})\ \Eprint
  {http://arxiv.org/abs/1905.11919} {arXiv:1905.11919 [gr-qc]} \BibitemShut
  {NoStop}%
\bibitem [{\citenamefont {Blixt}\ \emph {et~al.}(2021)\citenamefont {Blixt},
  \citenamefont {Guzm\'an}, \citenamefont {Hohmann},\ and\ \citenamefont
  {Pfeifer}}]{Blixt:2020ekl}%
  \BibitemOpen
  \bibfield  {author} {\bibinfo {author} {\bibfnamefont {Daniel}\ \bibnamefont
  {Blixt}}, \bibinfo {author} {\bibfnamefont {Mar\'\i{}a-Jos\'e}\ \bibnamefont
  {Guzm\'an}}, \bibinfo {author} {\bibfnamefont {Manuel}\ \bibnamefont
  {Hohmann}}, \ and\ \bibinfo {author} {\bibfnamefont {Christian}\ \bibnamefont
  {Pfeifer}},\ }\bibfield  {title} {\enquote {\bibinfo {title} {{Review of the
  Hamiltonian analysis in teleparallel gravity}},}\ }\href {\doibase
  10.1142/S0219887821300051} {\bibfield  {journal} {\bibinfo  {journal} {Int.
  J. Geom. Meth. Mod. Phys.}\ }\textbf {\bibinfo {volume} {18}},\ \bibinfo
  {pages} {2130005} (\bibinfo {year} {2021})},\ \Eprint
  {http://arxiv.org/abs/2012.09180} {arXiv:2012.09180 [gr-qc]} \BibitemShut
  {NoStop}%
\bibitem [{\citenamefont {Krasnov}\ and\ \citenamefont
  {Mitsou}(2021)}]{Krasnov:2021zen}%
  \BibitemOpen
  \bibfield  {author} {\bibinfo {author} {\bibfnamefont {Kirill}\ \bibnamefont
  {Krasnov}}\ and\ \bibinfo {author} {\bibfnamefont {Ermis}\ \bibnamefont
  {Mitsou}},\ }\bibfield  {title} {\enquote {\bibinfo {title} {{Pure Lorentz
  spin connection theories and uniqueness of general relativity}},}\ }\href
  {\doibase 10.1088/1361-6382/ac25e3} {\bibfield  {journal} {\bibinfo
  {journal} {Class. Quant. Grav.}\ }\textbf {\bibinfo {volume} {38}},\ \bibinfo
  {pages} {205009} (\bibinfo {year} {2021})},\ \Eprint
  {http://arxiv.org/abs/2106.05803} {arXiv:2106.05803 [gr-qc]} \BibitemShut
  {NoStop}%
\bibitem [{\citenamefont {Bahamonde}\ \emph {et~al.}(2023)\citenamefont
  {Bahamonde}, \citenamefont {Dialektopoulos}, \citenamefont
  {Escamilla-Rivera}, \citenamefont {Farrugia}, \citenamefont {Gakis},
  \citenamefont {Hendry}, \citenamefont {Hohmann}, \citenamefont {Levi~Said},
  \citenamefont {Mifsud},\ and\ \citenamefont
  {Di~Valentino}}]{Bahamonde:2021gfp}%
  \BibitemOpen
  \bibfield  {author} {\bibinfo {author} {\bibfnamefont {Sebastian}\
  \bibnamefont {Bahamonde}}, \bibinfo {author} {\bibfnamefont
  {Konstantinos~F.}\ \bibnamefont {Dialektopoulos}}, \bibinfo {author}
  {\bibfnamefont {Celia}\ \bibnamefont {Escamilla-Rivera}}, \bibinfo {author}
  {\bibfnamefont {Gabriel}\ \bibnamefont {Farrugia}}, \bibinfo {author}
  {\bibfnamefont {Viktor}\ \bibnamefont {Gakis}}, \bibinfo {author}
  {\bibfnamefont {Martin}\ \bibnamefont {Hendry}}, \bibinfo {author}
  {\bibfnamefont {Manuel}\ \bibnamefont {Hohmann}}, \bibinfo {author}
  {\bibfnamefont {Jackson}\ \bibnamefont {Levi~Said}}, \bibinfo {author}
  {\bibfnamefont {Jurgen}\ \bibnamefont {Mifsud}}, \ and\ \bibinfo {author}
  {\bibfnamefont {Eleonora}\ \bibnamefont {Di~Valentino}},\ }\bibfield  {title}
  {\enquote {\bibinfo {title} {{Teleparallel gravity: from theory to
  cosmology}},}\ }\href {\doibase 10.1088/1361-6633/ac9cef} {\bibfield
  {journal} {\bibinfo  {journal} {Rept. Prog. Phys.}\ }\textbf {\bibinfo
  {volume} {86}},\ \bibinfo {pages} {026901} (\bibinfo {year} {2023})},\
  \Eprint {http://arxiv.org/abs/2106.13793} {arXiv:2106.13793 [gr-qc]}
  \BibitemShut {NoStop}%
\bibitem [{\citenamefont {Delhom}\ \emph {et~al.}(2022)\citenamefont {Delhom},
  \citenamefont {Jim\'enez-Cano},\ and\ \citenamefont
  {Maldonado~Torralba}}]{Delhom:2022vae}%
  \BibitemOpen
  \bibfield  {author} {\bibinfo {author} {\bibfnamefont {Adri\`a}\ \bibnamefont
  {Delhom}}, \bibinfo {author} {\bibfnamefont {Alejandro}\ \bibnamefont
  {Jim\'enez-Cano}}, \ and\ \bibinfo {author} {\bibfnamefont
  {Francisco~Jos\'e}\ \bibnamefont {Maldonado~Torralba}},\ }\bibfield  {title}
  {\enquote {\bibinfo {title} {{Instabilities in field theories: Lecture notes
  with a view into modified gravity}},}\ \ }(\bibinfo {year} {2022})\ \Eprint
  {http://arxiv.org/abs/2207.13431} {arXiv:2207.13431 [gr-qc]} \BibitemShut
  {NoStop}%
\bibitem [{\citenamefont {Blagojevi\'c}\ and\ \citenamefont
  {Cvetkovi\'c}(2013{\natexlab{a}})}]{Blagojevic:2013dea}%
  \BibitemOpen
  \bibfield  {author} {\bibinfo {author} {\bibfnamefont {M.}~\bibnamefont
  {Blagojevi\'c}}\ and\ \bibinfo {author} {\bibfnamefont {B.}~\bibnamefont
  {Cvetkovi\'c}},\ }\bibfield  {title} {\enquote {\bibinfo {title}
  {{Three-dimensional gravity with propagating torsion: Hamiltonian structure
  of the scalar sector}},}\ }\href {\doibase 10.1103/PhysRevD.88.104032}
  {\bibfield  {journal} {\bibinfo  {journal} {Phys. Rev. D}\ }\textbf {\bibinfo
  {volume} {88}},\ \bibinfo {pages} {104032} (\bibinfo {year}
  {2013}{\natexlab{a}})},\ \Eprint {http://arxiv.org/abs/1309.0411}
  {arXiv:1309.0411 [gr-qc]} \BibitemShut {NoStop}%
\bibitem [{\citenamefont {Blagojevi\'c}\ and\ \citenamefont
  {Cvetkovi\'c}(2013{\natexlab{b}})}]{Blagojevic:2013taa}%
  \BibitemOpen
  \bibfield  {author} {\bibinfo {author} {\bibfnamefont {M.}~\bibnamefont
  {Blagojevi\'c}}\ and\ \bibinfo {author} {\bibfnamefont {B.}~\bibnamefont
  {Cvetkovi\'c}},\ }\bibfield  {title} {\enquote {\bibinfo {title} {{Poincar\'e
  gauge theory in 3D: canonical stability of the scalar sector}},}\ }\href@noop
  {} {\  (\bibinfo {year} {2013}{\natexlab{b}})},\ \Eprint
  {http://arxiv.org/abs/1310.8309} {arXiv:1310.8309 [gr-qc]} \BibitemShut
  {NoStop}%
\bibitem [{\citenamefont {Beltr\'an~Jim\'enez}\ and\ \citenamefont
  {Maldonado~Torralba}(2020{\natexlab{b}})}]{BeltranJimenez:2019hrm}%
  \BibitemOpen
  \bibfield  {author} {\bibinfo {author} {\bibfnamefont {Jose}\ \bibnamefont
  {Beltr\'an~Jim\'enez}}\ and\ \bibinfo {author} {\bibfnamefont
  {Francisco~Jos\'e}\ \bibnamefont {Maldonado~Torralba}},\ }\bibfield  {title}
  {\enquote {\bibinfo {title} {{Revisiting the stability of quadratic
  Poincar\'e gauge gravity}},}\ }\href {\doibase
  10.1140/epjc/s10052-020-8163-8} {\bibfield  {journal} {\bibinfo  {journal}
  {Eur. Phys. J. C}\ }\textbf {\bibinfo {volume} {80}},\ \bibinfo {pages} {611}
  (\bibinfo {year} {2020}{\natexlab{b}})},\ \Eprint
  {http://arxiv.org/abs/1910.07506} {arXiv:1910.07506 [gr-qc]} \BibitemShut
  {NoStop}%
\bibitem [{\citenamefont {Aoki}(2020)}]{Aoki:2020rae}%
  \BibitemOpen
  \bibfield  {author} {\bibinfo {author} {\bibfnamefont {Katsuki}\ \bibnamefont
  {Aoki}},\ }\bibfield  {title} {\enquote {\bibinfo {title} {{Nonlinearly
  ghost-free higher curvature gravity}},}\ }\href {\doibase
  10.1103/PhysRevD.102.124049} {\bibfield  {journal} {\bibinfo  {journal}
  {Phys. Rev. D}\ }\textbf {\bibinfo {volume} {102}},\ \bibinfo {pages}
  {124049} (\bibinfo {year} {2020})},\ \Eprint
  {http://arxiv.org/abs/2009.11739} {arXiv:2009.11739 [hep-th]} \BibitemShut
  {NoStop}%
\bibitem [{\citenamefont {Charmousis}\ and\ \citenamefont
  {Padilla}(2008)}]{Charmousis:2008ce}%
  \BibitemOpen
  \bibfield  {author} {\bibinfo {author} {\bibfnamefont {Christos}\
  \bibnamefont {Charmousis}}\ and\ \bibinfo {author} {\bibfnamefont {Antonio}\
  \bibnamefont {Padilla}},\ }\bibfield  {title} {\enquote {\bibinfo {title}
  {{The Instability of Vacua in Gauss-Bonnet Gravity}},}\ }\href {\doibase
  10.1088/1126-6708/2008/12/038} {\bibfield  {journal} {\bibinfo  {journal}
  {JHEP}\ }\textbf {\bibinfo {volume} {12}},\ \bibinfo {pages} {038} (\bibinfo
  {year} {2008})},\ \Eprint {http://arxiv.org/abs/0807.2864} {arXiv:0807.2864
  [hep-th]} \BibitemShut {NoStop}%
\bibitem [{\citenamefont {Charmousis}\ \emph {et~al.}(2009)\citenamefont
  {Charmousis}, \citenamefont {Niz}, \citenamefont {Padilla},\ and\
  \citenamefont {Saffin}}]{Charmousis:2009tc}%
  \BibitemOpen
  \bibfield  {author} {\bibinfo {author} {\bibfnamefont {Christos}\
  \bibnamefont {Charmousis}}, \bibinfo {author} {\bibfnamefont {Gustavo}\
  \bibnamefont {Niz}}, \bibinfo {author} {\bibfnamefont {Antonio}\ \bibnamefont
  {Padilla}}, \ and\ \bibinfo {author} {\bibfnamefont {Paul~M.}\ \bibnamefont
  {Saffin}},\ }\bibfield  {title} {\enquote {\bibinfo {title} {{Strong coupling
  in Horava gravity}},}\ }\href {\doibase 10.1088/1126-6708/2009/08/070}
  {\bibfield  {journal} {\bibinfo  {journal} {JHEP}\ }\textbf {\bibinfo
  {volume} {08}},\ \bibinfo {pages} {070} (\bibinfo {year} {2009})},\ \Eprint
  {http://arxiv.org/abs/0905.2579} {arXiv:0905.2579 [hep-th]} \BibitemShut
  {NoStop}%
\bibitem [{\citenamefont {Papazoglou}\ and\ \citenamefont
  {Sotiriou}(2010)}]{Papazoglou:2009fj}%
  \BibitemOpen
  \bibfield  {author} {\bibinfo {author} {\bibfnamefont {Antonios}\
  \bibnamefont {Papazoglou}}\ and\ \bibinfo {author} {\bibfnamefont
  {Thomas~P.}\ \bibnamefont {Sotiriou}},\ }\bibfield  {title} {\enquote
  {\bibinfo {title} {{Strong coupling in extended Horava-Lifshitz gravity}},}\
  }\href {\doibase 10.1016/j.physletb.2010.01.054} {\bibfield  {journal}
  {\bibinfo  {journal} {Phys. Lett. B}\ }\textbf {\bibinfo {volume} {685}},\
  \bibinfo {pages} {197--200} (\bibinfo {year} {2010})},\ \Eprint
  {http://arxiv.org/abs/0911.1299} {arXiv:0911.1299 [hep-th]} \BibitemShut
  {NoStop}%
\bibitem [{\citenamefont {Baumann}\ \emph {et~al.}(2011)\citenamefont
  {Baumann}, \citenamefont {Senatore},\ and\ \citenamefont
  {Zaldarriaga}}]{Baumann:2011dt}%
  \BibitemOpen
  \bibfield  {author} {\bibinfo {author} {\bibfnamefont {Daniel}\ \bibnamefont
  {Baumann}}, \bibinfo {author} {\bibfnamefont {Leonardo}\ \bibnamefont
  {Senatore}}, \ and\ \bibinfo {author} {\bibfnamefont {Matias}\ \bibnamefont
  {Zaldarriaga}},\ }\bibfield  {title} {\enquote {\bibinfo {title}
  {{Scale-Invariance and the Strong Coupling Problem}},}\ }\href {\doibase
  10.1088/1475-7516/2011/05/004} {\bibfield  {journal} {\bibinfo  {journal}
  {JCAP}\ }\textbf {\bibinfo {volume} {05}},\ \bibinfo {pages} {004} (\bibinfo
  {year} {2011})},\ \Eprint {http://arxiv.org/abs/1101.3320} {arXiv:1101.3320
  [hep-th]} \BibitemShut {NoStop}%
\bibitem [{\citenamefont {Wang}(2017)}]{Wang:2017brl}%
  \BibitemOpen
  \bibfield  {author} {\bibinfo {author} {\bibfnamefont {Anzhong}\ \bibnamefont
  {Wang}},\ }\bibfield  {title} {\enquote {\bibinfo {title} {{Ho\v{r}ava
  gravity at a Lifshitz point: A progress report}},}\ }\href {\doibase
  10.1142/S0218271817300142} {\bibfield  {journal} {\bibinfo  {journal} {Int.
  J. Mod. Phys. D}\ }\textbf {\bibinfo {volume} {26}},\ \bibinfo {pages}
  {1730014} (\bibinfo {year} {2017})},\ \Eprint
  {http://arxiv.org/abs/1701.06087} {arXiv:1701.06087 [gr-qc]} \BibitemShut
  {NoStop}%
\bibitem [{\citenamefont {Jim\'enez~Cano}(2021)}]{JimenezCano:2021rlu}%
  \BibitemOpen
  \bibfield  {author} {\bibinfo {author} {\bibfnamefont {Alejandro}\
  \bibnamefont {Jim\'enez~Cano}},\ }\emph {\bibinfo {title} {{Metric-affine
  Gauge theories of gravity. Foundations and new insights}}},\ \href@noop {}
  {Ph.D. thesis},\ \bibinfo  {school} {Granada U., Theor. Phys. Astrophys.}
  (\bibinfo {year} {2021}),\ \Eprint {http://arxiv.org/abs/2201.12847}
  {arXiv:2201.12847 [gr-qc]} \BibitemShut {NoStop}%
\bibitem [{\citenamefont {Barker}(2023)}]{Barker:2022kdk}%
  \BibitemOpen
  \bibfield  {author} {\bibinfo {author} {\bibfnamefont {W.~E.~V.}\
  \bibnamefont {Barker}},\ }\bibfield  {title} {\enquote {\bibinfo {title}
  {{Supercomputers against strong coupling in gravity with curvature and
  torsion}},}\ }\href {\doibase 10.1140/epjc/s10052-023-11179-6} {\bibfield
  {journal} {\bibinfo  {journal} {Eur. Phys. J. C}\ }\textbf {\bibinfo {volume}
  {83}},\ \bibinfo {pages} {228} (\bibinfo {year} {2023})},\ \Eprint
  {http://arxiv.org/abs/2206.00658} {arXiv:2206.00658 [gr-qc]} \BibitemShut
  {NoStop}%
\bibitem [{\citenamefont {D'Amico}\ \emph {et~al.}(2011)\citenamefont
  {D'Amico}, \citenamefont {de~Rham}, \citenamefont {Dubovsky}, \citenamefont
  {Gabadadze}, \citenamefont {Pirtskhalava},\ and\ \citenamefont
  {Tolley}}]{DAmico:2011eto}%
  \BibitemOpen
  \bibfield  {author} {\bibinfo {author} {\bibfnamefont {G.}~\bibnamefont
  {D'Amico}}, \bibinfo {author} {\bibfnamefont {C.}~\bibnamefont {de~Rham}},
  \bibinfo {author} {\bibfnamefont {S.}~\bibnamefont {Dubovsky}}, \bibinfo
  {author} {\bibfnamefont {G.}~\bibnamefont {Gabadadze}}, \bibinfo {author}
  {\bibfnamefont {D.}~\bibnamefont {Pirtskhalava}}, \ and\ \bibinfo {author}
  {\bibfnamefont {A.~J.}\ \bibnamefont {Tolley}},\ }\bibfield  {title}
  {\enquote {\bibinfo {title} {{Massive Cosmologies}},}\ }\href {\doibase
  10.1103/PhysRevD.84.124046} {\bibfield  {journal} {\bibinfo  {journal} {Phys.
  Rev. D}\ }\textbf {\bibinfo {volume} {84}},\ \bibinfo {pages} {124046}
  (\bibinfo {year} {2011})},\ \Eprint {http://arxiv.org/abs/1108.5231}
  {arXiv:1108.5231 [hep-th]} \BibitemShut {NoStop}%
\bibitem [{\citenamefont {Gumrukcuoglu}\ \emph {et~al.}(2012)\citenamefont
  {Gumrukcuoglu}, \citenamefont {Lin},\ and\ \citenamefont
  {Mukohyama}}]{Gumrukcuoglu:2012aa}%
  \BibitemOpen
  \bibfield  {author} {\bibinfo {author} {\bibfnamefont {A.~Emir}\ \bibnamefont
  {Gumrukcuoglu}}, \bibinfo {author} {\bibfnamefont {Chunshan}\ \bibnamefont
  {Lin}}, \ and\ \bibinfo {author} {\bibfnamefont {Shinji}\ \bibnamefont
  {Mukohyama}},\ }\bibfield  {title} {\enquote {\bibinfo {title} {{Anisotropic
  Friedmann-Robertson-Walker universe from nonlinear massive gravity}},}\
  }\href {\doibase 10.1016/j.physletb.2012.09.049} {\bibfield  {journal}
  {\bibinfo  {journal} {Phys. Lett. B}\ }\textbf {\bibinfo {volume} {717}},\
  \bibinfo {pages} {295--298} (\bibinfo {year} {2012})},\ \Eprint
  {http://arxiv.org/abs/1206.2723} {arXiv:1206.2723 [hep-th]} \BibitemShut
  {NoStop}%
\bibitem [{\citenamefont {Mazuet}\ \emph {et~al.}(2017)\citenamefont {Mazuet},
  \citenamefont {Mukohyama},\ and\ \citenamefont {Volkov}}]{Mazuet:2017rgq}%
  \BibitemOpen
  \bibfield  {author} {\bibinfo {author} {\bibfnamefont {Charles}\ \bibnamefont
  {Mazuet}}, \bibinfo {author} {\bibfnamefont {Shinji}\ \bibnamefont
  {Mukohyama}}, \ and\ \bibinfo {author} {\bibfnamefont {Mikhail~S.}\
  \bibnamefont {Volkov}},\ }\bibfield  {title} {\enquote {\bibinfo {title}
  {{Anisotropic deformations of spatially open cosmology in massive gravity
  theory}},}\ }\href {\doibase 10.1088/1475-7516/2017/04/039} {\bibfield
  {journal} {\bibinfo  {journal} {JCAP}\ }\textbf {\bibinfo {volume} {04}},\
  \bibinfo {pages} {039} (\bibinfo {year} {2017})},\ \Eprint
  {http://arxiv.org/abs/1702.04205} {arXiv:1702.04205 [hep-th]} \BibitemShut
  {NoStop}%
\bibitem [{\citenamefont {Beltr\'an~Jim\'enez}\ and\ \citenamefont
  {Jim\'enez-Cano}(2021)}]{BeltranJimenez:2020lee}%
  \BibitemOpen
  \bibfield  {author} {\bibinfo {author} {\bibfnamefont {Jose}\ \bibnamefont
  {Beltr\'an~Jim\'enez}}\ and\ \bibinfo {author} {\bibfnamefont {Alejandro}\
  \bibnamefont {Jim\'enez-Cano}},\ }\bibfield  {title} {\enquote {\bibinfo
  {title} {{On the strong coupling of Einsteinian Cubic Gravity and its
  generalisations}},}\ }\href {\doibase 10.1088/1475-7516/2021/01/069}
  {\bibfield  {journal} {\bibinfo  {journal} {JCAP}\ }\textbf {\bibinfo
  {volume} {01}},\ \bibinfo {pages} {069} (\bibinfo {year} {2021})},\ \Eprint
  {http://arxiv.org/abs/2009.08197} {arXiv:2009.08197 [gr-qc]} \BibitemShut
  {NoStop}%
\bibitem [{\citenamefont {Vainshtein}(1972)}]{Vainshtein:1972sx}%
  \BibitemOpen
  \bibfield  {author} {\bibinfo {author} {\bibfnamefont {A.~I.}\ \bibnamefont
  {Vainshtein}},\ }\bibfield  {title} {\enquote {\bibinfo {title} {{To the
  problem of nonvanishing gravitation mass}},}\ }\href {\doibase
  10.1016/0370-2693(72)90147-5} {\bibfield  {journal} {\bibinfo  {journal}
  {Phys. Lett. B}\ }\textbf {\bibinfo {volume} {39}},\ \bibinfo {pages}
  {393--394} (\bibinfo {year} {1972})}\BibitemShut {NoStop}%
\bibitem [{\citenamefont {Deffayet}\ \emph {et~al.}(2002)\citenamefont
  {Deffayet}, \citenamefont {Dvali}, \citenamefont {Gabadadze},\ and\
  \citenamefont {Vainshtein}}]{Deffayet:2001uk}%
  \BibitemOpen
  \bibfield  {author} {\bibinfo {author} {\bibfnamefont {Cedric}\ \bibnamefont
  {Deffayet}}, \bibinfo {author} {\bibfnamefont {G.~R.}\ \bibnamefont {Dvali}},
  \bibinfo {author} {\bibfnamefont {Gregory}\ \bibnamefont {Gabadadze}}, \ and\
  \bibinfo {author} {\bibfnamefont {Arkady~I.}\ \bibnamefont {Vainshtein}},\
  }\bibfield  {title} {\enquote {\bibinfo {title} {{Nonperturbative continuity
  in graviton mass versus perturbative discontinuity}},}\ }\href {\doibase
  10.1103/PhysRevD.65.044026} {\bibfield  {journal} {\bibinfo  {journal} {Phys.
  Rev. D}\ }\textbf {\bibinfo {volume} {65}},\ \bibinfo {pages} {044026}
  (\bibinfo {year} {2002})},\ \Eprint {http://arxiv.org/abs/hep-th/0106001}
  {arXiv:hep-th/0106001} \BibitemShut {NoStop}%
\bibitem [{\citenamefont {Deffayet}\ and\ \citenamefont
  {Rombouts}(2005)}]{Deffayet:2005ys}%
  \BibitemOpen
  \bibfield  {author} {\bibinfo {author} {\bibfnamefont {Cedric}\ \bibnamefont
  {Deffayet}}\ and\ \bibinfo {author} {\bibfnamefont {Jan-Willem}\ \bibnamefont
  {Rombouts}},\ }\bibfield  {title} {\enquote {\bibinfo {title} {{Ghosts,
  strong coupling and accidental symmetries in massive gravity}},}\ }\href
  {\doibase 10.1103/PhysRevD.72.044003} {\bibfield  {journal} {\bibinfo
  {journal} {Phys. Rev. D}\ }\textbf {\bibinfo {volume} {72}},\ \bibinfo
  {pages} {044003} (\bibinfo {year} {2005})},\ \Eprint
  {http://arxiv.org/abs/gr-qc/0505134} {arXiv:gr-qc/0505134} \BibitemShut
  {NoStop}%
\bibitem [{\citenamefont {de~Rham}(2014)}]{deRham:2014zqa}%
  \BibitemOpen
  \bibfield  {author} {\bibinfo {author} {\bibfnamefont {Claudia}\ \bibnamefont
  {de~Rham}},\ }\bibfield  {title} {\enquote {\bibinfo {title} {{Massive
  Gravity}},}\ }\href {\doibase 10.12942/lrr-2014-7} {\bibfield  {journal}
  {\bibinfo  {journal} {Living Rev. Rel.}\ }\textbf {\bibinfo {volume} {17}},\
  \bibinfo {pages} {7} (\bibinfo {year} {2014})},\ \Eprint
  {http://arxiv.org/abs/1401.4173} {arXiv:1401.4173 [hep-th]} \BibitemShut
  {NoStop}%
\bibitem [{\citenamefont {Deser}\ \emph {et~al.}(2014)\citenamefont {Deser},
  \citenamefont {Sandora}, \citenamefont {Waldron},\ and\ \citenamefont
  {Zahariade}}]{Deser:2014hga}%
  \BibitemOpen
  \bibfield  {author} {\bibinfo {author} {\bibfnamefont {S.}~\bibnamefont
  {Deser}}, \bibinfo {author} {\bibfnamefont {M.}~\bibnamefont {Sandora}},
  \bibinfo {author} {\bibfnamefont {A.}~\bibnamefont {Waldron}}, \ and\
  \bibinfo {author} {\bibfnamefont {G.}~\bibnamefont {Zahariade}},\ }\bibfield
  {title} {\enquote {\bibinfo {title} {{Covariant constraints for generic
  massive gravity and analysis of its characteristics}},}\ }\href {\doibase
  10.1103/PhysRevD.90.104043} {\bibfield  {journal} {\bibinfo  {journal} {Phys.
  Rev. D}\ }\textbf {\bibinfo {volume} {90}},\ \bibinfo {pages} {104043}
  (\bibinfo {year} {2014})},\ \Eprint {http://arxiv.org/abs/1408.0561}
  {arXiv:1408.0561 [hep-th]} \BibitemShut {NoStop}%
\bibitem [{\citenamefont {Percacci}\ and\ \citenamefont
  {Sezgin}(2020{\natexlab{b}})}]{Percacci:2020ddy}%
  \BibitemOpen
  \bibfield  {author} {\bibinfo {author} {\bibfnamefont {R.}~\bibnamefont
  {Percacci}}\ and\ \bibinfo {author} {\bibfnamefont {E.}~\bibnamefont
  {Sezgin}},\ }\bibfield  {title} {\enquote {\bibinfo {title} {{New class of
  ghost- and tachyon-free metric affine gravities}},}\ }\href {\doibase
  10.1103/PhysRevD.101.084040} {\bibfield  {journal} {\bibinfo  {journal}
  {Phys. Rev. D}\ }\textbf {\bibinfo {volume} {101}},\ \bibinfo {pages}
  {084040} (\bibinfo {year} {2020}{\natexlab{b}})},\ \Eprint
  {http://arxiv.org/abs/1912.01023} {arXiv:1912.01023 [hep-th]} \BibitemShut
  {NoStop}%
\bibitem [{\citenamefont {Piva}(2022)}]{Piva:2021nyj}%
  \BibitemOpen
  \bibfield  {author} {\bibinfo {author} {\bibfnamefont {Marco}\ \bibnamefont
  {Piva}},\ }\bibfield  {title} {\enquote {\bibinfo {title} {{Massive
  higher-spin multiplets and asymptotic freedom in quantum gravity}},}\ }\href
  {\doibase 10.1103/PhysRevD.105.045006} {\bibfield  {journal} {\bibinfo
  {journal} {Phys. Rev. D}\ }\textbf {\bibinfo {volume} {105}},\ \bibinfo
  {pages} {045006} (\bibinfo {year} {2022})},\ \Eprint
  {http://arxiv.org/abs/2110.09649} {arXiv:2110.09649 [hep-th]} \BibitemShut
  {NoStop}%
\bibitem [{\citenamefont {Iosifidis}\ \emph {et~al.}(2022)\citenamefont
  {Iosifidis}, \citenamefont {Myrzakulov}, \citenamefont {Ravera},
  \citenamefont {Yergaliyeva},\ and\ \citenamefont
  {Yerzhanov}}]{Iosifidis:2021xdx}%
  \BibitemOpen
  \bibfield  {author} {\bibinfo {author} {\bibfnamefont {Damianos}\
  \bibnamefont {Iosifidis}}, \bibinfo {author} {\bibfnamefont {Ratbay}\
  \bibnamefont {Myrzakulov}}, \bibinfo {author} {\bibfnamefont {Lucrezia}\
  \bibnamefont {Ravera}}, \bibinfo {author} {\bibfnamefont {Gulmira}\
  \bibnamefont {Yergaliyeva}}, \ and\ \bibinfo {author} {\bibfnamefont
  {Koblandy}\ \bibnamefont {Yerzhanov}},\ }\bibfield  {title} {\enquote
  {\bibinfo {title} {{Metric-Affine Vector\textendash{}Tensor correspondence
  and implications in F(R,T,Q,T,D) gravity}},}\ }\href {\doibase
  10.1016/j.dark.2022.101094} {\bibfield  {journal} {\bibinfo  {journal} {Phys.
  Dark Univ.}\ }\textbf {\bibinfo {volume} {37}},\ \bibinfo {pages} {101094}
  (\bibinfo {year} {2022})},\ \Eprint {http://arxiv.org/abs/2111.14214}
  {arXiv:2111.14214 [gr-qc]} \BibitemShut {NoStop}%
\bibitem [{\citenamefont {Jim\'enez-Cano}\ and\ \citenamefont
  {Maldonado~Torralba}(2022)}]{Jimenez-Cano:2022sds}%
  \BibitemOpen
  \bibfield  {author} {\bibinfo {author} {\bibfnamefont {Alejandro}\
  \bibnamefont {Jim\'enez-Cano}}\ and\ \bibinfo {author} {\bibfnamefont
  {Francisco~Jos\'e}\ \bibnamefont {Maldonado~Torralba}},\ }\bibfield  {title}
  {\enquote {\bibinfo {title} {{Vector stability in quadratic metric-affine
  theories}},}\ }\href {\doibase 10.1088/1475-7516/2022/09/044} {\bibfield
  {journal} {\bibinfo  {journal} {JCAP}\ }\textbf {\bibinfo {volume} {09}},\
  \bibinfo {pages} {044} (\bibinfo {year} {2022})},\ \Eprint
  {http://arxiv.org/abs/2205.05674} {arXiv:2205.05674 [gr-qc]} \BibitemShut
  {NoStop}%
\bibitem [{\citenamefont {Iosifidis}\ and\ \citenamefont
  {Pallikaris}(2023)}]{Iosifidis:2023pvz}%
  \BibitemOpen
  \bibfield  {author} {\bibinfo {author} {\bibfnamefont {Damianos}\
  \bibnamefont {Iosifidis}}\ and\ \bibinfo {author} {\bibfnamefont
  {Konstantinos}\ \bibnamefont {Pallikaris}},\ }\bibfield  {title} {\enquote
  {\bibinfo {title} {{Describing metric-affine theories anew: alternative
  frameworks, examples and solutions}},}\ }\href {\doibase
  10.1088/1475-7516/2023/05/037} {\bibfield  {journal} {\bibinfo  {journal}
  {JCAP}\ }\textbf {\bibinfo {volume} {05}},\ \bibinfo {pages} {037} (\bibinfo
  {year} {2023})},\ \Eprint {http://arxiv.org/abs/2301.11364} {arXiv:2301.11364
  [gr-qc]} \BibitemShut {NoStop}%
\bibitem [{\citenamefont {Carroll}\ and\ \citenamefont
  {Field}(1994)}]{Carroll:1994dq}%
  \BibitemOpen
  \bibfield  {author} {\bibinfo {author} {\bibfnamefont {Sean~M.}\ \bibnamefont
  {Carroll}}\ and\ \bibinfo {author} {\bibfnamefont {George~B.}\ \bibnamefont
  {Field}},\ }\bibfield  {title} {\enquote {\bibinfo {title} {{Consequences of
  propagating torsion in connection dynamic theories of gravity}},}\ }\href
  {\doibase 10.1103/PhysRevD.50.3867} {\bibfield  {journal} {\bibinfo
  {journal} {Phys. Rev. D}\ }\textbf {\bibinfo {volume} {50}},\ \bibinfo
  {pages} {3867--3873} (\bibinfo {year} {1994})},\ \Eprint
  {http://arxiv.org/abs/gr-qc/9403058} {arXiv:gr-qc/9403058} \BibitemShut
  {NoStop}%
\bibitem [{\citenamefont {Belyaev}\ and\ \citenamefont
  {Shapiro}(1998)}]{Belyaev:1997zv}%
  \BibitemOpen
  \bibfield  {author} {\bibinfo {author} {\bibfnamefont {A.~S.}\ \bibnamefont
  {Belyaev}}\ and\ \bibinfo {author} {\bibfnamefont {Ilya~L.}\ \bibnamefont
  {Shapiro}},\ }\bibfield  {title} {\enquote {\bibinfo {title} {{The Action for
  the (propagating) torsion and the limits on the torsion parameters from
  present experimental data}},}\ }\href {\doibase
  10.1016/S0370-2693(98)00258-5} {\bibfield  {journal} {\bibinfo  {journal}
  {Phys. Lett. B}\ }\textbf {\bibinfo {volume} {425}},\ \bibinfo {pages}
  {246--254} (\bibinfo {year} {1998})},\ \Eprint
  {http://arxiv.org/abs/hep-ph/9712503} {arXiv:hep-ph/9712503} \BibitemShut
  {NoStop}%
\bibitem [{\citenamefont {Vollick}(2006)}]{Vollick:2006uq}%
  \BibitemOpen
  \bibfield  {author} {\bibinfo {author} {\bibfnamefont {Dan~N.}\ \bibnamefont
  {Vollick}},\ }\bibfield  {title} {\enquote {\bibinfo {title}
  {{Einstein-Maxwell and Einstein-Proca theory from a modified gravitational
  action}},}\ }\href@noop {} {\  (\bibinfo {year} {2006})},\ \Eprint
  {http://arxiv.org/abs/gr-qc/0601016} {arXiv:gr-qc/0601016} \BibitemShut
  {NoStop}%
\bibitem [{\citenamefont {Barman}\ \emph {et~al.}(2020)\citenamefont {Barman},
  \citenamefont {Bhanja}, \citenamefont {Das},\ and\ \citenamefont
  {Maity}}]{Barman:2019mlj}%
  \BibitemOpen
  \bibfield  {author} {\bibinfo {author} {\bibfnamefont {Basabendu}\
  \bibnamefont {Barman}}, \bibinfo {author} {\bibfnamefont {Tapobroto}\
  \bibnamefont {Bhanja}}, \bibinfo {author} {\bibfnamefont {Debottam}\
  \bibnamefont {Das}}, \ and\ \bibinfo {author} {\bibfnamefont {Debaprasad}\
  \bibnamefont {Maity}},\ }\bibfield  {title} {\enquote {\bibinfo {title}
  {{Minimal model of torsion mediated dark matter}},}\ }\href {\doibase
  10.1103/PhysRevD.101.075017} {\bibfield  {journal} {\bibinfo  {journal}
  {Phys. Rev. D}\ }\textbf {\bibinfo {volume} {101}},\ \bibinfo {pages}
  {075017} (\bibinfo {year} {2020})},\ \Eprint
  {http://arxiv.org/abs/1912.09249} {arXiv:1912.09249 [hep-ph]} \BibitemShut
  {NoStop}%
\bibitem [{\citenamefont {Katanaev}(2021)}]{Katanaev:2020xdv}%
  \BibitemOpen
  \bibfield  {author} {\bibinfo {author} {\bibfnamefont {M.~O.}\ \bibnamefont
  {Katanaev}},\ }\bibfield  {title} {\enquote {\bibinfo {title} {{Gravity with
  dynamical torsion}},}\ }\href {\doibase 10.1088/1361-6382/abcbe0} {\bibfield
  {journal} {\bibinfo  {journal} {Class. Quant. Grav.}\ }\textbf {\bibinfo
  {volume} {38}},\ \bibinfo {pages} {015014} (\bibinfo {year} {2021})},\
  \Eprint {http://arxiv.org/abs/2109.09546} {arXiv:2109.09546 [gr-qc]}
  \BibitemShut {NoStop}%
\bibitem [{\citenamefont {Beltran~Jimenez}\ and\ \citenamefont
  {Koivisto}(2014)}]{BeltranJimenez:2014iie}%
  \BibitemOpen
  \bibfield  {author} {\bibinfo {author} {\bibfnamefont {Jose}\ \bibnamefont
  {Beltran~Jimenez}}\ and\ \bibinfo {author} {\bibfnamefont {Tomi~S.}\
  \bibnamefont {Koivisto}},\ }\bibfield  {title} {\enquote {\bibinfo {title}
  {{Extended Gauss-Bonnet gravities in Weyl geometry}},}\ }\href {\doibase
  10.1088/0264-9381/31/13/135002} {\bibfield  {journal} {\bibinfo  {journal}
  {Class. Quant. Grav.}\ }\textbf {\bibinfo {volume} {31}},\ \bibinfo {pages}
  {135002} (\bibinfo {year} {2014})},\ \Eprint {http://arxiv.org/abs/1402.1846}
  {arXiv:1402.1846 [gr-qc]} \BibitemShut {NoStop}%
\bibitem [{\citenamefont {Iosifidis}\ \emph {et~al.}(2018)\citenamefont
  {Iosifidis}, \citenamefont {Tsagas},\ and\ \citenamefont
  {Petkou}}]{Iosifidis:2018diy}%
  \BibitemOpen
  \bibfield  {author} {\bibinfo {author} {\bibfnamefont {Damianos}\
  \bibnamefont {Iosifidis}}, \bibinfo {author} {\bibfnamefont {Christos~G.}\
  \bibnamefont {Tsagas}}, \ and\ \bibinfo {author} {\bibfnamefont
  {Anastasios~C.}\ \bibnamefont {Petkou}},\ }\bibfield  {title} {\enquote
  {\bibinfo {title} {{Raychaudhuri equation in spacetimes with torsion and
  nonmetricity}},}\ }\href {\doibase 10.1103/PhysRevD.98.104037} {\bibfield
  {journal} {\bibinfo  {journal} {Phys. Rev. D}\ }\textbf {\bibinfo {volume}
  {98}},\ \bibinfo {pages} {104037} (\bibinfo {year} {2018})},\ \Eprint
  {http://arxiv.org/abs/1809.04992} {arXiv:1809.04992 [gr-qc]} \BibitemShut
  {NoStop}%
\bibitem [{\citenamefont {Iosifidis}\ and\ \citenamefont
  {Koivisto}(2019)}]{Iosifidis:2018zwo}%
  \BibitemOpen
  \bibfield  {author} {\bibinfo {author} {\bibfnamefont {Damianos}\
  \bibnamefont {Iosifidis}}\ and\ \bibinfo {author} {\bibfnamefont {Tomi}\
  \bibnamefont {Koivisto}},\ }\bibfield  {title} {\enquote {\bibinfo {title}
  {{Scale transformations in metric-affine geometry}},}\ }\href {\doibase
  10.3390/universe5030082} {\bibfield  {journal} {\bibinfo  {journal}
  {Universe}\ }\textbf {\bibinfo {volume} {5}},\ \bibinfo {pages} {82}
  (\bibinfo {year} {2019})},\ \Eprint {http://arxiv.org/abs/1810.12276}
  {arXiv:1810.12276 [gr-qc]} \BibitemShut {NoStop}%
\bibitem [{\citenamefont {Helpin}\ and\ \citenamefont
  {Volkov}(2020)}]{Helpin:2019vrv}%
  \BibitemOpen
  \bibfield  {author} {\bibinfo {author} {\bibfnamefont {Thomas}\ \bibnamefont
  {Helpin}}\ and\ \bibinfo {author} {\bibfnamefont {Mikhail~S.}\ \bibnamefont
  {Volkov}},\ }\bibfield  {title} {\enquote {\bibinfo {title} {{A metric-affine
  version of the Horndeski theory}},}\ }\href {\doibase
  10.1142/S0217751X20400102} {\bibfield  {journal} {\bibinfo  {journal} {Int.
  J. Mod. Phys. A}\ }\textbf {\bibinfo {volume} {35}},\ \bibinfo {pages}
  {2040010} (\bibinfo {year} {2020})},\ \Eprint
  {http://arxiv.org/abs/1911.12768} {arXiv:1911.12768 [hep-th]} \BibitemShut
  {NoStop}%
\bibitem [{\citenamefont
  {Orejuela~Garc\'\i{}a}(2020)}]{OrejuelaGarcia:2020viw}%
  \BibitemOpen
  \bibfield  {author} {\bibinfo {author} {\bibfnamefont {Jos\'e~Alberto}\
  \bibnamefont {Orejuela~Garc\'\i{}a}},\ }\emph {\bibinfo {title} {{Lovelock
  Theories as extensions to General Relativity}}},\ \href@noop {} {Ph.D.
  thesis},\ \bibinfo  {school} {U. Granada (main)} (\bibinfo {year}
  {2020})\BibitemShut {NoStop}%
\bibitem [{\citenamefont {Ghilencea}(2020)}]{Ghilencea:2020piz}%
  \BibitemOpen
  \bibfield  {author} {\bibinfo {author} {\bibfnamefont {D.~M.}\ \bibnamefont
  {Ghilencea}},\ }\bibfield  {title} {\enquote {\bibinfo {title} {{Palatini
  quadratic gravity: spontaneous breaking of gauged scale symmetry and
  inflation}},}\ }\href {\doibase 10.1140/epjc/s10052-020-08722-0} {\bibfield
  {journal} {\bibinfo  {journal} {Eur. Phys. J. C}\ }\textbf {\bibinfo {volume}
  {80}},\ \bibinfo {pages} {1147} (\bibinfo {year} {2020})},\ \Eprint
  {http://arxiv.org/abs/2003.08516} {arXiv:2003.08516 [hep-th]} \BibitemShut
  {NoStop}%
\bibitem [{\citenamefont {Beltr\'an~Jim\'enez}\ \emph
  {et~al.}(2020{\natexlab{b}})\citenamefont {Beltr\'an~Jim\'enez},
  \citenamefont {Heisenberg},\ and\ \citenamefont
  {Koivisto}}]{BeltranJimenez:2020sih}%
  \BibitemOpen
  \bibfield  {author} {\bibinfo {author} {\bibfnamefont {Jose}\ \bibnamefont
  {Beltr\'an~Jim\'enez}}, \bibinfo {author} {\bibfnamefont {Lavinia}\
  \bibnamefont {Heisenberg}}, \ and\ \bibinfo {author} {\bibfnamefont {Tomi}\
  \bibnamefont {Koivisto}},\ }\bibfield  {title} {\enquote {\bibinfo {title}
  {{The coupling of matter and spacetime geometry}},}\ }\href {\doibase
  10.1088/1361-6382/aba31b} {\bibfield  {journal} {\bibinfo  {journal} {Class.
  Quant. Grav.}\ }\textbf {\bibinfo {volume} {37}},\ \bibinfo {pages} {195013}
  (\bibinfo {year} {2020}{\natexlab{b}})},\ \Eprint
  {http://arxiv.org/abs/2004.04606} {arXiv:2004.04606 [hep-th]} \BibitemShut
  {NoStop}%
\bibitem [{\citenamefont {Xu}\ \emph {et~al.}(2020)\citenamefont {Xu},
  \citenamefont {Harko}, \citenamefont {Shahidi},\ and\ \citenamefont
  {Liang}}]{Xu:2020yeg}%
  \BibitemOpen
  \bibfield  {author} {\bibinfo {author} {\bibfnamefont {Yixin}\ \bibnamefont
  {Xu}}, \bibinfo {author} {\bibfnamefont {Tiberiu}\ \bibnamefont {Harko}},
  \bibinfo {author} {\bibfnamefont {Shahab}\ \bibnamefont {Shahidi}}, \ and\
  \bibinfo {author} {\bibfnamefont {Shi-Dong}\ \bibnamefont {Liang}},\
  }\bibfield  {title} {\enquote {\bibinfo {title} {{Weyl type $f(Q,T)$ gravity,
  and its cosmological implications}},}\ }\href {\doibase
  10.1140/epjc/s10052-020-8023-6} {\bibfield  {journal} {\bibinfo  {journal}
  {Eur. Phys. J. C}\ }\textbf {\bibinfo {volume} {80}},\ \bibinfo {pages} {449}
  (\bibinfo {year} {2020})},\ \Eprint {http://arxiv.org/abs/2005.04025}
  {arXiv:2005.04025 [gr-qc]} \BibitemShut {NoStop}%
\bibitem [{\citenamefont {Yang}\ \emph {et~al.}(2021)\citenamefont {Yang},
  \citenamefont {Shahidi}, \citenamefont {Harko},\ and\ \citenamefont
  {Liang}}]{Yang:2021fjy}%
  \BibitemOpen
  \bibfield  {author} {\bibinfo {author} {\bibfnamefont {Jin-Zhao}\
  \bibnamefont {Yang}}, \bibinfo {author} {\bibfnamefont {Shahab}\ \bibnamefont
  {Shahidi}}, \bibinfo {author} {\bibfnamefont {Tiberiu}\ \bibnamefont
  {Harko}}, \ and\ \bibinfo {author} {\bibfnamefont {Shi-Dong}\ \bibnamefont
  {Liang}},\ }\bibfield  {title} {\enquote {\bibinfo {title} {{Geodesic
  deviation, Raychaudhuri equation, Newtonian limit, and tidal forces in
  Weyl-type $f(Q,T)$ gravity}},}\ }\href {\doibase
  10.1140/epjc/s10052-021-08910-6} {\bibfield  {journal} {\bibinfo  {journal}
  {Eur. Phys. J. C}\ }\textbf {\bibinfo {volume} {81}},\ \bibinfo {pages} {111}
  (\bibinfo {year} {2021})},\ \Eprint {http://arxiv.org/abs/2101.09956}
  {arXiv:2101.09956 [gr-qc]} \BibitemShut {NoStop}%
\bibitem [{\citenamefont {Quiros}(2022)}]{Quiros:2021eju}%
  \BibitemOpen
  \bibfield  {author} {\bibinfo {author} {\bibfnamefont {Israel}\ \bibnamefont
  {Quiros}},\ }\bibfield  {title} {\enquote {\bibinfo {title} {{Nonmetricity
  theories and aspects of gauge symmetry}},}\ }\href {\doibase
  10.1103/PhysRevD.105.104060} {\bibfield  {journal} {\bibinfo  {journal}
  {Phys. Rev. D}\ }\textbf {\bibinfo {volume} {105}},\ \bibinfo {pages}
  {104060} (\bibinfo {year} {2022})},\ \Eprint
  {http://arxiv.org/abs/2111.05490} {arXiv:2111.05490 [gr-qc]} \BibitemShut
  {NoStop}%
\bibitem [{\citenamefont {Quiros}(2023)}]{Quiros:2022uns}%
  \BibitemOpen
  \bibfield  {author} {\bibinfo {author} {\bibfnamefont {Israel}\ \bibnamefont
  {Quiros}},\ }\bibfield  {title} {\enquote {\bibinfo {title}
  {{Phenomenological signatures of gauge invariant theories of gravity with
  vectorial nonmetricity}},}\ }\href {\doibase 10.1103/PhysRevD.107.104028}
  {\bibfield  {journal} {\bibinfo  {journal} {Phys. Rev. D}\ }\textbf {\bibinfo
  {volume} {107}},\ \bibinfo {pages} {104028} (\bibinfo {year} {2023})},\
  \Eprint {http://arxiv.org/abs/2208.10048} {arXiv:2208.10048 [gr-qc]}
  \BibitemShut {NoStop}%
\bibitem [{\citenamefont {Yang}\ \emph {et~al.}(2022)\citenamefont {Yang},
  \citenamefont {Shahidi},\ and\ \citenamefont {Harko}}]{Yang:2022icz}%
  \BibitemOpen
  \bibfield  {author} {\bibinfo {author} {\bibfnamefont {Jin-Zhao}\
  \bibnamefont {Yang}}, \bibinfo {author} {\bibfnamefont {Shahab}\ \bibnamefont
  {Shahidi}}, \ and\ \bibinfo {author} {\bibfnamefont {Tiberiu}\ \bibnamefont
  {Harko}},\ }\bibfield  {title} {\enquote {\bibinfo {title} {{Black hole
  solutions in the quadratic Weyl conformal geometric theory of gravity}},}\
  }\href {\doibase 10.1140/epjc/s10052-022-11131-0} {\bibfield  {journal}
  {\bibinfo  {journal} {Eur. Phys. J. C}\ }\textbf {\bibinfo {volume} {82}},\
  \bibinfo {pages} {1171} (\bibinfo {year} {2022})},\ \Eprint
  {http://arxiv.org/abs/2212.05542} {arXiv:2212.05542 [gr-qc]} \BibitemShut
  {NoStop}%
\bibitem [{\citenamefont {Burikham}\ \emph {et~al.}(2023)\citenamefont
  {Burikham}, \citenamefont {Harko}, \citenamefont {Pimsamarn},\ and\
  \citenamefont {Shahidi}}]{Burikham:2023bil}%
  \BibitemOpen
  \bibfield  {author} {\bibinfo {author} {\bibfnamefont {Piyabut}\ \bibnamefont
  {Burikham}}, \bibinfo {author} {\bibfnamefont {Tiberiu}\ \bibnamefont
  {Harko}}, \bibinfo {author} {\bibfnamefont {Kulapant}\ \bibnamefont
  {Pimsamarn}}, \ and\ \bibinfo {author} {\bibfnamefont {Shahab}\ \bibnamefont
  {Shahidi}},\ }\bibfield  {title} {\enquote {\bibinfo {title} {{Dark matter as
  a Weyl geometric effect}},}\ }\href {\doibase 10.1103/PhysRevD.107.064008}
  {\bibfield  {journal} {\bibinfo  {journal} {Phys. Rev. D}\ }\textbf {\bibinfo
  {volume} {107}},\ \bibinfo {pages} {064008} (\bibinfo {year} {2023})},\
  \Eprint {http://arxiv.org/abs/2302.08289} {arXiv:2302.08289 [gr-qc]}
  \BibitemShut {NoStop}%
\bibitem [{\citenamefont {Haghani}\ and\ \citenamefont
  {Harko}(2023)}]{Haghani:2023nrm}%
  \BibitemOpen
  \bibfield  {author} {\bibinfo {author} {\bibfnamefont {Zahra}\ \bibnamefont
  {Haghani}}\ and\ \bibinfo {author} {\bibfnamefont {Tiberiu}\ \bibnamefont
  {Harko}},\ }\bibfield  {title} {\enquote {\bibinfo {title} {{Compact stellar
  structures in Weyl geometric gravity}},}\ }\href {\doibase
  10.1103/PhysRevD.107.064068} {\bibfield  {journal} {\bibinfo  {journal}
  {Phys. Rev. D}\ }\textbf {\bibinfo {volume} {107}},\ \bibinfo {pages}
  {064068} (\bibinfo {year} {2023})},\ \Eprint
  {http://arxiv.org/abs/2303.10339} {arXiv:2303.10339 [gr-qc]} \BibitemShut
  {NoStop}%
\bibitem [{\citenamefont {Aringazin}\ and\ \citenamefont
  {Mikhailov}(1991)}]{Aringazin:1991}%
  \BibitemOpen
  \bibfield  {author} {\bibinfo {author} {\bibfnamefont {A~K}\ \bibnamefont
  {Aringazin}}\ and\ \bibinfo {author} {\bibfnamefont {A~L}\ \bibnamefont
  {Mikhailov}},\ }\bibfield  {title} {\enquote {\bibinfo {title} {Matter fields
  in spacetime with vector nonmetricity},}\ }\href {\doibase
  10.1088/0264-9381/8/9/004} {\bibfield  {journal} {\bibinfo  {journal} {Class.
  Quant. Grav.}\ }\textbf {\bibinfo {volume} {8}},\ \bibinfo {pages} {1685}
  (\bibinfo {year} {1991})}\BibitemShut {NoStop}%
\bibitem [{\citenamefont {Vitagliano}\ \emph {et~al.}(2010)\citenamefont
  {Vitagliano}, \citenamefont {Sotiriou},\ and\ \citenamefont
  {Liberati}}]{Vitagliano:2010pq}%
  \BibitemOpen
  \bibfield  {author} {\bibinfo {author} {\bibfnamefont {Vincenzo}\
  \bibnamefont {Vitagliano}}, \bibinfo {author} {\bibfnamefont {Thomas~P.}\
  \bibnamefont {Sotiriou}}, \ and\ \bibinfo {author} {\bibfnamefont {Stefano}\
  \bibnamefont {Liberati}},\ }\bibfield  {title} {\enquote {\bibinfo {title}
  {{The dynamics of generalized Palatini Theories of Gravity}},}\ }\href
  {\doibase 10.1103/PhysRevD.82.084007} {\bibfield  {journal} {\bibinfo
  {journal} {Phys. Rev. D}\ }\textbf {\bibinfo {volume} {82}},\ \bibinfo
  {pages} {084007} (\bibinfo {year} {2010})},\ \Eprint
  {http://arxiv.org/abs/1007.3937} {arXiv:1007.3937 [gr-qc]} \BibitemShut
  {NoStop}%
\bibitem [{\citenamefont {Beltran~Jimenez}\ \emph {et~al.}(2016)\citenamefont
  {Beltran~Jimenez}, \citenamefont {Heisenberg},\ and\ \citenamefont
  {Koivisto}}]{BeltranJimenez:2016wxw}%
  \BibitemOpen
  \bibfield  {author} {\bibinfo {author} {\bibfnamefont {Jose}\ \bibnamefont
  {Beltran~Jimenez}}, \bibinfo {author} {\bibfnamefont {Lavinia}\ \bibnamefont
  {Heisenberg}}, \ and\ \bibinfo {author} {\bibfnamefont {Tomi~S.}\
  \bibnamefont {Koivisto}},\ }\bibfield  {title} {\enquote {\bibinfo {title}
  {{Cosmology for quadratic gravity in generalized Weyl geometry}},}\ }\href
  {\doibase 10.1088/1475-7516/2016/04/046} {\bibfield  {journal} {\bibinfo
  {journal} {JCAP}\ }\textbf {\bibinfo {volume} {04}},\ \bibinfo {pages} {046}
  (\bibinfo {year} {2016})},\ \Eprint {http://arxiv.org/abs/1602.07287}
  {arXiv:1602.07287 [hep-th]} \BibitemShut {NoStop}%
\bibitem [{\citenamefont {Iosifidis}(2019)}]{Iosifidis:2018jwu}%
  \BibitemOpen
  \bibfield  {author} {\bibinfo {author} {\bibfnamefont {Damianos}\
  \bibnamefont {Iosifidis}},\ }\bibfield  {title} {\enquote {\bibinfo {title}
  {{Exactly Solvable Connections in Metric-Affine Gravity}},}\ }\href {\doibase
  10.1088/1361-6382/ab0be2} {\bibfield  {journal} {\bibinfo  {journal} {Class.
  Quant. Grav.}\ }\textbf {\bibinfo {volume} {36}},\ \bibinfo {pages} {085001}
  (\bibinfo {year} {2019})},\ \Eprint {http://arxiv.org/abs/1812.04031}
  {arXiv:1812.04031 [gr-qc]} \BibitemShut {NoStop}%
\bibitem [{\citenamefont {{Barker}}\ and\ \citenamefont
  {{Zell}}({\natexlab{a}})}]{FieldEquations}%
  \BibitemOpen
  \bibfield  {author} {\bibinfo {author} {\bibfnamefont {W.}~\bibnamefont
  {{Barker}}}\ and\ \bibinfo {author} {\bibfnamefont {S.}~\bibnamefont
  {{Zell}}},\ }\href@noop {} {\emph {\bibinfo {title} {{Supplemental materials:
  Field equations}}}},\ \bibinfo {note} {{See Supplemental Material at
  \href{https://github.com/wevbarker/SupplementalMaterials-2306}{www.github.com/wevbarker/SupplementalMaterials-2306}.}}\BibitemShut
  {Stop}%
\bibitem [{\citenamefont {Henneaux}\ and\ \citenamefont
  {Teitelboim}(1986)}]{Henneaux:1986ht}%
  \BibitemOpen
  \bibfield  {author} {\bibinfo {author} {\bibfnamefont {M.}~\bibnamefont
  {Henneaux}}\ and\ \bibinfo {author} {\bibfnamefont {C.}~\bibnamefont
  {Teitelboim}},\ }\bibfield  {title} {\enquote {\bibinfo {title} {{P FORM
  ELECTRODYNAMICS}},}\ }\href {\doibase 10.1007/BF01889624} {\bibfield
  {journal} {\bibinfo  {journal} {Found. Phys.}\ }\textbf {\bibinfo {volume}
  {16}},\ \bibinfo {pages} {593--617} (\bibinfo {year} {1986})}\BibitemShut
  {NoStop}%
\bibitem [{\citenamefont {{Barker}}\ and\ \citenamefont
  {{Zell}}({\natexlab{b}})}]{NonTrivialMultipliers}%
  \BibitemOpen
  \bibfield  {author} {\bibinfo {author} {\bibfnamefont {W.}~\bibnamefont
  {{Barker}}}\ and\ \bibinfo {author} {\bibfnamefont {S.}~\bibnamefont
  {{Zell}}},\ }\href@noop {} {\emph {\bibinfo {title} {{Supplemental materials:
  Multipliers are not trivial}}}},\ \bibinfo {note} {{See Supplemental Material
  at
  \href{https://github.com/wevbarker/SupplementalMaterials-2306}{www.github.com/wevbarker/SupplementalMaterials-2306}.}}\BibitemShut
  {Stop}%
\bibitem [{\citenamefont {Beltr\'an~Jim\'enez}\ \emph
  {et~al.}(2018)\citenamefont {Beltr\'an~Jim\'enez}, \citenamefont
  {Heisenberg},\ and\ \citenamefont {Koivisto}}]{BeltranJimenez:2018vdo}%
  \BibitemOpen
  \bibfield  {author} {\bibinfo {author} {\bibfnamefont {Jose}\ \bibnamefont
  {Beltr\'an~Jim\'enez}}, \bibinfo {author} {\bibfnamefont {Lavinia}\
  \bibnamefont {Heisenberg}}, \ and\ \bibinfo {author} {\bibfnamefont
  {Tomi~S.}\ \bibnamefont {Koivisto}},\ }\bibfield  {title} {\enquote {\bibinfo
  {title} {{Teleparallel Palatini theories}},}\ }\href {\doibase
  10.1088/1475-7516/2018/08/039} {\bibfield  {journal} {\bibinfo  {journal}
  {JCAP}\ }\textbf {\bibinfo {volume} {08}},\ \bibinfo {pages} {039} (\bibinfo
  {year} {2018})},\ \Eprint {http://arxiv.org/abs/1803.10185} {arXiv:1803.10185
  [gr-qc]} \BibitemShut {NoStop}%
\bibitem [{\citenamefont {Anderson}\ and\ \citenamefont
  {Bergmann}(1951)}]{Anderson:1951ta}%
  \BibitemOpen
  \bibfield  {author} {\bibinfo {author} {\bibfnamefont {James~L.}\
  \bibnamefont {Anderson}}\ and\ \bibinfo {author} {\bibfnamefont {Peter~G.}\
  \bibnamefont {Bergmann}},\ }\bibfield  {title} {\enquote {\bibinfo {title}
  {{Constraints in covariant field theories}},}\ }\href {\doibase
  10.1103/PhysRev.83.1018} {\bibfield  {journal} {\bibinfo  {journal} {Phys.
  Rev.}\ }\textbf {\bibinfo {volume} {83}},\ \bibinfo {pages} {1018--1025}
  (\bibinfo {year} {1951})}\BibitemShut {NoStop}%
\bibitem [{\citenamefont {Bergmann}\ and\ \citenamefont
  {Goldberg}(1955)}]{Bergmann:1954tc}%
  \BibitemOpen
  \bibfield  {author} {\bibinfo {author} {\bibfnamefont {Peter~G.}\
  \bibnamefont {Bergmann}}\ and\ \bibinfo {author} {\bibfnamefont {Irwin}\
  \bibnamefont {Goldberg}},\ }\bibfield  {title} {\enquote {\bibinfo {title}
  {{Dirac bracket transformations in phase space}},}\ }\href {\doibase
  10.1103/PhysRev.98.531} {\bibfield  {journal} {\bibinfo  {journal} {Phys.
  Rev.}\ }\textbf {\bibinfo {volume} {98}},\ \bibinfo {pages} {531--538}
  (\bibinfo {year} {1955})}\BibitemShut {NoStop}%
\bibitem [{\citenamefont {Castellani}(1982)}]{Castellani:1981us}%
  \BibitemOpen
  \bibfield  {author} {\bibinfo {author} {\bibfnamefont {Leonardo}\
  \bibnamefont {Castellani}},\ }\bibfield  {title} {\enquote {\bibinfo {title}
  {{Symmetries in Constrained Hamiltonian Systems}},}\ }\href {\doibase
  10.1016/0003-4916(82)90031-8} {\bibfield  {journal} {\bibinfo  {journal}
  {Annals Phys.}\ }\textbf {\bibinfo {volume} {143}},\ \bibinfo {pages} {357}
  (\bibinfo {year} {1982})}\BibitemShut {NoStop}%
\bibitem [{\citenamefont {Henneaux}\ and\ \citenamefont
  {Teitelboim}(1992)}]{Henneaux:1992ig}%
  \BibitemOpen
  \bibfield  {author} {\bibinfo {author} {\bibfnamefont {M.}~\bibnamefont
  {Henneaux}}\ and\ \bibinfo {author} {\bibfnamefont {C.}~\bibnamefont
  {Teitelboim}},\ }\href {\doibase https://doi.org/10.1515/9780691213866}
  {\emph {\bibinfo {title} {{Quantization of gauge systems}}}}\ (\bibinfo
  {publisher} {Princeton University Press},\ \bibinfo {address} {Princeton},\
  \bibinfo {year} {1992})\BibitemShut {NoStop}%
\bibitem [{\citenamefont {Golovnev}(2023)}]{Golovnev:2022rui}%
  \BibitemOpen
  \bibfield  {author} {\bibinfo {author} {\bibfnamefont {Alexey}\ \bibnamefont
  {Golovnev}},\ }\bibfield  {title} {\enquote {\bibinfo {title} {{On the Role
  of Constraints and Degrees of Freedom in the Hamiltonian Formalism}},}\
  }\href {\doibase 10.3390/universe9020101} {\bibfield  {journal} {\bibinfo
  {journal} {Universe}\ }\textbf {\bibinfo {volume} {9}},\ \bibinfo {pages}
  {101} (\bibinfo {year} {2023})},\ \Eprint {http://arxiv.org/abs/2212.11260}
  {arXiv:2212.11260 [hep-th]} \BibitemShut {NoStop}%
\bibitem [{\citenamefont {Kopczynski}(1982)}]{Kopczynski:1982}%
  \BibitemOpen
  \bibfield  {author} {\bibinfo {author} {\bibfnamefont {W}~\bibnamefont
  {Kopczynski}},\ }\bibfield  {title} {\enquote {\bibinfo {title} {Problems
  with metric-teleparallel theories of gravitation},}\ }\href {\doibase
  10.1088/0305-4470/15/2/020} {\bibfield  {journal} {\bibinfo  {journal}
  {Journal of Physics A: Mathematical and General}\ }\textbf {\bibinfo {volume}
  {15}},\ \bibinfo {pages} {493--506} (\bibinfo {year} {1982})}\BibitemShut
  {NoStop}%
\bibitem [{\citenamefont {Boyarsky}\ \emph {et~al.}(2019)\citenamefont
  {Boyarsky}, \citenamefont {Drewes}, \citenamefont {Lasserre}, \citenamefont
  {Mertens},\ and\ \citenamefont {Ruchayskiy}}]{Boyarsky:2018tvu}%
  \BibitemOpen
  \bibfield  {author} {\bibinfo {author} {\bibfnamefont {A.}~\bibnamefont
  {Boyarsky}}, \bibinfo {author} {\bibfnamefont {M.}~\bibnamefont {Drewes}},
  \bibinfo {author} {\bibfnamefont {T.}~\bibnamefont {Lasserre}}, \bibinfo
  {author} {\bibfnamefont {S.}~\bibnamefont {Mertens}}, \ and\ \bibinfo
  {author} {\bibfnamefont {O.}~\bibnamefont {Ruchayskiy}},\ }\bibfield  {title}
  {\enquote {\bibinfo {title} {{Sterile neutrino Dark Matter}},}\ }\href
  {\doibase 10.1016/j.ppnp.2018.07.004} {\bibfield  {journal} {\bibinfo
  {journal} {Prog. Part. Nucl. Phys.}\ }\textbf {\bibinfo {volume} {104}},\
  \bibinfo {pages} {1--45} (\bibinfo {year} {2019})},\ \Eprint
  {http://arxiv.org/abs/1807.07938} {arXiv:1807.07938 [hep-ph]} \BibitemShut
  {NoStop}%
\bibitem [{\citenamefont {Shaposhnikov}\ \emph {et~al.}(2020)\citenamefont
  {Shaposhnikov}, \citenamefont {Shkerin}, \citenamefont {Timiryasov},\ and\
  \citenamefont {Zell}}]{Shaposhnikov:2020frq}%
  \BibitemOpen
  \bibfield  {author} {\bibinfo {author} {\bibfnamefont {Mikhail}\ \bibnamefont
  {Shaposhnikov}}, \bibinfo {author} {\bibfnamefont {Andrey}\ \bibnamefont
  {Shkerin}}, \bibinfo {author} {\bibfnamefont {Inar}\ \bibnamefont
  {Timiryasov}}, \ and\ \bibinfo {author} {\bibfnamefont {Sebastian}\
  \bibnamefont {Zell}},\ }\bibfield  {title} {\enquote {\bibinfo {title}
  {{Einstein-Cartan gravity, matter, and scale-invariant generalization~}},}\
  }\href {\doibase 10.1007/JHEP08(2021)162} {\bibfield  {journal} {\bibinfo
  {journal} {JHEP}\ }\textbf {\bibinfo {volume} {10}},\ \bibinfo {pages} {177}
  (\bibinfo {year} {2020})},\ \Eprint {http://arxiv.org/abs/2007.16158}
  {arXiv:2007.16158 [hep-th]} \BibitemShut {NoStop}%
\bibitem [{\citenamefont {Karananas}\ \emph {et~al.}(2021)\citenamefont
  {Karananas}, \citenamefont {Shaposhnikov}, \citenamefont {Shkerin},\ and\
  \citenamefont {Zell}}]{Karananas:2021zkl}%
  \BibitemOpen
  \bibfield  {author} {\bibinfo {author} {\bibfnamefont {Georgios~K.}\
  \bibnamefont {Karananas}}, \bibinfo {author} {\bibfnamefont {Mikhail}\
  \bibnamefont {Shaposhnikov}}, \bibinfo {author} {\bibfnamefont {Andrey}\
  \bibnamefont {Shkerin}}, \ and\ \bibinfo {author} {\bibfnamefont {Sebastian}\
  \bibnamefont {Zell}},\ }\bibfield  {title} {\enquote {\bibinfo {title}
  {{Matter matters in Einstein-Cartan gravity}},}\ }\href {\doibase
  10.1103/PhysRevD.104.064036} {\bibfield  {journal} {\bibinfo  {journal}
  {Phys. Rev. D}\ }\textbf {\bibinfo {volume} {104}},\ \bibinfo {pages}
  {064036} (\bibinfo {year} {2021})},\ \Eprint
  {http://arxiv.org/abs/2106.13811} {arXiv:2106.13811 [hep-th]} \BibitemShut
  {NoStop}%
\bibitem [{\citenamefont {Garzilli}\ \emph {et~al.}(2019)\citenamefont
  {Garzilli}, \citenamefont {Magalich}, \citenamefont {Theuns}, \citenamefont
  {Frenk}, \citenamefont {Weniger}, \citenamefont {Ruchayskiy},\ and\
  \citenamefont {Boyarsky}}]{Garzilli:2018jqh}%
  \BibitemOpen
  \bibfield  {author} {\bibinfo {author} {\bibfnamefont {Antonella}\
  \bibnamefont {Garzilli}}, \bibinfo {author} {\bibfnamefont {Andrii}\
  \bibnamefont {Magalich}}, \bibinfo {author} {\bibfnamefont {Tom}\
  \bibnamefont {Theuns}}, \bibinfo {author} {\bibfnamefont {Carlos~S.}\
  \bibnamefont {Frenk}}, \bibinfo {author} {\bibfnamefont {Christoph}\
  \bibnamefont {Weniger}}, \bibinfo {author} {\bibfnamefont {Oleg}\
  \bibnamefont {Ruchayskiy}}, \ and\ \bibinfo {author} {\bibfnamefont {Alexey}\
  \bibnamefont {Boyarsky}},\ }\bibfield  {title} {\enquote {\bibinfo {title}
  {{The Lyman-$\alpha$ forest as a diagnostic of the nature of the dark
  matter}},}\ }\href {\doibase 10.1093/mnras/stz2188} {\bibfield  {journal}
  {\bibinfo  {journal} {Mon. Not. Roy. Astron. Soc.}\ }\textbf {\bibinfo
  {volume} {489}},\ \bibinfo {pages} {3456--3471} (\bibinfo {year} {2019})},\
  \Eprint {http://arxiv.org/abs/1809.06585} {arXiv:1809.06585 [astro-ph.CO]}
  \BibitemShut {NoStop}%
\bibitem [{\citenamefont {Garzilli}\ \emph {et~al.}(2021)\citenamefont
  {Garzilli}, \citenamefont {Magalich}, \citenamefont {Ruchayskiy},\ and\
  \citenamefont {Boyarsky}}]{Garzilli:2019qki}%
  \BibitemOpen
  \bibfield  {author} {\bibinfo {author} {\bibfnamefont {Antonella}\
  \bibnamefont {Garzilli}}, \bibinfo {author} {\bibfnamefont {Andrii}\
  \bibnamefont {Magalich}}, \bibinfo {author} {\bibfnamefont {Oleg}\
  \bibnamefont {Ruchayskiy}}, \ and\ \bibinfo {author} {\bibfnamefont {Alexey}\
  \bibnamefont {Boyarsky}},\ }\bibfield  {title} {\enquote {\bibinfo {title}
  {{How to constrain warm dark matter with the Lyman-$\alpha$ forest}},}\
  }\href {\doibase 10.1093/mnras/stab192} {\bibfield  {journal} {\bibinfo
  {journal} {Mon. Not. Roy. Astron. Soc.}\ }\textbf {\bibinfo {volume} {502}},\
  \bibinfo {pages} {2356--2363} (\bibinfo {year} {2021})},\ \Eprint
  {http://arxiv.org/abs/1912.09397} {arXiv:1912.09397 [astro-ph.CO]}
  \BibitemShut {NoStop}%
\bibitem [{\citenamefont {Rasanen}(2019)}]{Rasanen:2018ihz}%
  \BibitemOpen
  \bibfield  {author} {\bibinfo {author} {\bibfnamefont {Syksy}\ \bibnamefont
  {Rasanen}},\ }\bibfield  {title} {\enquote {\bibinfo {title} {{Higgs
  inflation in the Palatini formulation with kinetic terms for the metric}},}\
  }\href {\doibase 10.21105/astro.1811.09514} {\bibfield  {journal} {\bibinfo
  {journal} {Open J. Astrophys.}\ }\textbf {\bibinfo {volume} {2}},\ \bibinfo
  {pages} {1} (\bibinfo {year} {2019})},\ \Eprint
  {http://arxiv.org/abs/1811.09514} {arXiv:1811.09514 [gr-qc]} \BibitemShut
  {NoStop}%
\bibitem [{\citenamefont {Raatikainen}\ and\ \citenamefont
  {Rasanen}(2019)}]{Raatikainen:2019qey}%
  \BibitemOpen
  \bibfield  {author} {\bibinfo {author} {\bibfnamefont {Sami}\ \bibnamefont
  {Raatikainen}}\ and\ \bibinfo {author} {\bibfnamefont {Syksy}\ \bibnamefont
  {Rasanen}},\ }\bibfield  {title} {\enquote {\bibinfo {title} {{Higgs
  inflation and teleparallel gravity}},}\ }\href {\doibase
  10.1088/1475-7516/2019/12/021} {\bibfield  {journal} {\bibinfo  {journal}
  {JCAP}\ }\textbf {\bibinfo {volume} {12}},\ \bibinfo {pages} {021} (\bibinfo
  {year} {2019})},\ \Eprint {http://arxiv.org/abs/1910.03488} {arXiv:1910.03488
  [gr-qc]} \BibitemShut {NoStop}%
\bibitem [{\citenamefont {Battista}\ and\ \citenamefont
  {De~Falco}(2021)}]{Battista:2021rlh}%
  \BibitemOpen
  \bibfield  {author} {\bibinfo {author} {\bibfnamefont {Emmanuele}\
  \bibnamefont {Battista}}\ and\ \bibinfo {author} {\bibfnamefont {Vittorio}\
  \bibnamefont {De~Falco}},\ }\bibfield  {title} {\enquote {\bibinfo {title}
  {{First post-Newtonian generation of gravitational waves in Einstein-Cartan
  theory}},}\ }\href {\doibase 10.1103/PhysRevD.104.084067} {\bibfield
  {journal} {\bibinfo  {journal} {Phys. Rev. D}\ }\textbf {\bibinfo {volume}
  {104}},\ \bibinfo {pages} {084067} (\bibinfo {year} {2021})},\ \Eprint
  {http://arxiv.org/abs/2109.01384} {arXiv:2109.01384 [gr-qc]} \BibitemShut
  {NoStop}%
\bibitem [{\citenamefont {Melichev}\ and\ \citenamefont
  {Percacci}(2023)}]{Melichev:2023lwj}%
  \BibitemOpen
  \bibfield  {author} {\bibinfo {author} {\bibfnamefont {Oleg}\ \bibnamefont
  {Melichev}}\ and\ \bibinfo {author} {\bibfnamefont {Roberto}\ \bibnamefont
  {Percacci}},\ }\bibfield  {title} {\enquote {\bibinfo {title} {{On the
  renormalization of Poincar\'e gauge theories}},}\ }\href@noop {} {\
  (\bibinfo {year} {2023})},\ \Eprint {http://arxiv.org/abs/2307.02336}
  {arXiv:2307.02336 [hep-th]} \BibitemShut {NoStop}%
\bibitem [{\citenamefont {{Barker}}\ and\ \citenamefont
  {{Zell}}({\natexlab{c}})}]{HamiltonianAnalysis}%
  \BibitemOpen
  \bibfield  {author} {\bibinfo {author} {\bibfnamefont {W.}~\bibnamefont
  {{Barker}}}\ and\ \bibinfo {author} {\bibfnamefont {S.}~\bibnamefont
  {{Zell}}},\ }\href@noop {} {\emph {\bibinfo {title} {{Supplemental materials:
  Hamiltonian analysis}}}},\ \bibinfo {note} {{See Supplemental Material at
  \href{https://github.com/wevbarker/SupplementalMaterials-2306}{www.github.com/wevbarker/SupplementalMaterials-2306}.}}\BibitemShut
  {Stop}%
\end{thebibliography}%

\end{document}